\documentclass{aa}  
\usepackage[colorlinks=true, allcolors=blue]{hyperref}
\usepackage{stfloats}
\usepackage{booktabs} 
\usepackage{hyperref}
\usepackage{natbib}
\usepackage{graphicx}
\usepackage{subcaption}
\usepackage[utf8]{inputenc}
\usepackage{txfonts}
\usepackage{amsmath}
\DeclareMathOperator{\logit}{logit}
\usepackage[dvipsnames]{xcolor}
\definecolor{SFRBlue}{RGB}{00,102,205}
\usepackage{caption} 
\authorrunning{Kumar et al.}

\usepackage{xcolor}

\begin{document} 
   \title{Orbital decomposition and evolution of \texttt{LEGA-C}-like galaxies in \texttt{IllustrisTNG} from $z=1$ to $z=0$}
   
   \author{Prem Kumar\thanks{Email: prem.kumar@univie.ac.at}\inst{1}
          \and 
          Alina Boecker\inst{1}
          \and
          Glenn van de Ven\inst{1}
          \and 
          Stefanie Reiter\inst{1}
          \and
          Ling Zhu\inst{2}
          \and
          Annalisa Pillepich\inst{3}
          \and
          Prashin Jethwa\inst{1}
          \and
          Sabine Thater\inst{1}
          }
   \institute{Institute for Astronomy (IfA), University of Vienna,
              T\"urkenschanzstrasse 17, A-1180 Vienna, Austria
              \and
              Shanghai Astronomical Observatory, Chinese Academy of Sciences, 80 Nandan Road, Shanghai 200030, China
              \and
              Max-Planck-Institut f\"ur Astronomie, K\"onigstuhl 17, D-69117 Heidelberg, Germany
              }

   \date{Received September 15, 1996; accepted March 16, 1997}

 
  \abstract
   {Stellar orbits encode the dynamical history of galaxies and provide a physically motivated link between galaxy structures and assembly processes. Although orbit-based decompositions are well established in the Local Universe, their evolution across cosmic time, especially at higher redshifts, remains poorly explored.}
   {We investigate the orbital structure of \texttt{LEGA-C-like} galaxies in the \texttt{TNG100} simulation from $z=1$ to $z=0$ to identify how stellar mass, star formation activity, environment, merger history, gas content, and ex situ stellar mass growth shape their orbital transformation.}
   {We decomposed the stellar orbital structure of simulated galaxies into cold, warm, hot, and counter-rotating (CR) components using time-averaged orbital circularities. This provides a theoretical basis for future comparisons with Schwarzschild-based orbital decompositions of observed galaxies. We analyzed the galaxy populations at $z=1$ and $z=0$, then tracked the descendants of individual $z=1$ galaxies to quantify their orbital evolution over this time.}
  {We find a strong mass dependence of orbital structure at $z=0$ as well as at $z=1$. Galaxies below $\log(M_\star/M_\odot)\sim11.2$ have a combined warm and cold orbital fraction of $\sim70\%$, whereas more massive systems have a hot-orbit fraction of $\sim60\%$. These trends are consistent with assembly pathways driven predominantly by wet and dry major mergers, respectively. Tracking the same galaxies to $z=0$ reveals a systematic, but mass-dependent, evolution toward lower rotational support: At both low and high stellar masses, the hot-orbit fraction increases, whereas the relative change is greater in low-mass galaxies. The survival of cold structure depends strongly on the star formation state at $z=1$; star-forming progenitors retain substantially larger cold fractions than the quiescent progenitors. At both redshifts, the orbital fractions show a rise (decline) in hot and CR  (cold and warm) orbits with the ex situ stellar mass fraction at a fixed stellar mass.}
  {These results provide the first evolutionary framework for interpreting orbit-based decompositions of galaxies at the \texttt{LEGA-C} epoch. Stellar orbital fractions quantify the morphological and dynamical transformation of galaxies and show systematic evolutionary trends at the population level. Individual galaxies, however, follow diverse evolutionary pathways. Comparisons to future Schwarzschild-based decompositions of observed high-redshift galaxies can use these results to interpret the internal physical structure of galaxy populations and constrain theoretical models of galaxy evolution.}


   \maketitle
%
\section{Introduction}
\label{Introduction}

Galaxies in a lambda cold dark matter ($\Lambda$CDM) Universe are thought to evolve primarily through two processes: in situ star formation, where stars form from gas within the galaxy itself (dominant until $z \sim 2$), and ex situ star formation, where stars form outside the galaxy and are later accreted through mergers with other galaxies \citep{Oser2010, Lackner2012, wellons2015formation}. Several cosmological, hydrodynamical simulations \citep[e.g.][]{Rodriguez2016} and semi-analytic models \citep[e.g.][]{Guo2008MNRAS, Cora2018MNRAS, Lee2013ApJ} show that the ex situ fraction increases significantly with galaxy mass, reaching over 60\% for the most massive systems, with major mergers having a significant contribution. Given this strong mass dependence, mergers naturally emerge as one of the key drivers of galaxy growth and transformation. They not only contribute to stellar mass growth, but also influence internal processes, often triggering new episodes of star formation. Importantly, mergers can dramatically alter a galaxy’s structure and dynamics: transforming stellar discs into spheroids \citep{Schweizer1982, Barnes1992, Mihos1996}, changing orbital configurations \citep{Naab2009, Dokkum2010, Newman2012, Ferreras2014}, redistributing chemical elements \citep{Torrey2012, Barrera2015, Bustamante2018, Buck2023, Pan2025}, and fueling central supermassive black holes \citep{Treister2012, Ellison2019}

However, mergers alone are not sufficient to explain the long-term structural evolution of galaxies. A crucial additional ingredient is the quenching of star formation, which determines whether dynamically cold discs can be rebuilt after merger events. In the absence of quenching, gas accretion and ongoing star formation can regenerate rotationally supported structures, thereby counteracting merger-driven heating. In the current paradigm of galaxy formation, quenching arises from both internal and external processes. In massive galaxies, feedback from active galactic nuclei (AGNs) plays a central role by heating or expelling gas and preventing further cooling and star formation \citep{Springel2005MNRAS, Bower2006MNRAS, Croton2006MNRAS, Weinberger2017a, Somerville2015ARA&A}. In contrast, for lower-mass systems, particularly satellite galaxies, environmental processes such as ram-pressure stripping, tidal interactions, and strangulation can efficiently suppress star formation \citep{Gunn1972ApJ, Merritt1983ApJ, Larson1980ApJ, Wetzel2012MNRAS, Wright2022MNRAS, Boselli2022, Xie2025, Rodrguez2025A&A}. These processes not only regulate the stellar mass growth of galaxies, but also their dynamical evolution by controlling the availability of cold gas required to sustain or rebuild rotationally supported stellar discs. As a result, while mergers and dynamical heating have the capacity to transform galaxy structures, quenching plays a key role in determining whether such transformations are long-lasting.

The stellar orbits and their distribution provide a powerful diagnostic in probing galaxy evolution, as they act as a fossil record of the dynamical history of a galaxy. The orbital distribution of stars in a galaxy can reveal whether stars formed in situ or were accreted from external sources. Stars that formed in situ under the secular phase tend to reside on dynamically cold, near-circular orbits, giving rise to fast-rotating stellar discs \citep{White1978, Fall1980, Lagos2017}. In contrast, stars that formed in turbulent or perturbed environments, or those subsequently heated dynamically by mergers or internal instabilities, occupy hotter, more randomized orbits \citep{Leaman2017}. Importantly, stars that formed together often retain coherent motions over long timescales, preserving signatures of their common formation \citep{Kamdar2019, Bird2013}. Overall, both star formation episodes and merging events play a fundamental role in shaping the structural and kinematic properties of galaxies, as well as their stellar populations.

To understand how the stellar orbital distribution translates into the structural components of galaxies we observe, it is crucial to decompose galaxies in a physically meaningful manner. A widely used approach over the past few decades is photometric bulge–disk decomposition \citep[e.g.][]{Peng2002AJ, Simard2011ApJS, Meert2015MNRAS}. In this framework, the bulge is commonly taken as a proxy for a dispersion-supported stellar component, whereas the disk is assumed to trace a rotationally supported component. A galaxy’s two-dimensional surface brightness distribution is modeled as a combination of analytic functions: typically a Sérsic profile \citep{Sérsic1968} for the bulge and an exponential profile \citep{Freeman1970ApJ} for the disk. However, purely photometric methods often face limitations, especially in disentangling overlapping structural components, which can lead to ambiguous or even misleading classifications of bulges and discs. Hydrodynamical simulations further highlight these limitations. They show, for instance, galaxies with exponential surface brightness profiles but whose stars occupy kinematically hot orbits \citep{Teklu2015ApJ, Obreja2016MNRAS}. Moreover, comparisons between photometric and kinematic decompositions have revealed significant discrepancies \citep{Scannapieco2010MNRAS, Martig2012ApJ, Obreja2016MNRAS}. \citet{Zhu2018MNRAS} showed that photometric bulge-disk decompositions often differ from orbit-based dynamical decompositions, which classify stars into cold, warm, and hot components. In particular, low-mass late-type galaxies tend to exhibit a larger photometric disk fraction compared to their dynamically derived cold-orbit fraction, whereas high-mass early-type galaxies show a larger photometric bulge fraction than their corresponding dynamical hot-orbit fraction.

An alternative and more physically motivated approach for understanding galaxy dynamics that has been used is dynamical modeling. Among them, one of the most widely used dynamical modeling techniques is the Schwarzschild orbit-superposition method  \citep{Schwarzschild1979ApJ}. This method constructs a self-consistent model of a galaxy by assigning weights to a large library of stellar orbits computed within a given gravitational potential. Over recent decades, the Schwarzschild approach has been successfully applied across a broad range of astrophysical contexts, including the study of supermassive black holes and their masses \citep[e.g.,][]{Marel1998ApJ, Verolme2002MNRA, Gebhardt2003ApJ, Valluri2004ApJ, Krajnovic2009MNRAS, Rusli2013AJ, Seth2014Natur, Thater2017A&A, Thater2019A&A, Liepold2020ApJ, Quenneville2021ApJS}, as well as early-type galaxies (ETGs)  \citep{Cappellari2016ARA, Thomas2007MNRAS, van2008MNRAS, Thomas2014ApJ, Fahrion2019A, Poci2019MNRAS, Jin2020MNRAS, Thater2022MNRAS, Santucci2022ApJ}, while also allowing us to recover the internal orbital structures of globular clusters \citep{vande2006A, Feldmeier2017MNRAS, Zocchi2012A&A}.

Over the past few decades, the emergence of large and efficient integral field spectroscopy (IFS) surveys, such as \texttt{}those provided by Spectroscopic Areal Unit for Research on Optical Nebulae (SAURON; \citealt{deZeeuw2002MNRAS}), \texttt{ATLAS$^{3\text{D}}$} \citep{Cappellari2011MNRAS}, \texttt{} Calar Alto Legacy Integral Field Area (CALIFA; \citealt{Sanchez2023MNRAS}), \texttt{}Sydney–Australian Astronomical Observatory Multi-object Integral Field Spectrograph (SAMI) \citep{Croom2012MNRAS, Bryant2015MNRAS, Croom2021MNRAS}, \texttt{}Mapping Nearby Galaxies at Apache Point Observatory (MaNGA; \citealt{Bundy2015ApJ}) and \texttt{Fornax3D} \citep{Sarzi2018}, have revolutionized our understanding of the stellar structures of galaxies. By providing spatially resolved stellar kinematic maps for thousands of nearby galaxies, these surveys, in combination with advanced orbit-based dynamical modeling techniques, have enabled detailed studies of the internal orbital structures of galaxies from observational data. \cite{Zhu2018MNRAS2} used triaxial Schwarzschild orbit-superposition models to decompose 300 galaxies from the CALIFA survey into four distinct orbital components: a cold component with nearly circular orbits, a hot component with predominantly radial orbits, a warm component with intermediate orbital characteristics, and a counter-rotating (CR) component (with orbits opposite to the cold component). They found that within one effective radius, the relative fractions of these orbital components correlate with stellar mass and that galaxies tend to host more stars on warm orbits than on either cold or hot ones. Consistent results were reported by \cite{Jin2020MNRAS} for 149 early-type galaxies (ETGs) from the \texttt{MaNGA} survey and by \cite{Santucci2022ApJ} for galaxies in the \texttt{SAMI} survey. Recent efforts have also been extended to nearby early-type galaxies from the \texttt{ATLAS$^{3\text{D}}$} survey using the \texttt{DYNAMITE} framework \citep{Thater2025IAUS}.

Comparable efforts have been made to decompose galaxies in cosmological simulations. 
For instance, \cite{Xu2019MNRAS} demonstrated that the orbital fractions of galaxies from the \texttt{TNG100} simulation are consistent with those derived from \texttt{CALIFA} galaxies \citep{Zhu2018b}. Extending this comparison, \cite{Zhang2025} found that \texttt{TNG} galaxies are also consistent with the luminosity fractions of orbital components derived from Schwarzschild modelling, thus further supporting the ability of \texttt{IllustrisTNG} to capture the dynamical diversity of galaxy populations. A similar level of agreement has been found for galaxies in the \texttt{EAGLE} (Evolution and Assembly of GaLaxies and their Environments) simulation \citealt{McAlpine2016}.

While these studies offer valuable insights into the distribution of dynamical orbits, they are largely confined to the Local Universe, with surveys such as \texttt{CALIFA} \citep{Walcher2014A} spanning $0.005 < z < 0.03$, \texttt{MaNGA} \citep{Bundy2015ApJ} covering approximately $0.01 \lesssim z \lesssim 0.15$, and \texttt{SAMI} probing $0.004 < z < 0.113$ \citep{Croom2012MNRAS, Bryant2015MNRAS, Croom2021MNRAS}. In contrast, orbital decomposition studies at high redshift remain scarce, despite their crucial importance for understanding galaxy evolution. At earlier cosmic epochs, galaxies were significantly more gas-rich \citep{Obreschkow2009, Lagos2011MNRAS}, merger rates were substantially higher \citep{Conselice2009MNRAS, Rodriguez2015}, and star formation activity was significantly higher \citep{Madau1996MNRAS}. Since stellar orbits preserve the dynamical imprint of past events, performing orbital decompositions at high redshift provides a unique window into the cumulative effects of mergers, gas accretion, and quenching processes over cosmic time.

Observationally, extending such analyses to higher redshifts has remained challenging. The \texttt{LEGA-C} Survey provides a unique opportunity in this context, delivering high signal-to-noise (S/N) spectra for thousands of galaxies at $z \sim 1$ \citep{van20161, Straatman2018, van2021}. This epoch marks a critical phase in galaxy evolution, characterized by elevated star formation rates, higher gas fractions, and more frequent merger activity compared to the Local Universe. By combining these observations with state-of-the-art cosmological hydrodynamical simulations, we are now uniquely positioned to investigate how stellar orbits were assembled during this key epoch $z \sim 1$ and to link these orbital structures to the physical processes that have shaped galaxies across cosmic time.

In this paper, we present the first study of the dynamical evolution of \texttt{LEGA-C}–like galaxies in the \texttt{IllustrisTNG} suite of cosmological simulations by linking their orbit-based decompositions from $z=1$ to $z=0$. This approach provides a unique opportunity to follow the evolution of galaxy structure and internal dynamics across a critical phase of galaxy assembly. We decomposed the galaxies into cold, warm, hot, and CR orbital components following the methodology of \citet{Zhu2018b}. Our primary goal is to characterize the intrinsic properties of these orbital components, trace their evolution over cosmic time, and identify the physical mechanisms responsible for their transformation. To this end, we employed the \texttt{LEGA-C} analogs constructed by \citet{Wu2021}, which connect galaxies in the \texttt{TNG100} simulation at $z=1$ to their surviving descendants at $z=0$ via merger trees, enabling a direct link between orbital structure and assembly history.

This paper is organized as follows. In Section \ref{sec:tools_and_methods}, we introduce the IllustrisTNG simulations, the merger tree algorithm, the sample selection of \texttt{LEGA-C} analogs, and our orbital decomposition methodology. In Section \ref{sec:orbitalfraction}, we present the orbital fractions of galaxies at $z=1$ and $z=0$ and trace their evolution across cosmic time. Section \ref{sec:mergers} investigates the role of galaxy mergers, distinguishing between major and minor, as well as wet and dry, events. In Section \ref{sec:ExsituMass}, we explore the connection between orbital structure and the ex situ stellar mass fraction. In Section \ref{sec:discussion}, we discuss our results in the context of numerical resolution effects, including tests using \texttt{TNG50} . Finally, in Section \ref{sec:Summary_Conclusions}, we summarize our main findings and outline the broader implications and future observational prospects.

\begin{figure*}[t]
    \centering
    
    \includegraphics[width=0.8\textwidth]{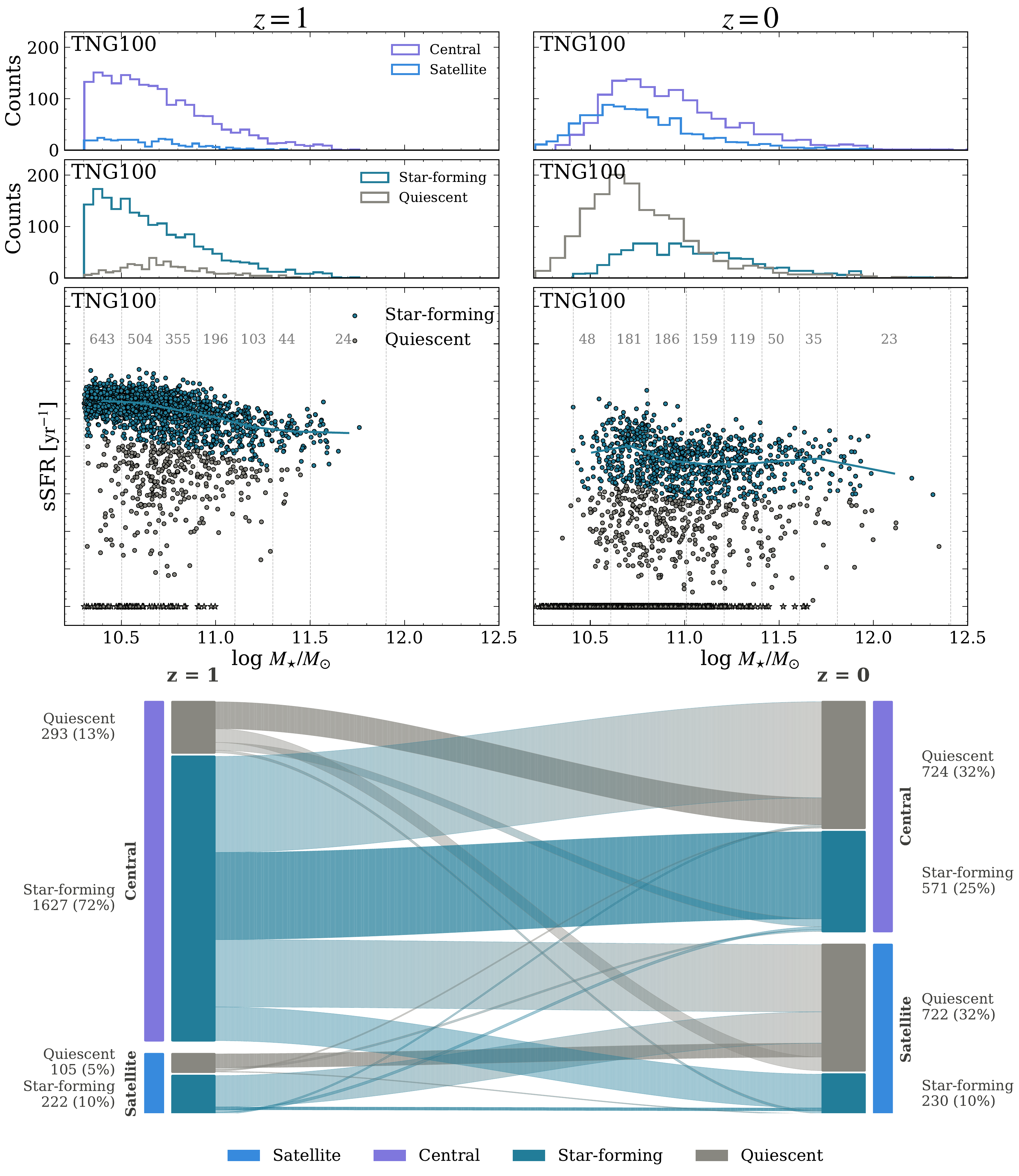}
    \caption{\texttt{TNG100} galaxy populations from $z=1$ to $z=0$.
Middle panels: sSFR versus stellar mass at $z=1$ (left) and $z=0$ (right). Galaxies are classified relative to the SFMS as star-forming or quiescent. The solid blue line shows the median SFMS in 0.2 dex stellar-mass bins; galaxies with zero SFR are placed at $\mathrm{sSFR}=10^{-15},\mathrm{yr}^{-1}$ and marked with black stars.
Top panels: Stellar-mass distributions separated into \texttt{central} and \texttt{satellite} galaxies (top row) and star-forming and quiescent galaxies (second row). At $z=1$, the sample is dominated by star-forming centrals, while by $z=0$ quiescent galaxies and satellites are substantially more prevalent. The SFMS normalization also declines, with more galaxies lying below it. This transition is strongest among satellites, where quiescent systems dominate by $z=0$, consistent with environmental quenching after infall. The simultaneous growth of the quiescent central population, however, indicates that quenching also involves processes independent of environment.
}
    \label{fig:star_formation}
\end{figure*}

\section{Data and methodology}
\label{sec:tools_and_methods}

\subsection{The IllustrisTNG simulation}
\label{sec:IllustrisTNG}

In this study, we used data from the \texttt{IllustrisTNG} project (\texttt{Illustris} -- The Next Generation: \citep{Springel2018a, Pillepich2018a, Nelson2018a, Naiman2018a, Marinacci2018a}), a suite of state-of-the-art cosmological magneto-hydrodynamical simulations\footnote{https://www.tng-project.org}. These simulations were performed using the \texttt{AREPO} moving-mesh code \citep{Springel2010a,Pakmor2011a, Pakmor2013a,Pakmor2016a} and \cite{Planck2016} cosmological parameters: matter density $\Omega_{\mathrm{m}} = 0.3089$, baryon density $\Omega_{\mathrm{b}} = 0.0486$, dark energy density $\Omega_{\Lambda} = 0.6911$, Hubble constant $H_{0} = 67.74\,\mathrm{km\,s^{-1}\,Mpc^{-1}}$, power spectrum normalization $\sigma_{8} = 0.8159$, and primordial spectral index $n_{\mathrm{s}} = 0.9667$. The \texttt{IllustrisTNG} is built upon the foundation laid by its predecessor, \texttt{Illustris} \citep{Vogelsberger2013b, Vogelsberger2014a,Vogelsberger2014b,Genel2014a, Nelson2015A&C,Sijacki2015a}, with substantial improvements in the modeling of key baryonic processes, such as supernova (SN) feedback and AGN feedback, as well as chemical enrichment.

\texttt{IllustrisTNG} has proven highly successful in reproducing a wide range of observed galaxy properties and scaling relations across cosmic time. These include the galaxy stellar mass function at z < 4 \citep{Pillepich2018a}, the bimodality in galaxy colors \citep{Nelson2018a}, spatially resolved star formation in star-forming galaxies \citep{Nelson2021a}, the fraction of quiescent galaxies \citep{Donnari2019, Donnari2021}, and the mass–size relation for both early- and late-type galaxies \cite{Genel2018, Huertas2019a, Rodriguez2019a}.

The \texttt{IllustrisTNG} simulation suite consists of three principal runs with varying resolutions and volumes: \texttt{TNG50}, \texttt{TNG100}, and \texttt{TNG300}. Each run evolves from redshift $z = 20$ to the present day and spans comoving box volumes of $51.7^3$ cMpc$^3$, $110.7^3$ cMpc$^3$, and $302.6^3$ cMpc$^3$, respectively. The corresponding dark matter particle mass resolutions are $4.5 \times 10^5$ $\mathrm{M}\odot$, $7.5 \times 10^6$ $\mathrm{M}\odot$, and $5.9 \times 10^7$ $\mathrm{M}\odot$, while the baryonic mass resolutions are $8.5 \times 10^4$ $\mathrm{M}\odot$, $1.4 \times 10^6$ $\mathrm{M}_\odot$, and $1.1 \times 10^7$ $\mathrm{M}_\odot$, respectively. \texttt{TNG50}, with its high resolution and smaller volume, allows for detailed studies of internal galaxy structure \citep{Pillepich2019MNRAS, Nelson2019MNRAS}. In contrast, \texttt{TNG300} offers a significantly larger volume at lower resolution, making it ideal for statistical analyzes of galaxy populations and large-scale structures. Bridging these two regimes, \texttt{TNG100} provides an optimal balance between resolution and volume, making it well-suited for studying processes on intermediate scales. For further details and a comprehensive understanding of the numerical and physical models implemented in the \texttt{IllustrisTNG} simulations, we encourage readers to refer to methodological papers by \cite{Weinberger2017a} and \cite{Pillepich2018b}.

In this work, we primarily used \texttt{TNG100}, which provides a good balance between simulation volume and numerical resolution, making it well suited for studying the orbital structure and evolution of statistically representative galaxy populations. Where relevant, we compare our results with \texttt{TNG50} to assess the impact of numerical resolution.

\begin{figure*}[t]
    \centering
    \includegraphics[width=0.9\textwidth]{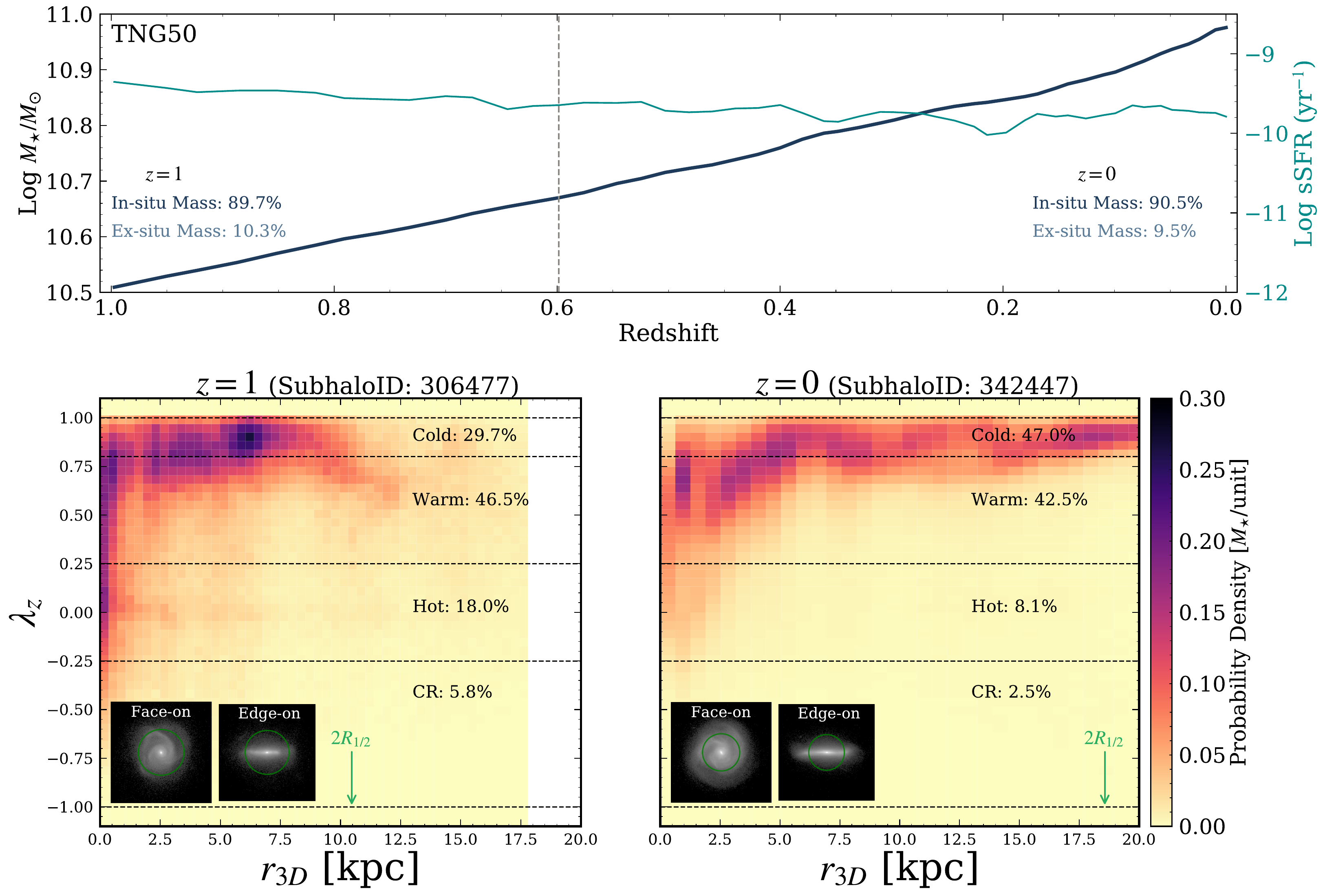}
\caption{Evolution of an individual galaxy in the \texttt{TNG50} simulation, traced from \texttt{SubfindID 306477} at $z=1$ to its descendant \texttt{SubfindID 342447} at $z=0$. Top panel: Evolution of the stellar mass (dark blue) and sSFR (teal) from $z=1$ to $z=0$. The galaxy grows from $\log(M_\star/M_\odot)\simeq10.5$ at $z=1$ to $\simeq11.0$ at $z=0$, while remaining dominated by in situ stellar mass growth. The in situ stellar mass fraction changes only slightly, from $89.7\%$ at $z=1$ to $90.5\%$ at $z=0$, with correspondingly small ex situ contributions of $10.3\%$ and $9.5\%$, respectively. Bottom panels: Circularity distributions at $z=1$ (left) and $z=0$ (right), shown as a function of the time-averaged three-dimensional radius. The colour scale indicates the stellar-mass probability density, while the horizontal dashed lines mark the boundaries between the cold, warm, hot, and CR orbital components. From $z=1$ to $z=0$, the cold fraction increases from $29.7\%$ to $47.0\%$, while the hot and CR fractions decrease from $18.0\%$ to $8.1\%$ and from $5.8\%$ to $2.5\%$, respectively; the warm component changes more modestly from $46.5\%$ to $42.5\%$. The evolution therefore illustrates the progressive build-up of a more strongly rotation-supported stellar structure in a galaxy whose mass growth is dominated by in situ star formation rather than accretion. Insets show face-on and edge-on stellar projections, with the green circle indicating twice the stellar half-mass radius, $2R_{1/2}$.}        \label{fig:CircularityMap_SubhaloID_342447}
\end{figure*}

\begin{figure*}[t]
    \centering
    \includegraphics[width=0.9\textwidth]{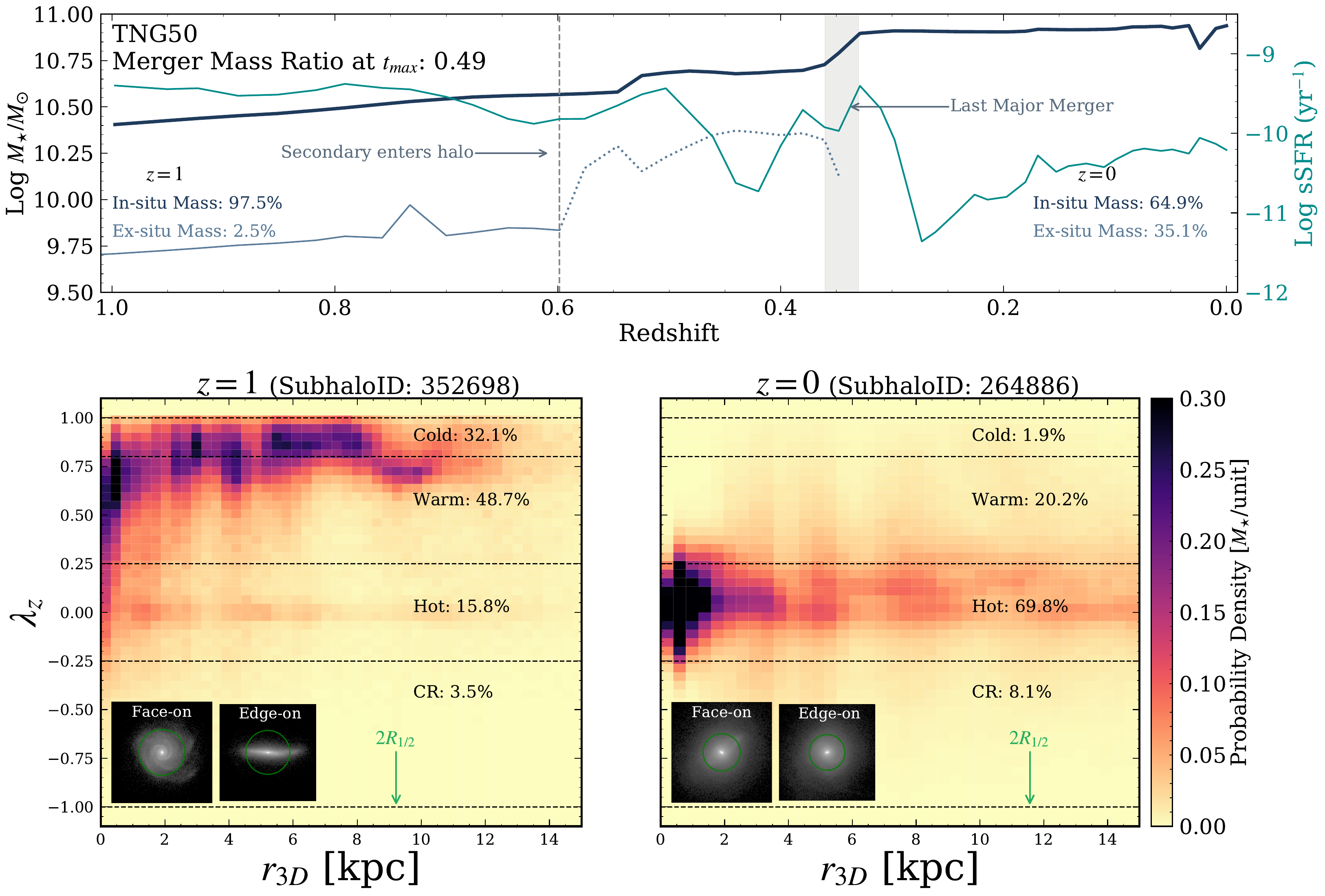}
\caption{Similar to Fig.~\ref{fig:CircularityMap_SubhaloID_342447}, but for a \texttt{TNG50} galaxy that undergoes a late major merger. The galaxy is traced from \texttt{SubfindID 352698} at $z=1$ to its descendant \texttt{SubfindID 264886} at $z=0$. At $z=1$, it has $\log(M_\star/M_\odot)\simeq 10.4$ and is strongly dominated by in situ stellar mass growth ($97.5\%$ in situ and $2.5\%$ ex situ). The top panel shows that the most massive secondary progenitor enters the halo at $z\approx0.6$, followed by a major merger at $z\approx0.35$ (shaded region) with merger mass ratio $\mu=0.49$. After this event, the ex situ mass fraction rises substantially, reaching $35.1\%$ by $z=0$. The bottom panels show the corresponding transformation in orbital structure: from a rotation-supported system at $z=1$, with cold, warm, hot, and CR fractions of $32.1\%$, $48.7\%$, $15.8\%$, and $3.5\%$, respectively, to a dispersion-dominated remnant at $z=0$, where the hot component increases to $69.8\%$ while the cold and warm fractions decline to $1.9\%$ and $20.2\%$. The insets illustrate the associated morphological evolution from a disk-like galaxy to a much more spheroidal system.}
    \label{fig:CircMap_SNAAP99_SubhaloID264886}
\end{figure*}

\subsection{Merger trees and galaxy tracking}
\label{sec:IllustrisTNG_MergerTree}

In the \texttt{IllustrisTNG} simulations, dark matter halos are identified using the \texttt{friends-of-friends(FoF)} algorithm \citep{Davis1985}, with baryonic particles assigned to the same \texttt{FoF} group as their nearest dark matter particle. Substructures within these halos, known as subhalos, are cataloged using the \texttt{SUBFIND} algorithm \citep{Springel2001a, Dolag2009}. The time evolution of subhalos is then tracked with the \texttt{SUBLINK} merger tree algorithm \citep{Rodriguez2015}, which reconstructs their formation history by linking subhalos across snapshots. This is achieved by assigning each subhalo, when possible, a unique descendant in the subsequent snapshot based on a scoring system that considers the number and binding-energy ranking of shared particles.

Within each \texttt{FoF} group, the most massive subhalo is classified as the central galaxy and all others as satellites. A subhalo is a progenitor of another if the latter is its unique descendant. While a subhalo may have multiple progenitors, it has only one descendant, consistent with hierarchical structure growth in $\Lambda$CDM cosmologies. The first progenitor is defined as the branch with the most massive merger history \citep{Lucia2007}. Mergers are identified when secondary galaxies coalesce with a primary galaxy and subsequently disappear.

We identified mergers from the \texttt{IllustrisTNG} merger trees and quantified them using the stellar mass ratio defined by \citet{Rodriguez2015}, with both progenitor masses measured at $t_{\mathrm{max}}$, the snapshot at which the secondary progenitor reaches its maximum stellar mass before merging. We classified mergers with $\mu>1/4$ as major and those with $1/10<\mu<1/4$ as minor.

\subsection{\texttt{LEGA-C } analogs in \texttt{IllustrisTNG}}
\label{sec:LEGAC_Aanalogues}

The \texttt{LEGA-C }survey \citep{van20161, Straatman2018, van2021} is a deep spectroscopic program carried out with the VIMOS instrument on the ESO Very Large Telescope. The survey targets $K_\mathrm{s}$-band-selected galaxies in the COSMOS field \citep{Scoville2007} over the redshift range $0.6 < z < 1.0$, employing a redshift-dependent magnitude limit designed to approximate a stellar-mass selection. The final data release by \citet{van2021} includes 4,081 spectra, of which 3,741 are unique galaxies, predominantly with stellar masses above $10^{10}\,\mathrm{M}_\odot$. Each target was observed with a total integration time of $\sim$20 hours, yielding rest-frame optical spectra at a resolution of $R \sim 3500$ and a typical continuum S/N of $\sim$20\,\AA$^{-1}$. The parent photometric catalog is drawn from UltraVISTA \citep{Muzzin2013}, ensuring consistency in photometric and spectroscopic coverage. The resulting high-quality dataset enables robust measurements of stellar kinematics, emission and absorption line properties, and stellar population parameters for a representative sample of massive galaxies at intermediate redshifts.

However, in the present study, we did not use these mock spectra directly. Instead, we employed the galaxy catalog provided by \citet{Wu2021} to identify \texttt{LEGA-C-like} systems within \texttt{TNG100}. These analogs were selected at $z = 1$ to match the stellar-mass distribution of the observed \texttt{LEGA-C} sample ($10.3 < \log(M_\star/M_\odot) < 11.8$ within a 30~pkpc aperture).

\subsubsection{Primary sample and descendant sample}
Our primary sample was constructed by matching \texttt{SubfindIDs} from the \cite{Wu2021} to the corresponding \texttt{SubhaloIDs} in the \texttt{TNG100} snapshots. This results in an initial sample of 3,058 subhalos at $z = 1$, hereafter referred to as the \texttt{whole LEGA-C-like} sample, covering a stellar-mass range of $\log(M_\star/M_\odot) = 10.3$--$11.8$.  To follow their evolution, we trace each galaxy forward along its main descendant branch in the merger trees from $z = 1$ to $z = 0$. Galaxies that remain identifiable as distinct subhalos at $z = 0$ are classified as survivors. This procedure yields a final sample of 2,281 unique galaxies, which we denote as the \texttt{survived LEGA-C-like} sample. Unless otherwise stated, all evolutionary trends shown are derived by following galaxies along these main descendant branches. 

For completeness, when considering properties at $z = 0$, we also included galaxies that are not descendants of the initial $z = 1$ sample, but fall within the same stellar-mass range at the present day. This defines the \texttt{extended LEGA-C-like} sample, which consists of the \texttt{survived LEGA-C-like} galaxies together with additional systems that enter the mass range by $z = 0$.

\subsubsection{Galaxy classifications}
\label{sec:galaxy_classification}

\paragraph{Central and satellite classification:}
The central galaxy is defined as the primary subhalo of the FoF group, corresponding to the entry \texttt{GroupFirstSub} in the group catalogue. All remaining subhalos within the FoF group are classified as satellite galaxies. We adopted this standard \texttt{IllustrisTNG} classification throughout this work.

Accordingly, our \texttt{LEGA-C-like} sample includes both central and satellite galaxies. In the top row of Fig. \ref{fig:star_formation}, we show the stellar mass distributions of centrals and satellites at both redshifts. At $z=1$, the sample is predominantly composed of central galaxies across the full mass range. By $z=0$, the number of satellite galaxies has increased significantly, leading to a more substantial contribution from satellites.
\paragraph{Star formation classification:}
We classify galaxies as star-forming or quiescent based on their offset from the star-forming main sequence (SFMS) \citep{Pillepich2019MNRAS,Donnari2019,Donnari2021}. Throughout this work, we use the terms quenched and quiescent interchangeably to describe galaxies with suppressed or no star formation. We do not distinguish starburst galaxies from the broader star-forming population, nor green-valley galaxies from the quiescent population. The middle row of Fig.~\ref{fig:star_formation} shows the distribution of the \texttt{survived LEGA-C-like} sample in the sSFR--stellar mass plane at $z=1$ (left) and $z=0$ (right), together with the derived SFMS.
To obtain an SFMS estimate that is not biased by quiescent galaxies, we must first determine the relation iteratively in stellar-mass bins of width 0.2 dex. We computed the median sSFR in each bin using all galaxies and classify them according to their offset from this initial relation. We then recomputed the median using only galaxies identified as star-forming and repeat the procedure until the SFMS and classification converge. Relative to the converged SFMS, galaxies with $\log \Delta \mathrm{sSFR}>-1$ were classified as star-forming, while those with $\log \Delta \mathrm{sSFR}<-1$ were classified as quiescent. Galaxies with zero SFR, which arise partly from the finite numerical resolution of \texttt{TNG100}, were included in the quiescent population and shown as star symbols at a fixed value of $\mathrm{sSFR}=10^{-15},\mathrm{yr}^{-1}$ for visualization. As expected, the SFMS is seen to evolve with redshift \citep{Daddi2007ApJ,Noeske2007ApJ,Whitaker2012ApJ,Tomczak2016ApJ,Popesso2023MNRAS}, with galaxies at $z=1$ exhibiting systematically higher sSFR values than those at $z=0$.
The Sankey diagram in the bottom row of Fig.~\ref{fig:star_formation} illustrates the evolutionary pathways of galaxies from $z=1$ to $z=0$, tracking changes in both star formation activity and central/satellite status. At $z=1$, the population is dominated by star-forming centrals (72\%), with smaller contributions from quiescent centrals (13\%) and satellites. By $z=0$, the quiescent fraction has increased substantially among both centrals (32\%) and satellites (32\%), accompanied by a redistribution of galaxies into the satellite population. A significant fraction of initially star-forming centrals remain centrals but quench, while others become satellites and subsequently transition to the quiescent population. The growth of the quiescent central population, however, indicates that quenching also involves processes independent of environment. This can be explained by internal processes such as AGN feedback, which suppresses gas cooling and star formation in massive halos \citep{Weinberger2017a,Pillepich2018a}. For satellites, environmental processes such as ram-pressure stripping, tidal interactions, and starvation (or strangulation) may remove or heat the cold-gas reservoir following falling into a more massive halo. Both galaxies that become satellites after $z=1$ and those already classified as satellites at $z=1$ show a high probability of being quiescent by $z=0$. Galaxies that are already quiescent at $z=1$ also largely remain so, irrespective of their central or satellite status, while rejuvenation represents only a minor evolutionary pathway.
Overall, the probability of transitioning to the quiescent population between $z=1$ and $z=0$ is $\sim47\%$, while the probability of transitioning from a central to a satellite is $\sim33\%$.

\subsection{Orbital decomposition}
\label{sec:orbitaldecomposition}

We characterized the orbital structure of simulated galaxies using the snapshot-averaged (or time averaged) stellar circularity parameter, $\overline{\lambda_z}$, which is defined as the ratio between the specific angular momentum of a stellar particle along the galaxy’s  z-axis and the angular momentum of a circular orbit with the same binding energy. This is given by
\begin{equation}
\overline{\lambda_z}
= \frac{\overline{L}_z}{J_c(E)} .
\end{equation}

Here, $\overline{L}_z = \overline{x v_y - y v_x}$ is the snapshot-averaged specific angular momentum component along the $z$-axis. The quantity $J_c(E)$ represents the maximum angular momentum attainable by a circular orbit at a given binding energy $E$ \citep{Abadi2003ApJ}.

Here, we used the snapshot-averaged circularity instead of the instantaneous circularity because it better traces the actual orbit followed by a stellar particle. Instantaneous circularity can misclassify the particles, because at a given snapshot, a particle on a radial or box orbit may momentarily exhibit a high $L_{z}$, comparable to that of a cold or warm orbit, thereby misleading its true orbital classification. In Appendix \ref{instantaneous_time-averagedCircularity}, we show an example of the instantaneous circularity and the snapshot-averaged circularity for one of the galaxies, demonstrating the effectiveness of this approach. 

In principle, the snapshot-averaged circularity of each stellar particle in a galaxy can be computed by evaluating the circularity at equal time intervals along its orbit and averaging over the corresponding snapshots. The quantity $J_{c}(E)$ at each snapshot can be constructed from the gravitational potential by combining the mass distributions of stars, dark matter, and gas, and by integrating particle orbits in the resulting potential. However, reconstructing reliable gravitational potentials for all \texttt{TNG} galaxies across all snapshots is computationally expensive. Therefore, following recent implementations of Schwarzschild dynamical models \citep{Zhu2018MNRAS2}, we approximate $J_c(E)$ using the estimator,

\begin{equation}
J_c(E) \simeq \overline{\mathbf{r}} \times \overline{V_{\mathrm{rms}}} ,
\end{equation}

where $\overline{\mathbf{r}}$ is the time-averaged three-dimensional galactocentric radius of the orbit, and $\overline{V_{\mathrm{rms}}}$ is the time-averaged second velocity moment, defined as
\begin{equation}
\overline{V_{\mathrm{rms}}} \equiv \sqrt{\overline{(v_x + v_y + v_z)^{2}}} .
\end{equation}

Using this approximation, we compute the snapshot-averaged circularity, $\overline{\lambda_z}$, by averaging the circularity values over 11 snapshots, similar to \cite{Santucci2024MNRAS} and  Reiter et al. (in prep.), which can be written as

\begin{equation}
\overline{\lambda_z}
= \frac{\overline{L_z}
}
       {\overline{\mathbf{r}} \times \overline{V_{\mathrm{rms}}}} .
\end{equation}

This alternative definition of circularity has been widely used in observations \citep{Zhu2018MNRAS2, Xu2019MNRAS, Jin2019MNRAS, Jin2020MNRAS}, and is particularly well suited to large cosmological simulations where detailed potential reconstruction is not always feasible.

 Following \cite{Zhu2018MNRAS2, Zhu2018MNRAS2, Xu2019MNRAS, Jin2019MNRAS, Jin2020MNRAS, Santucci2022ApJ}, we classify stellar orbits into four distinct dynamical components based on their circularity parameter, $\overline{\lambda_z}$: the cold component ($0.8 \leq \overline{\lambda_z} \leq 1$), the warm component ($0.25 < \overline{\lambda_z} < 0.8$), the hot component ($-0.25 \leq \overline{\lambda_z} \leq 0.25$), and the CR component ($-1 \leq \overline{\lambda_z} < -0.25$). Stellar particles in the cold component are on nearly circular orbits that co-rotate with the galaxy’s net angular momentum $\textbf{L}_{gal}$. These stars predominantly form a flattened, rotationally supported disk. Stars in the hot component occupy radially biased or boxy orbits that can reach the central regions of the galaxy, giving rise to a spheroidal, dispersion-dominated bulge component. Stars in warm components move on somewhat disk-like orbits but reach larger vertical heights above the disk midplane. In CR components, stars rotate in the opposite direction to $\textbf{L}_{gal}$. Theoretically, these orbits are expected to trace different formation and evolutionary pathways, including in situ disk growth, dynamical heating processes, and accretion or merger-driven assembly.

In Fig.~\ref{fig:CircularityMap_SubhaloID_342447} and Fig.~\ref{fig:CircMap_SNAAP99_SubhaloID264886}, we present two galaxies from the \texttt{TNG50} simulation as illustrative examples of the evolution of their circularity distributions from $z=1$ to $z=0$.\footnote{Videos showing the evolution of these galaxies from $z=1$ to $z=0$ are available at \href{https://www.youtube.com/playlist?list=PLN2R7fI-dgiQ3o_GHC7gZIEB02DerEPEU}{Galactic Orbital Decomposition}.} (These examples are used only for visualization; we have verified that galaxies in \texttt{TNG100} exhibit circularity maps of comparable quality. We further compare the orbital distributions obtained from \texttt{TNG50} and \texttt{TNG100} in Sec.~\ref{sec:TNG50}). The galaxies are selected at $z=1$ and followed to $z=0$ along their main descendant branches in the SubLink merger trees. Although they have similar stellar masses and orbital distributions at $z=1$, their subsequent evolutionary pathways differ substantially, illustrating how assembly history can shape galaxy dynamics. Both are centrals at $z=1$ and become satellites by $z=0$, providing additional environmental context for their evolution. The top panels also show the evolution of their specific star formation rates (sSFRs).

In Fig.~\ref{fig:CircularityMap_SubhaloID_342447}, at $z=1$, the galaxy has a stellar mass of $\log(M_\star/M_\odot)=10.5$, with the bulk of its mass assembled through in situ star formation. At this epoch, more than $90\%$ of the stellar mass is formed in situ, while only a minor fraction ($\sim10\%$) originates from ex situ accretion, which may arise from mergers, flybys, or other interactions. The sSFR evolution indicates relatively steady star formation over time. Beyond $z=1$, the galaxy experiences smooth, quiet growth, allowing its internal dynamical structure to evolve largely undisturbed. The bottom panels show the circularity maps at $z=1$ and $z=0$. At both epochs, the orbital structure is dominated by stars on cold and warm orbits, indicating a high degree of ordered motion. At $z=1$, the warm component dominates, with $\sim30\%$ of the orbits in the cold component, reflecting the presence of a rotationally supported disk, along with a moderate fraction of hot orbits. By $z=0$, the galaxy maintains a similarly ordered orbital configuration, with an increase in the relative contribution of the cold component and a corresponding decrease in the hot and CR fractions. The hot orbital component remains subdominant at both epochs, and the CR component is negligible, consistent with the absence of significant merger-driven dynamical heating. Despite becoming a satellite by $z=0$, its sSFR does not show a dramatic suppression, indicating that environmental quenching effects are weak in this system.

In contrast, Fig.~\ref{fig:CircMap_SNAAP99_SubhaloID264886} shows a relatively more merger driven mass growth history. Similar to Fig.~\ref{fig:CircularityMap_SubhaloID_342447}, at $z=1$, this galaxy has a stellar mass of $\log(M_\star/M_\odot)=10.45$, with the bulk of its mass assembled through in situ star formation. However, instead of quiet growth, this galaxy experiences a major merger at $z \sim 0.35$ with a mass ratio of $\mu=0.49$. The mass growth of the secondary galaxy is shown in red lines in the plot. The sSFR evolution shows significant fluctuations, including a sharp decline following the last major merger, indicating merger-driven disruption of star formation activity. The mass growth of the secondary galaxy is shown in red lines in the plot. The circularity map at $z=1$ indicates that the orbital structure is dominated by warm and cold orbits, with a combined fraction of $\sim80\%$, including $\sim15.8\%$ in hot orbits and only $\sim3.5\%$ in CR orbits. At this epoch, the galaxy appears to be rotation-supported and disk-like, as shown by the face-on and edge-on projections together with the circularity distribution. By contrast, the circularity distribution at $z=0$ shows a markedly different orbital structure. The hot component becomes dominant, contributing $\sim70\%$ of the stellar mass, while the cold component is nearly absent. The circularity map is strongly concentrated around $\overline{\lambda_z} \approx 0$, indicating that stellar orbits are primarily dispersion-supported and largely randomized. This transformation suggests significant dynamical heating and angular momentum redistribution over cosmic time. Furthermore, after transitioning from a central galaxy at $z=1$ to a satellite at $z=0$, the combined effects of the merger and environmental processes likely contribute to the strong suppression of star formation activity observed in its sSFR evolution.

Overall, these two examples demonstrate that galaxies with roughly similar stellar masses and orbital distributions at a given epoch can follow markedly different evolutionary pathways. While non-violent mass growth preserves and even enhances ordered rotational structures, major mergers can efficiently destroy disk components and transform galaxies into dispersion-dominated systems. This comparison highlights the critical role of late-time assembly history in determining the present-day dynamical state of galaxies, even when their early properties appear remarkably similar at a given redshift.
\begin{figure*}[t]
    \sidecaption
    \includegraphics[width=12cm]{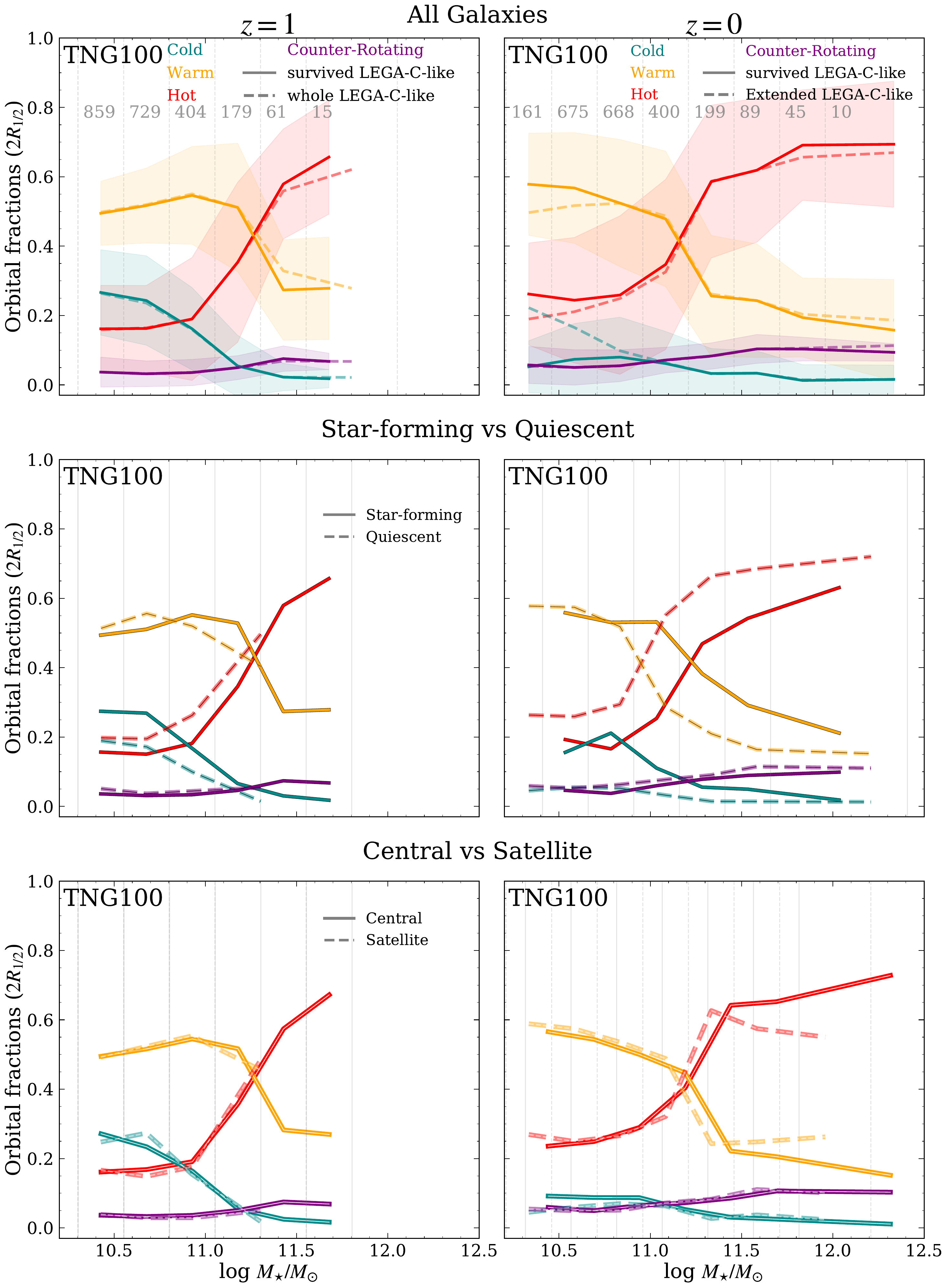}
    \caption{Stellar orbital fractions within $2R_{1/2}$ as a function of stellar mass for \texttt{TNG100} galaxies at $z=1$ (left) and $z=0$ (right), for the \texttt{survived LEGA-C-like} (solid) and \texttt{whole/extended LEGA-C-like} (dashed) samples. The lines show the median orbital fractions, with shaded regions indicating the $1\sigma$ scatter. Top row: At both redshifts, orbital structure depends strongly on stellar mass: lower-mass galaxies are dominated by warm and cold orbits, whereas massive systems become dominated by hot orbits. From $z=1$ to $z=0$, the hot fraction increases, while the cold and warm components decline, indicating a systematic evolution toward dynamically hotter stellar structures. Middle row: Evolution is also shown to be dependent strongly on star formation activity: at fixed stellar mass, \texttt{quiescent} galaxies generally have higher (lower) hot (cold) fractions than \texttt{star-forming} galaxies, with the difference becoming particularly pronounced by $z=0$. Bottom row: In contrast, \texttt{central} and \texttt{satellite} galaxies follow broadly similar mass-dependent trends.}
    \label{fig:fraction_plot_z1_z0}
\end{figure*}

\section{Orbital structure of \texttt{LEGA-C-like} galaxies in \texttt{TNG100}}
\label{sec:orbitalfraction}

In this section, we examine how the orbital structure of \texttt{LEGA-C-like} galaxies depends on stellar mass, star formation activity, environment, and cosmic time. We first characterize the orbital fractions at $z=1$ and $z=0$ as a function of stellar mass (Sect.~\ref{sec:orbitalfraction_stellar_mass}), before separating galaxies by star formation activity and central/satellite status. In Sect.~\ref{sec:orbitalevolution}, we then trace individual galaxies from $z=1$ to $z=0$ to determine how their orbital structures evolve with time.

\subsection{Orbital structure as a function of stellar mass}
\label{sec:orbitalfraction_stellar_mass}
The top row of Fig.~\ref{fig:fraction_plot_z1_z0} shows the median stellar orbital fractions as a function of stellar mass at $z=1$ (left) and $z=0$ (right). We measure the orbital fractions within $2R_{1/2}$, where $R_{1/2}$ is the 3D stellar half-mass radius, and decompose the stellar orbits components. Solid lines show the medians for the \texttt{survived LEGA-C-like} galaxies that can be individually tracked from $z=1$ to $z=0$, while dashed lines show the corresponding medians for the \texttt{whole sample}, containing all galaxies with $\log(M_\star/M_\odot)>10.3$ at each redshift. We calculated the medians in stellar-mass bins of 0.25~dex, requiring at least ten galaxies per bin and merging sparsely populated bins with adjacent ones. Shaded regions indicate the $1\sigma$ scatter.

\subsubsection{All galaxies}
\label{sec:all_galaxies}

Orbital structure at $z = 1$: the orbital structure depends strongly on stellar mass (Fig.~\ref{fig:fraction_plot_z1_z0}, left). Warm and hot orbits dominate over the full mass range, although lower-mass galaxies retain substantial rotational support. At $\log(M_\star/M_\odot)\sim10.3$, warm orbits contribute $\sim50\%$, cold orbits $\sim25\%$, and hot orbits $\sim18\%$, while CR orbits contribute $\lesssim5\%$. Thus, lower-mass galaxies at $z=1$ contain significant cold and warm rotationally supported structures. The orbital structure changes markedly above $\log(M_\star/M_\odot)\sim11$. The warm fraction initially rises to $\sim55\%$ while the cold fraction falls below $\sim20\%$, before the warm component declines rapidly at higher masses. At $\log(M_\star/M_\odot)\gtrsim11.2$, hot orbits become dominant and reach more than $\sim60\%$ at the massive end, showing that the most massive galaxies are already predominantly dispersion-supported at $z=1$. The CR component remains subdominant but increases mildly with stellar mass. This transition may reflect the increasing importance of mergers and other dynamically disruptive processes at high stellar masses, which we investigate further in Sect.~\ref{sec:mergers}. The \texttt{survived LEGA-C-like} galaxies closely follow the orbital fractions of the \texttt{whole sample} at $z=1$, indicating that the progenitors surviving to $z=0$ are representative of the broader galaxy population at this redshift.

Orbital structure at $z = 0$: by this point, galaxies retain the same broad mass dependence but have evolved toward dynamically hotter orbital structures (Fig.~\ref{fig:fraction_plot_z1_z0}, right). Below $\log(M_\star/M_\odot)\sim11$, the cold fraction decreases from $\sim25\%$ at $z=1$ to only $\sim5$--$10\%$ at $z=0$, while the hot fraction increases and remains approximately constant at $\sim24\%$. The cold component shows a modest maximum around $\log(M_\star/M_\odot)\sim10.7$, whereas the warm fraction decreases more monotonically with stellar mass than at $z=1$. These changes indicate a general reduction in rotational support between the two epochs.
At higher stellar masses, the transition to hot-orbit dominance remains pronounced. The hot fraction rises rapidly above $\log(M_\star/M_\odot)\sim11$ and reaches $\sim65$--$70\%$ at the massive end, while the cold fraction falls below $\sim5\%$ and the warm component continues to decline. The CR fraction remains small, at $\sim5$--$10\%$, but increases modestly toward higher masses. The increasing hot fraction and corresponding reduction in cold and warm orbits are consistent with cumulative dynamical heating from mergers and other perturbations \citep{Sellwood2014R}. We examine the physical drivers of this evolution in Sect.~\ref{sec:mergers}.

The difference between the \texttt{survived LEGA-C-like} and \texttt{extended LEGA-C-like} samples at $z=0$ is strongest below $\log(M_\star/M_\odot)\sim11$. In this regime, the \texttt{extended LEGA-C-like} sample shows higher cold fractions and lower warm and hot fractions, more closely resembling the trends seen at $z=1$. Galaxies that enter the sample after $z=1$, either through later formation or growth above the stellar-mass threshold, therefore tend to be more rotationally supported than the descendants of the $z=1$ population. This distinction motivates our direct analysis of the orbital evolution of individual galaxies in Sect.~\ref{sec:orbitalevolution}.

Finally, in Appendix~\ref{ComparisonPhase-Space}, we test the robustness of our orbital decomposition by comparing these fractions with those obtained using the phase-space circularity definition for the \texttt{survived LEGA-C-like} galaxies at both redshifts \citep{Xu2019MNRAS, Zhang2025}. The two methods recover the same overall trends, with only minor differences in the inferred orbital fractions.

\subsubsection{Star-forming versus quiescent galaxies}
\label{sec:star_forming_vs_quiescent_galaxies}

Orbital structure at $z = 1$: both populations exhibit a broadly similar global trend, in which warm and cold orbits dominate the low-mass regime, while hot orbits become important toward higher stellar masses. However, at fixed stellar mass, clear differences emerge between star-forming and quiescent galaxies. Star-forming galaxies consistently exhibit larger fractions of cold orbits than quiescent galaxies, indicating a higher degree of ordered rotation and dynamically colder stellar structures. A similar trend is observed for the hot component: at fixed stellar mass, quiescent galaxies host higher fractions of hot orbits across nearly the entire mass range. The warm orbital component dominates both populations over a broad stellar-mass range, but its behavior differs between the two classes. At stellar masses below $\log(M_\star/M_\odot)\approx10.9$, quiescent galaxies exhibit relatively higher warm fractions than star-forming galaxies. Above this mass scale, however, star-forming galaxies become increasingly dominated by warm orbits, whereas CR orbits remain relatively similar in both populations.

Orbital structure at $z = 0$: by this point, the dynamical distinction between star-forming and quiescent galaxies becomes more pronounced. Star-forming galaxies retain substantially higher cold-orbit fractions across nearly the full stellar-mass range, whereas quiescent galaxies contain little contribution from dynamically cold orbits. This difference helps explain the suppressed cold fraction of the overall \texttt{survived LEGA-C-like} population at $z=0$ relative to $z=1$ (Sect.~\ref{sec:all_galaxies}), as the galaxy population becomes increasingly quiescent with time.
The persistence of cold and warm orbits in star-forming galaxies suggests that continued in situ star formation and gas accretion can maintain or rebuild rotationally supported stellar structures. In contrast, the weak cold component of quiescent galaxies indicates that such disk growth becomes inefficient after quenching. The connection between star formation activity and orbital evolution therefore suggests that changes in the stellar dynamical structure are closely linked to the processes that regulate star formation.

\subsubsection{Central versus satellite galaxies}
\label{sec:central_vs_satellite_galaxies}

In contrast to the dependence on star formation activity, we find no significant difference between the orbital fractions of central and satellite galaxies at either $z=1$ or $z=0$. Within the statistical uncertainties, both populations follow the same stellar-mass-dependent trends in their cold, warm, hot, and CR components.

\begin{figure*}[t]
    \centering
    \includegraphics[width=1\textwidth]{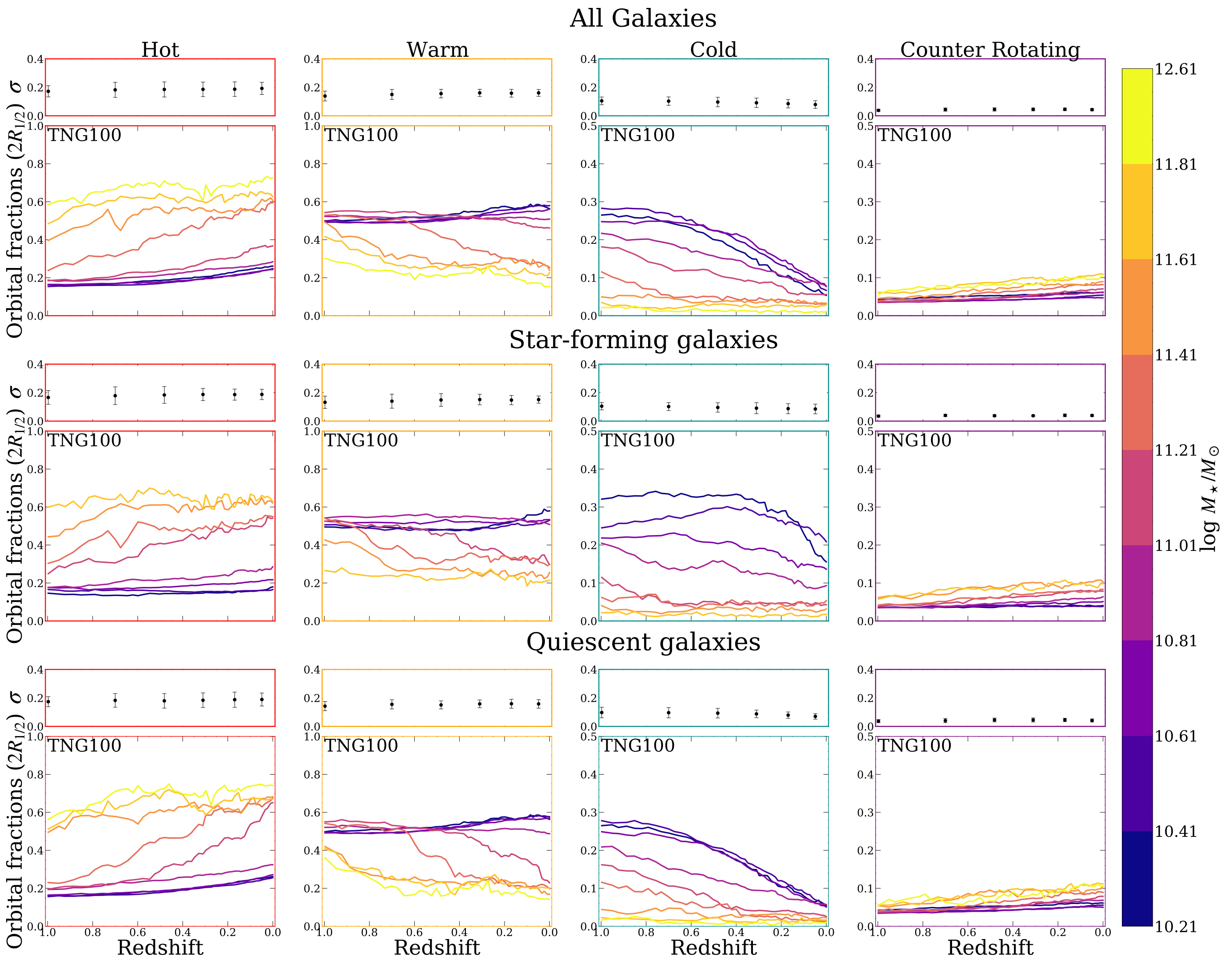}
    \caption{Evolution of the stellar orbital fractions from $z=1$ to $z=0$ for all galaxies (top), star-forming galaxies (middle), and quiescent galaxies (bottom) binned by their $z=0$ stellar mass. The upper row shows the corresponding mean standard deviation values ($\sigma$) of the orbital fractions (see Appendix~\ref{OrbitalEvolution_Stellar mass_Sigma}) . The orbital components show strongly mass dependent evolution: in the most massive bins, the hot (warm) component shows an increase (decline) from $z=1$ to $z=0$, whereas in the less massive bins, the warm components remain higher than the hot components across redshift. Meanwhile, the cold components continue to decline from $z=1$ to $z=0$, from their highest to their lowest fractions. Comparing star-forming and quiescent galaxies, we find that star-forming galaxies retain larger cold fractions, particularly at lower stellar masses.}

        \label{fig:orbital_evolution}
\end{figure*}

\subsection{Evolution of orbital structures}
\label{sec:orbitalevolution}

We go on to investigate the evolution of galaxy orbital structure from $z=1$ to $z=0$ by binning galaxies according to their stellar mass at $z=0$. This links present-day dynamical states to earlier orbital configurations and allows the trends to be interpreted in terms of cumulative assembly histories. Such direct evolutionary tracking is only possible in simulations, where individual galaxies can be uniquely followed across cosmic time. We therefore dynamically decompose the \texttt{survived LEGA-C-like} sample at every snapshot between $z=1$ and $z=0$.

Figure~\ref{fig:orbital_evolution} shows the evolution of the median orbital fractions from $z=1$ to $z=0$ for \texttt{all galaxies}, \texttt{star-forming} galaxies, and \texttt{quiescent} galaxies, grouped by their descendant stellar mass at $z=0$. The panels display the orbital components as a function of redshift, divided by stellar mass bins. For each galaxy population, the upper panel associated with each orbital component shows the mean standard deviation of the orbital fractions within each stellar mass bin at six different redshifts. This is included to quantify and visualize the typical scatter among individual galaxies in each mass bin as a function of redshift. We also provide the reader with a more detailed $1\sigma$ scatter in the orbital fractions, as well as the evolution of the median stellar mass (with $1\sigma$ scatter) for each mass bin at $z=0$ in Appendix~\ref{OrbitalEvolution_Stellar mass_Sigma}. Note that the classification of galaxies as star-forming or quiescent is performed only at $z=1$. We do not reclassify galaxies at each redshift, since galaxies transitioning between the two populations would otherwise move between bins over time, which would complicate the interpretation of the mass-bin evolution for the same set of galaxies.

\subsubsection{All galaxies}
\label{sec:all_galaxies_evo}

The evolution of the orbital structure changes around a present-day stellar mass of $\log(M_\star/M_\odot)\sim11.2$.  Galaxies in mass bins below this threshold have their highest cold-orbit fractions at $z=1$, indicating substantial rotational support at the beginning of the evolutionary tracks. The cold fraction then subsequently declines toward $z=0$, while the hot and CR components increase. Interestingly, over this entire period, the warm component remains dominant. Overall, the decrease in the cold component can be largely attributed to the rise in the other three components.

In contrast, galaxies above stellar mass $\log(M_\star/M_\odot) \approx 11.2$  at $z = 0$ already exhibit a substantial hot orbital component at $z = 1$, which increases steadily and monotonically, typically accounting for $\gtrsim 60\%$ of their stellar orbits. From $z = 1$ to $z = 0$, meanwhile, the warm fraction in these systems declines from $\sim 35$--$55\%$ at $z = 1$ to $\sim 15$--$25\%$ at $z = 0$, reflecting a systematic conversion of mildly non-circular orbits into dispersion-dominated motion. The cold component in these systems is already very low and remains relatively stable and subdominant throughout this period, maintaining a median fraction of $\lesssim 5\%$, while the CR component shows only a slight increase from $\sim 3\%$ to $\sim 8\%$. 

The mass-dependent evolutionary trends are insensitive to whether galaxies are grouped by their stellar mass at $z=0$ or $z=1$. We repeated the analysis using stellar-mass bins defined at $z=1$ and followed the same galaxies to $z=0$, recovering the same qualitative evolution of the orbital components across the full stellar-mass range. The transition near $\log(M_\star/M_\odot)\sim11.2$ therefore does not arise from our choice to bin galaxies by their descendant mass.

The already high hot-orbit fractions of the most massive progenitors at $z=1$ suggest that much of their dynamical transformation occurred at earlier epochs. This is consistent with an early assembly history involving substantial merger activity and dynamical heating at $z\gtrsim1$, as found for massive galaxies in \texttt{IllustrisTNG} \citep{Rodriguez2016}. Their persistently low cold fractions further suggest that the formation or regrowth of dynamically cold stellar discs becomes inefficient in these systems at later times. We investigate the role of mergers in establishing these orbital structures in Sect.~\ref{sec:mergers}.

Lower-mass galaxies instead undergo a more gradual loss of rotational support between $z=1$ and $z=0$. Their larger cold fractions at $z=1$ indicate more prominent dynamically cold structures, while the subsequent decline of the cold component and increase of dynamically hotter orbits point to progressive heating over cosmic time. Processes such as mergers, secular evolution, environmental interactions, and disk thickening may all contribute to this evolution, and we explore their relative importance in the following sections.

\subsubsection{Star-forming versus quiescent galaxies}
\label{sec:star_forming_vs_quiescent_galaxies_evo}

The strongest difference between galaxies classified as star-forming and quiescent at $z=1$ lies in the evolution of their cold-orbit fractions (second and third rows of Fig.~\ref{fig:orbital_evolution}). Among the lower-mass galaxies that are star-forming at $z=1$, the cold fraction remains comparatively stable toward $z=0$. In contrast, galaxies that are already quiescent at $z=1$ show a pronounced decline in their cold component, which becomes less than 5\% by $z=0$ across nearly all stellar-mass bins.

The persistence of cold orbits in the $z=1$ star-forming population could be linked  with the continued gas accretion and in situ star formation, helping to maintain or rebuild rotationally supported stellar structures. Conversely, the decline of cold orbits among galaxies already quiescent at $z=1$ could be explained by the reduced replenishment of dynamically cold stellar material, allowing dynamical heating to progressively diminish the cold component. These results therefore link the evolution of rotational support to the star formation state of galaxy progenitors, although the fixed $z=1$ classification means that we do not directly isolate the dynamical effect of the quenching process itself.

We also repeat the evolutionary analysis after separating galaxies into centrals and satellites. As in the instantaneous comparison in Sect.~\ref{sec:central_vs_satellite_galaxies}, we find no significant differences in their overall orbital evolution.

\begin{figure*}[t]
    \sidecaption
    \includegraphics[width=12cm]{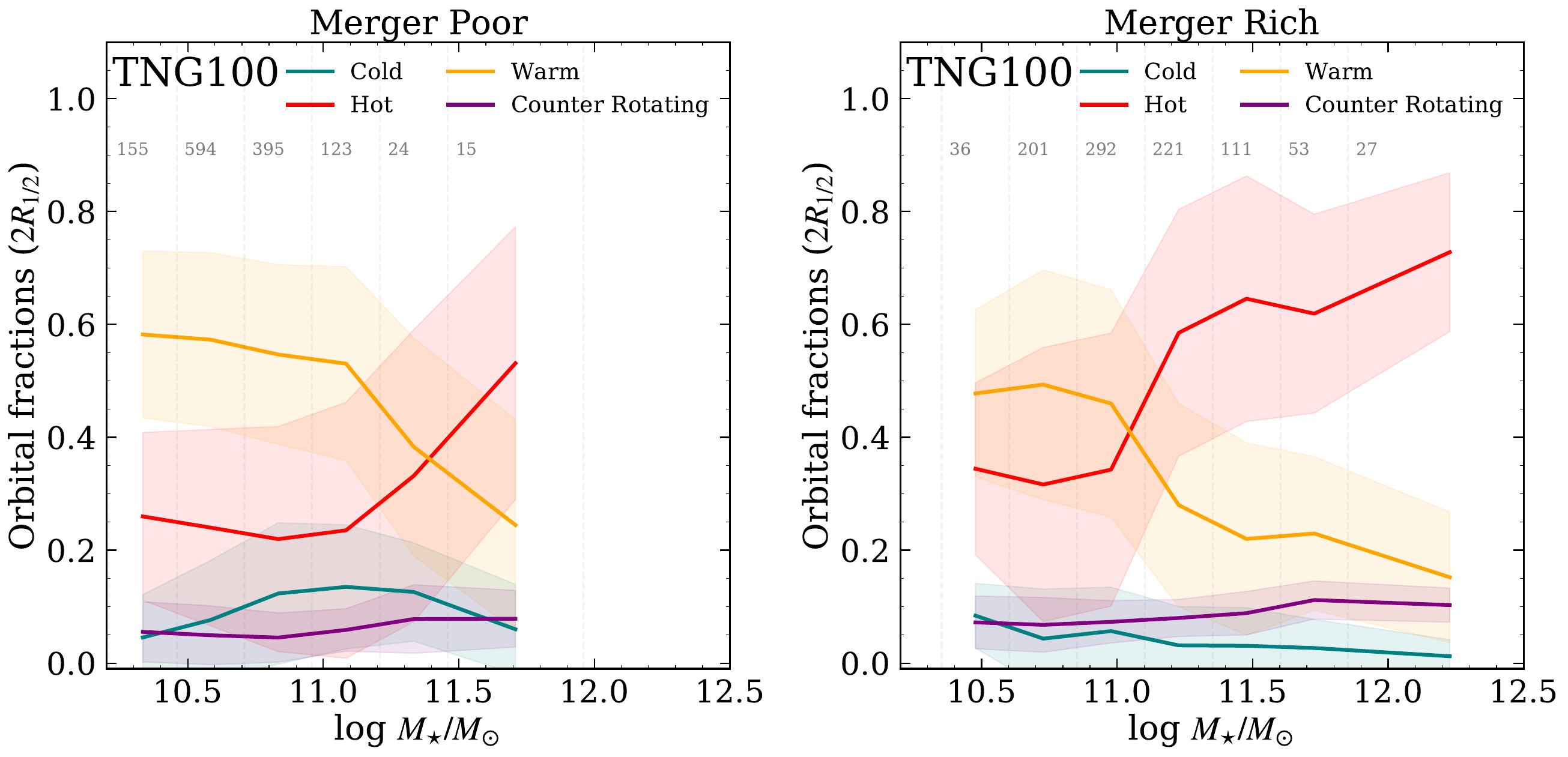}
    \caption{Median orbital mass fractions at $z=0$ as a function of stellar mass for \texttt{merger-poor} (left) and \texttt{merger-rich} (right) galaxies. \texttt{Merger-rich} galaxies experienced at least one merger with mass ratio $\mu>1/10$ between $z=1$ and $z=0$, whereas \texttt{merger-poor} galaxies experienced no merger above this threshold. \texttt{Merger-poor} galaxies remain predominantly warm-orbit dominated across most of the stellar-mass range and retain higher cold fractions than their \texttt{merger-rich} counterparts. In contrast, \texttt{merger-rich} galaxies undergo a pronounced transition from warm- to hot-orbit dominance around $\log(M_\star/M_\odot)\sim11$--$11.2$, with the hot fraction reaching $\sim70\%$ at the high-mass end. Their cold fraction remains below $\sim10\%$ and decreases with stellar mass, while the CR fraction increases mildly.}
    \label{fig:Combined_mergers_comparison}
\end{figure*}

\section{Impact of galaxy mergers on orbital components }
\label{sec:mergers}

The orbital evolution identified in Sect.~\ref{sec:orbitalevolution} raises the question of which physical processes drive the decline of cold orbits and the growth of dynamically hotter components. One potential mechanism for producing such changes is galaxy merging, which can redistribute angular momentum and dynamically heat stellar orbits. The strength of this effect depends on properties such as the merger mass ratio, gas fraction, and orbital configuration \citep{Read2008MNRAS, Pillepich2015ApJ, Zhu2022A&A, Wu2025}. We therefore investigate the role of mergers in shaping stellar orbital structure by comparing galaxies with different merger histories between $z=1$ and $z=0$.

\subsection{\texttt{Merger-rich} and \texttt{merger-poor} galaxies}
\label{sec:with_without_mergers}

Figure~\ref{fig:Combined_mergers_comparison} compares the orbital fractions at $z=0$ of galaxies with and without significant merger activity since $z=1$. We define \texttt{merger-rich} galaxies as those that experienced at least one merger with a stellar mass ratio $\mu>1/10$ between $z=1$ and $z=0$, while \texttt{merger-poor} galaxies experienced no mergers above this threshold during the same period.

Galaxies in the \texttt{merger-rich} sample display a small peak in the warm component and a corresponding dip in the hot fraction at the low-mass end around $\log(M_\star/M_\odot) \sim 10.6$, suggesting that lower-mass galaxies that have experienced mergers between $z=1$ and $z=0$ can still retain a degree of ordered motion at this epoch. Beyond $\log(M_\star/M_\odot) \sim 10.6$, the warm component declines while the hot fraction rises sharply, with a clear warm-to-hot transition occurring around $\log(M_\star/M_\odot) \sim 11$--$11.2$. Above this threshold, the hot component becomes dominant, reaching fractions of $\sim 70\%$ at the high-mass end. The cold component remains under 10\% throughout and declines continuously from the low- to high-mass end, while the CR fraction shows a mild increase toward higher masses. This behavior reflects the increasing importance of merger-driven heating in massive galaxies.

In contrast, \texttt{merger-poor} galaxies remain predominantly warm-orbit dominated across most of the stellar-mass range. The warm fraction contributes $\sim60\%$ at the low-mass end and decreases only gradually with increasing mass. The hot fraction remains comparatively modest at $\sim 25\%$ at the low-mass end and increases only mildly with stellar mass, with a more noticeable rise visible above $\log(M_\star/M_\odot) \sim 11-11.2$, where warm orbits begin to decline. The cold component shows a subtle peak around $\log(M_\star/M_\odot) \sim 11.2$ before declining, while the CR fraction remains low and nearly flat throughout. Overall, the absence of significant mergers between $z=1$ and $z=0$ allows these systems to preserve a substantial degree of ordered rotation. 

The contrast between the \texttt{merger-rich} and \texttt{merger-poor} populations at $z=0$ shows that even mergers occurring in the relatively recent past, between $z=1$ and $z=0$, play a significant role in driving the transformation from warm orbital configurations to hot, dispersion-dominated structures. This is particularly evident in massive galaxies above $\log(M_\star/M_\odot) \sim 11-11.2$, where the hot fraction in the \texttt{merger-rich} sample substantially exceeds that of their \texttt{merger-poor} counterparts at the same stellar mass.

Our \texttt{merger-poor} classification does not imply that these galaxies are completely merger-free. They may experience mini mergers with $\mu<1/10$ between $z=1$ and $z=0$ or mergers at $z>1$, neither of which enters our classification. We therefore use \texttt{merger-poor} specifically to denote galaxies without mergers above $\mu=1/10$ during the redshift interval considered here. Despite this distinction, their systematically warmer orbital structures compared with the \texttt{merger-rich} population further support an important role for significant mergers in driving dynamical heating.

\begin{figure*}[t]
    \centering
    \includegraphics[width=1\textwidth]{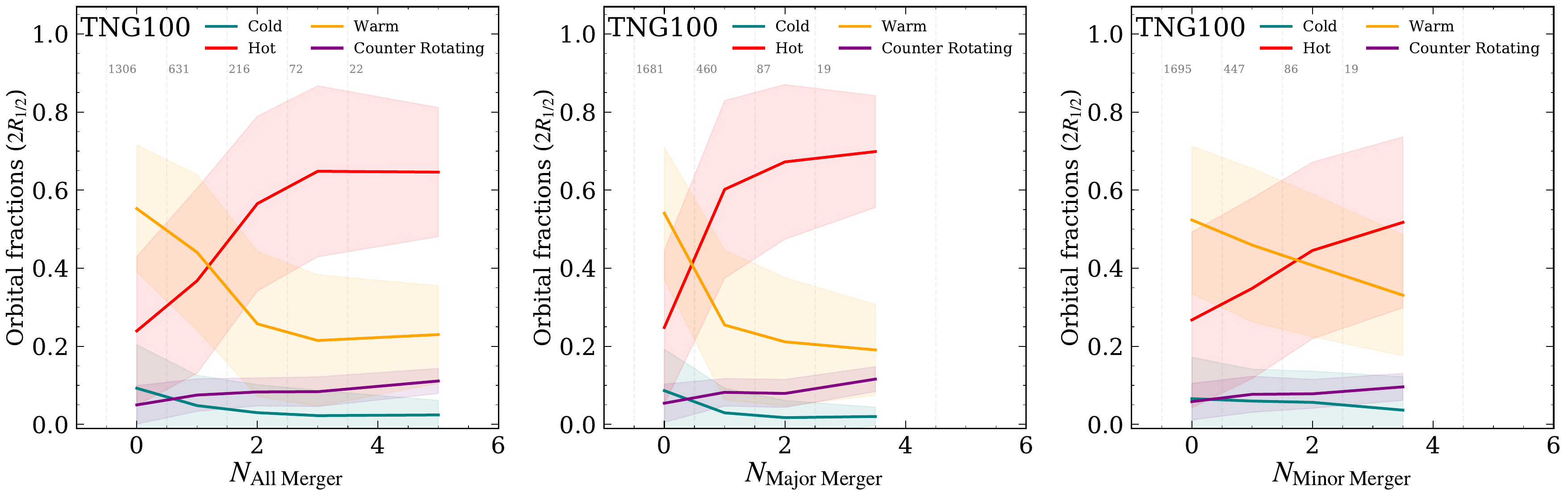}
    \caption{Median stellar orbital fractions at $z=0$ as a function of the number of mergers experienced between $z=2$ and $z=0$ for galaxies in the \texttt{merger-rich} sample. Left, middle, and right panels: Dependence on the total number of mergers, number of major mergers ($\mu \geq 1/4$), and the number of minor mergers ($1/10 \leq \mu < 1/4$), respectively. The orbital structure changes systematically with merger incidence: galaxies that experience more mergers become increasingly dominated by hot orbits, while the warm and, to a lesser extent, cold components decline. This transformation is strongest for major mergers, for which the hot fraction rises rapidly and becomes dominant after only a small number of events. }
    \label{fig:Num_major_minor_mergers_z0}
\end{figure*}

\subsection{Effects of major and minor mergers}
The dynamical impact of a merger is expected to depend strongly on its mass ratio. To examine this dependence, we subdivide the \texttt{merger-rich} sample into galaxies that experienced major mergers ($\mu \geq 1/4$) and those that underwent minor mergers ($1/10 \leq \mu < 1/4$) between $z=1$ and $z=0$. In Fig.~\ref{fig:Num_major_minor_mergers_z0}, we present the median orbital fractions at $z=0$ as a function of the total number of mergers and, separately, the numbers of major and minor mergers.

Across all merger types, the orbital components show a clear dependence on merger frequency. Galaxies that experienced more mergers exhibit lower cold and warm fractions and correspondingly higher hot and CR fractions. However, the rate of this transformation differs substantially between major and minor mergers. Major mergers produce a more rapid change in orbital structure: even a single major merger significantly suppresses the warm and cold components while enhancing the hot fraction. In contrast, galaxies undergoing minor mergers show a more gradual evolution with increasing merger number. Although the qualitative trends are similar, a comparable dynamical transformation requires substantially more minor-merger events. For example, reaching a hot-orbit fraction of $\sim50\%$ typically requires approximately twice as many minor mergers as major mergers.

To further assess this difference, Appendix~\ref{MajorMinorMerger} examines the stellar-mass dependence of the orbital fractions when the \texttt{merger-rich} sample is separated according to major- and minor-merger contributions (Fig.~\ref{fig:Combined_mergers_comparison}). Across the full stellar-mass range, galaxies that experienced major mergers have systematically higher hot and CR fractions and lower warm and cold fractions than those dominated by minor mergers. The same qualitative difference remains when restricting the comparison to galaxies with exclusively major- or exclusively minor-merger histories.

These results show that the mass ratio regulates how efficiently mergers alter stellar orbital structure. However, merger mass ratio alone does not determine the outcome: the amount of gas involved in the interaction can strongly influence whether rotational support is destroyed, preserved, or subsequently rebuilt. We therefore next examine how the orbital structure depends on the gas content of mergers.
\begin{figure*}[t]
    \sidecaption
    \includegraphics[width=12cm]{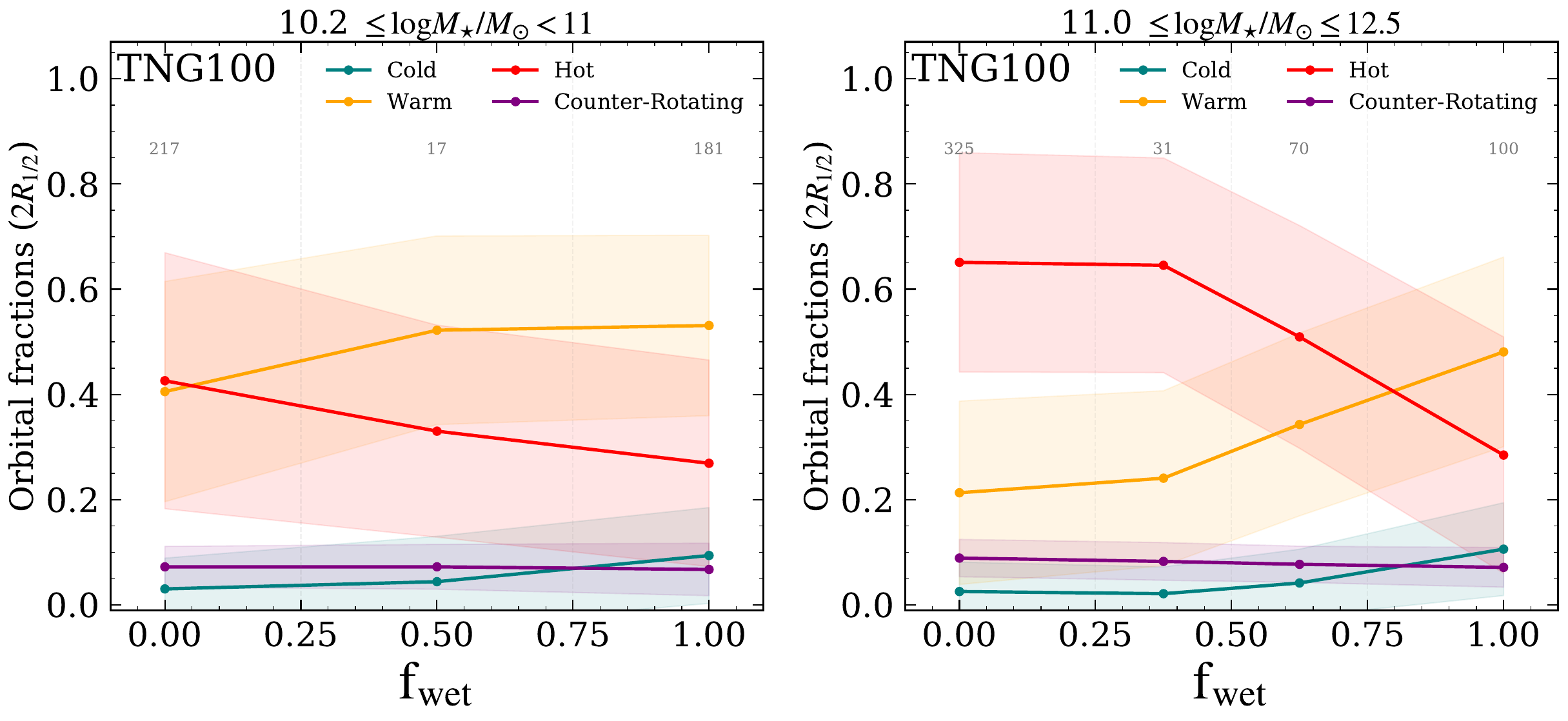}
    \caption{Median orbital fractions, at $z=0$, as a function of the wet-to-total merger ratio, 
    $f_{\rm wet} = N_{\rm wet}/(N_{\rm wet}+N_{\rm dry})$. Mergers are classified as wet or dry using a merger gas fraction threshold of $f_{\rm gas,\,merger}=0.2$. The warm and cold orbital fraction generally increases with $f_{\rm wet}$, while the hot and CR fraction decreases. Data are divided into two stellar mass bins: $10.2 \leq \log M_\star/M_\odot < 11$ and $11 \leq \log M_\star/M_\odot < 12.5$. At fixed $f_{\rm wet}$, the more massive galaxies exhibit higher hot fractions and lower warm and cold fractions than lower-mass galaxies.}
    \label{fig:WetToDryMerger}
\end{figure*}

\subsection{Impact of wet and dry mergers}
\label{sec:WetandDry_Mergers}

In addition to mass ratio, the gas content of merging galaxies can strongly influence the dynamical outcome of a merger. Gas-rich (wet) mergers are dissipative: gas can radiatively cool and subsequently settle into rotationally supported configurations, while merger-induced star formation can contribute to the survival or regrowth of a disk \citep{Barnes1996ApJ, Springel2003, Robertson2006A}. Gas-poor (dry) mergers, in contrast, are dominated by collisionless stellar dynamics and violent relaxation, favoring the formation of dynamically hotter and more slowly rotating remnants \citep{Naab2006MNRAS, Hopkins2009MNRA, Hoffman2009}.

To quantify the role of merger gas content, we classify mergers as wet or dry using the combined gas fraction of the merging system. Following \citet{Lagos2018MNRAS}, we define

\begin{equation}
f_{\rm gas,merger}
= \frac{M_{\rm gas}^{\rm primary} + M_{\rm gas}^{\rm secondary}}
{M_{\star}^{\rm primary} + M_{\star}^{\rm secondary}},
\end{equation}

where the gas and stellar masses are measured at the snapshot when the secondary progenitor reaches its maximum stellar mass prior to merging, consistent with our definition of the merger mass ratio. Mergers with $f_{\rm gas,merger}<0.2$ are classified as dry, while those with $f_{\rm gas,merger}\geq0.2$ are classified as wet.
Figure~\ref{fig:WetToDryMerger} presents the median orbital fractions at $z=0$ as a function of the wet-to-total merger ratio.

\begin{equation}
f_{\rm wet} =
\frac{N_{\rm wet}}
{N_{\rm wet}+N_{\rm dry}},
\end{equation}

where $f_{\rm wet}=0$ corresponds to a purely dry merger history and $f_{\rm wet}=1$ to a purely wet merger history. The results are shown for two stellar-mass bins, $10.2 \leq \log(M_\star/M_\odot)<11.0$ and $11.0 \leq \log(M_\star/M_\odot)\leq12.5$ at $z=0$.

Across both mass bins, the orbital structure varies systematically with $f_{\rm wet}$. The warm fraction increases with increasing $f_{\rm wet}$, while the hot fraction declines. This trend is particularly pronounced at high stellar masses, where galaxies with predominantly dry merger histories are strongly dominated by hot orbits, whereas galaxies with increasingly wet merger histories contain substantially larger warm components. The cold fraction also increases with $f_{\rm wet}$, although it remains subdominant to the warm component. Conversely, the CR fraction shows a mild decline toward increasingly wet merger histories.

These trends are consistent with the different dynamical effects expected from dissipative and dissipation-less merging. Using the \texttt{EAGLE} simulations, \citet{Lagos2018MNRAS} similarly found that dry mergers lead to stronger reductions in stellar specific angular momentum and produce higher Sérsic indices and kinematically defined bulge-to-total ratios than wet mergers.

At fixed $f_{\rm wet}$, massive galaxies ($\log(M_\star/M_\odot)\gtrsim11$) exhibit systematically higher hot fractions and lower warm and cold fractions than their lower-mass counterparts. Thus, although merger gas content is strongly associated with the resulting orbital structure, stellar mass introduces an additional dependence, consistent with the different assembly histories of low- and high-mass galaxies \citep{Bundy2009ApJ}.

Galaxies that experienced both wet and dry mergers contribute to both $N_{\rm wet}$ and $N_{\rm dry}$ in the definition of $f_{\rm wet}$. We verified that restricting the analysis to galaxies with exclusively wet or exclusively dry merger histories does not qualitatively change the trends described above.

\begin{figure}[t]
    \centering
    \includegraphics[width=0.4\textwidth]{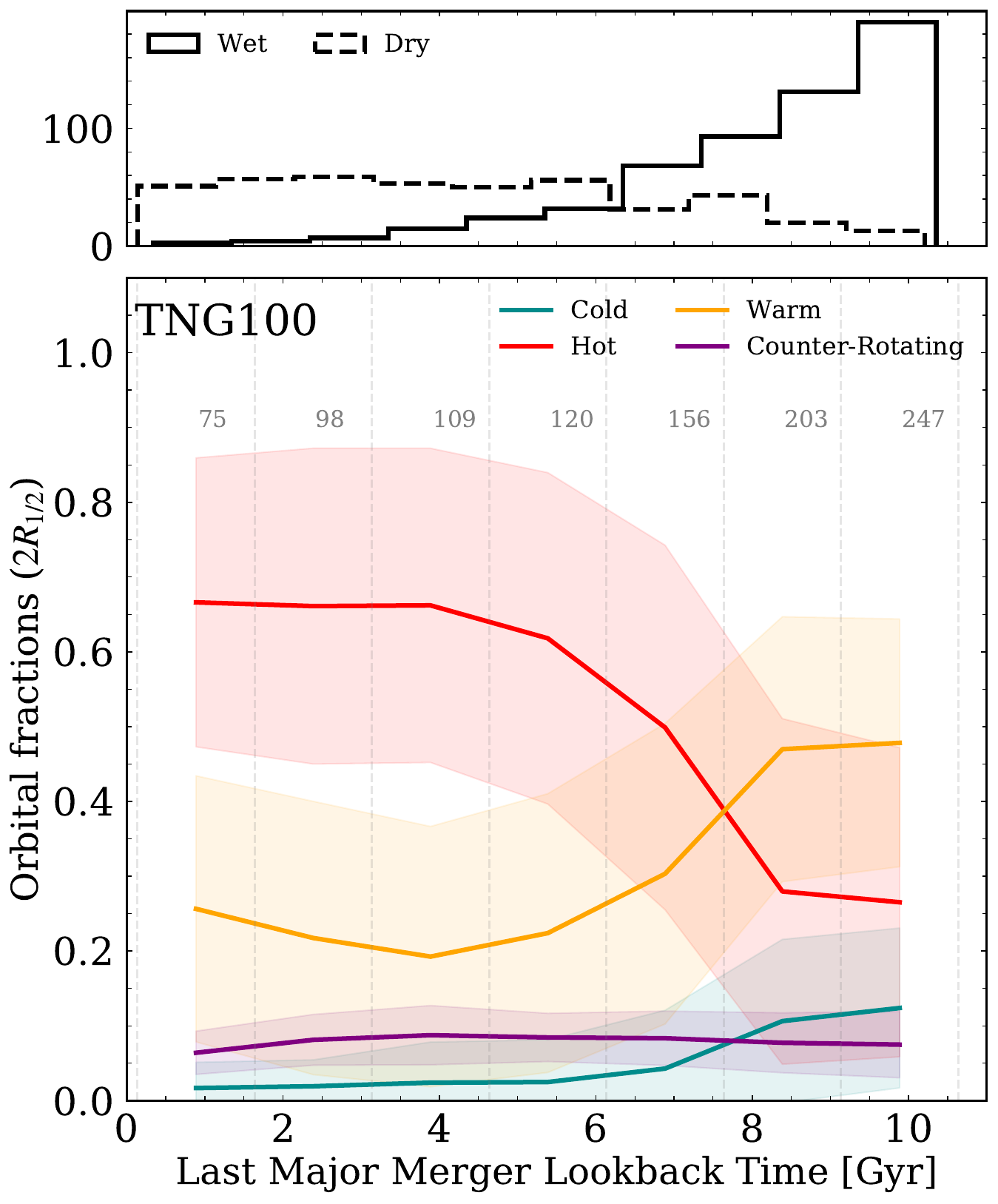}
    \caption{Top: Number of \texttt{TNG100} galaxies as a function of the look-back time to the last major merger, subdivided into wet (gas-rich; solid line) and dry (gas-poor; dashed line) mergers. Mergers occurring at earlier epochs are predominantly wet,  while the dry-merger fraction increases towards more recent times.
    Bottom: Median stellar orbital fractions at $z=0$ as a function of the look-back time to the last major merger, defined as the most recent merger with stellar mass ratio of $\mu \geq 1/4$. Cold, warm, hot, and CR orbital components are shown, with the shaded regions indicating the $1\sigma$ scatter. Galaxies with recent major mergers are dominated by hot orbits, whereas systems with early last major mergers exhibit higher warm and cold orbital fractions, indicating rotational support.}
    \label{fig:LastMajorMerger}
\end{figure}

\subsection{Orbital component and time since last major merger}
\label{sec:LastMajorMerger}

Having established that both merger mass ratio and gas content influence the resulting orbital structure, we next examine whether the timing of the merger also plays a role. The look-back time to the last major merger provides a measure of how long a galaxy has had to dynamically evolve after a major interaction. Recent major mergers are expected to leave a stronger imprint on the present-day orbital structure, whereas signatures of mergers occurring at earlier epochs may be modified by subsequent star formation, disk regrowth, and dynamical evolution. We therefore examine the orbital fractions at $z=0$ as a function of the look-back time to the last major merger ($\mu \geq 1/4$), identified from the \texttt{SUBLINK} merger trees.

Figure~\ref{fig:LastMajorMerger} shows a clear dependence of the present-day orbital structure on the timing of the last major merger. The upper panel additionally shows that major mergers occurring at earlier epochs were predominantly wet, whereas the relative contribution of dry mergers increases toward more recent times.

Galaxies whose last major merger occurred recently ($\lesssim 4$--$5~\mathrm{Gyr}$ ago) are strongly dominated by hot, dispersion-supported orbits, with median hot fractions exceeding $\sim60\%$. The warm component is substantially reduced, while the cold fraction contributes only a few percent of the stellar mass. The predominance of relatively dry mergers at these recent epochs is consistent with the limited recovery of rotational support, since little cold gas is available for subsequent disk regrowth. This interpretation is also consistent with the broader population evolution shown by the Sankey diagram in Fig.~\ref{fig:star_formation}, where the surviving LEGA-C-like sample evolves from being predominantly star-forming at $z=1$ to predominantly quiescent at $z=0$.

For intermediate look-back times ($\sim5$--$8~\mathrm{Gyr}$), the orbital composition changes progressively. The hot fraction declines, while the warm component increases and becomes comparable to or exceeds the hot fraction by $\sim7$--$8~\mathrm{Gyr}$. The cold fraction also increases modestly. These trends indicate that galaxies with a longer interval since their last major merger have, on average, regained a larger degree of rotational support.

Galaxies whose last major merger occurred at earlier epochs ($\gtrsim9~\mathrm{Gyr}$ ago) show a markedly different present-day orbital structure. The warm component becomes dominant, contributing $\sim40$--$45\%$ of the stellar mass, while the hot fraction decreases to $\sim25$--$30\%$. The cold fraction reaches its highest median values, although it remains subdominant. These systems also predominantly experienced wet major mergers, suggesting that the combination of a gas-rich interaction and a long post-merger evolutionary interval favors the preservation or rebuilding of rotationally supported stellar components. Thus, the dynamical imprint of an early major merger can be substantially reduced by subsequent evolution. In contrast, the CR component shows only a weak dependence on the timing of the last major merger, remaining at the $\sim5$--$10\%$ level across the full range of look-back times.

\begin{figure*}[t]
    \centering
    \includegraphics[width=1\textwidth]{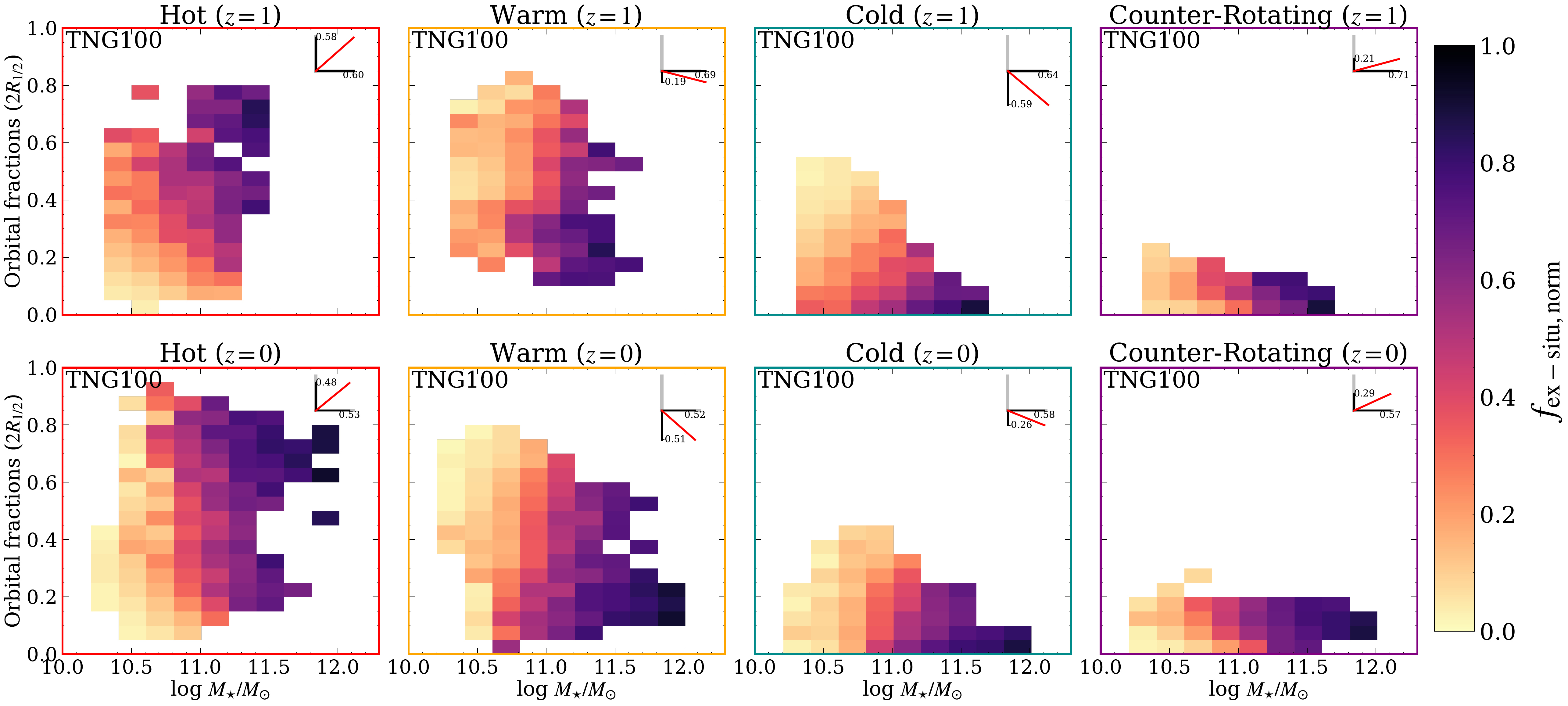}
    \caption{Fractions of orbital components a function of stellar mass. Top row: Results at $z = 1$. Bottom row: Corresponding distributions at $z = 0$. In each panel, the color scale indicates the normalized median ex situ stellar mass fraction of galaxies within each bin. Only bins containing at least four galaxies are shown. At both redshifts, the orbital fractions appears to have double dependence on stellar mass and ex situ fraction. The inset in each panel summarizes the partial Pearson correlation coefficients (PCCs), with the $x$-axis showing $\rho_{f_{\rm ex\text{-}situ},M_\star| f_{\rm orb}}$ and the $y$-axis showing $\rho_{f_{\rm ex\text{-}situ},f_{\rm orb} | M_\star}$. The red line indicates the combined direction of the two partial correlations, while the black lines on each axis show the individual PCC values corresponding to the stellar-mass and orbital-fraction directions.}
    \label{fig:exsitu_correlation}
\end{figure*}

\section{Orbital components and ex situ growth}
\label{sec:ExsituMass}

The previous analysis showed that the present-day orbital structure depends on several aspects of merger history, including merger mass ratio, gas content, and timing. These quantities characterize individual merger events, but do not directly measure the cumulative contribution of externally assembled stars to a galaxy. We therefore complement the merger-based analysis by examining the ex situ stellar mass fraction, $f_{\rm ex\text{-}situ}$, which provides an integrated measure of stellar mass acquired outside the main progenitor.

In the \texttt{IllustrisTNG} framework, $f_{\rm ex\text{-}situ}$ is defined as the fraction of a galaxy's stellar mass that formed outside its main progenitor branch and was subsequently accreted. Stellar particles are traced to their formation sites through the merger trees and classified as in situ if they formed within the main progenitor branch, and ex situ otherwise \citep{Rodriguez2016, Rodriguez2015, Rodriguez2017MNRAS}. The ex situ component includes stars accreted through completed and ongoing mergers, as well as contributions from flybys or stripping of surviving systems and a small fraction of stars formed outside any identified bound galaxy. The ex situ fraction therefore traces the cumulative importance of external stellar assembly, although it is not a direct proxy for the number of mergers, since mergers may also supply gas that subsequently forms stars in situ.

\subsection{Orbital structure–ex situ fraction correlations}

Figure~\ref{fig:exsitu_correlation} shows two-dimensional histograms of the median ex situ stellar mass fraction, ($f_{\rm ex\text{-}situ}$), in orbital fraction–stellar mass space. Each bin is color-coded according to the normalized median ($f_{\rm ex\text{-}situ}$) measured within ($2R_{1/2}$).
 At both redshifts, $z=1$ and $z=0$, galaxies with $\log(M_\star/M_\odot) > 11$ exhibit higher ex situ fractions compared to lower-mass galaxies. In general, at fixed $M_\star$, galaxies with lower $f_{\rm ex\text{-}situ}$ tend to have higher cold and warm orbital fractions. This suggests a clear dependence of the ex situ mass fraction on both stellar mass and the distribution of orbital components. To disentangle these coupled dependencies, we use Partial Correlation Coefficient (PCC), which allows us to remove the mutual inter-correlations between variables and isolate the intrinsic dependence of one parameter on another while controlling for a third variable \citep{Bait2017MNRAS, Bluck2020MNRAS, Bluck2020MNRASB, Baker2022MNRAS}. The PCC is given by

\begin{equation}
\rho_{xy,z} =
\frac{\rho_{xy} - \rho_{xz}\rho_{yz}}
{\sqrt{(1-\rho_{xz}^{2})(1-\rho_{yz}^{2})}},
\end{equation}
where $\rho_{xy,z}$ is the partial correlation coefficient between variables $x$ and $y$ after controlling for a third variable $z$, and $\rho_{xy}$, $\rho_{xz}$, and $\rho_{yz}$ are the corresponding Pearson correlation coefficients.
For each orbital component, we compute two PCCs. Here, $x$, $y$, and $z$ represent the stellar mass, ex situ stellar mass fraction, and orbital fraction, with the role assigned to each variable depending on the specific correlation being evaluated.

In the inset of each panel in Fig.~\ref{fig:exsitu_correlation}, the $x$-axis shows 
$\rho_{f_{\rm ex\text{-}situ},\,M_\star \,|\, f_{\rm orb}}$, 
while the $y$-axis shows 
$\rho_{f_{\rm ex\text{-}situ},\,f_{\rm orb} \,|\, M_\star}$. 
The red line therefore indicates the combined direction of the two partial correlations, summarizing whether the ex situ fraction is primarily driven by stellar mass, by the orbital fraction, or by both. The grey lines mark the coordinate axes, and the black markers on each axis show the individual PCC values along the stellar-mass and orbital-fraction directions, respectively.

A consistent result across all orbital components is that the ex situ stellar mass fraction correlates strongly with stellar mass at both redshifts, with partial correlation coefficients of $\rho \sim 0.6$--$0.7$ at $z=1$ and $\rho \sim 0.5$--$0.6$ at $z=0$. Thus, independently of orbital structure, more massive galaxies generally contain larger accreted stellar fractions. At fixed stellar mass, the orbital components show distinct correlations with the ex situ fraction. The cold-orbit fraction is negatively correlated with the ex situ fraction at both redshifts, with a stronger relation at $z=1$ ($\rho=-0.59$) than at $z=0$ ($\rho=-0.26$). In contrast, the hot-orbit fraction shows a positive correlation of similar strength at both redshifts ($\rho\sim0.5$), while the CR fraction exhibits a weaker positive correlation ($\rho\sim0.2$--$0.3$). The warm component shows a redshift evolution: its weak anti-correlation at $z=1$ ($\rho=-0.19$) becomes substantially stronger by $z=0$ ($\rho=-0.51$). A comparison with the \texttt{extended LEGA-C-like} sample shows a correlation of the ex situ fraction with warm orbit, with $\rho = -0.19 \quad$ $(z=1)$, and $\rho = -0.41 \quad$ $(z=0)$. This indicates that the discrepancy primarily originates from sample-selection effects associated with the surviving target population.  Overall, larger ex situ fractions are associated with reduced cold and warm components and enhanced hot and CR components.

\begin{figure*}[t]
    \sidecaption
    \includegraphics[width=12cm]{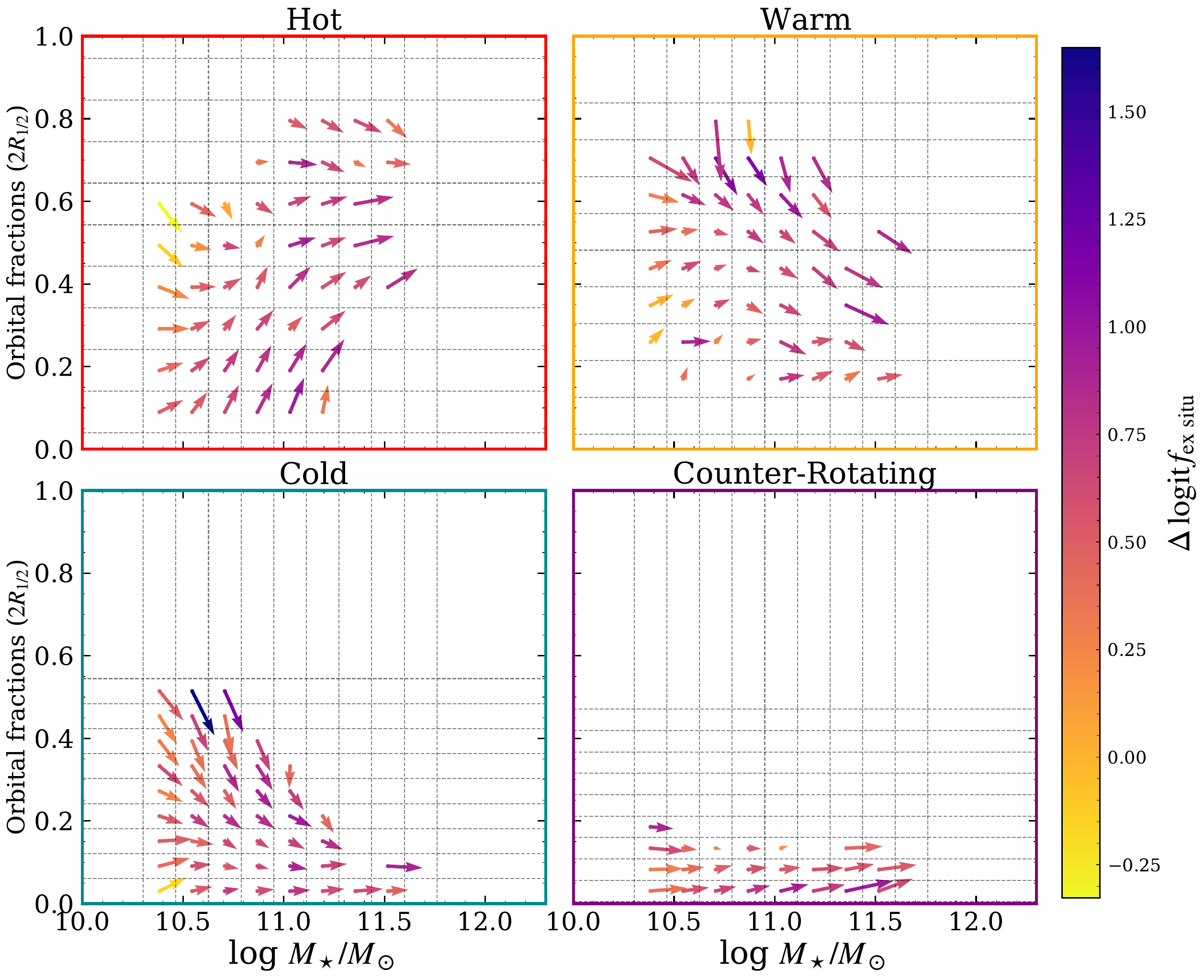}
    \caption{
    Statistical evolution of galaxies in orbital-fraction–stellar-mass space from $z=1$ to $z=0$.
    Galaxies are binned by stellar mass and orbital fraction at $z=1$, with each bin containing at least five systems.
    Arrows connect the median location of each bin at $z=1$ (tail) to that at $z=0$ (head), indicating the typical evolutionary direction.
    The arrow colour encodes the change in the logit-transformed ex situ stellar mass fraction,
    $\Delta\logit(f_{\rm ex situ}) = \logit(f_{\rm ex situ}^{z=0}) - \logit(f_{\rm ex situ}^{z=1})$.
    Larger positive values correspond to stronger relative growth of the ex situ component between $z=1$ and $z=0$, while values near zero indicate little net change.}
        \label{fig:exsitu_orbital_tracks_arrow}
\end{figure*}

\begin{figure*}[t]
    \sidecaption
    \includegraphics[width=12cm]{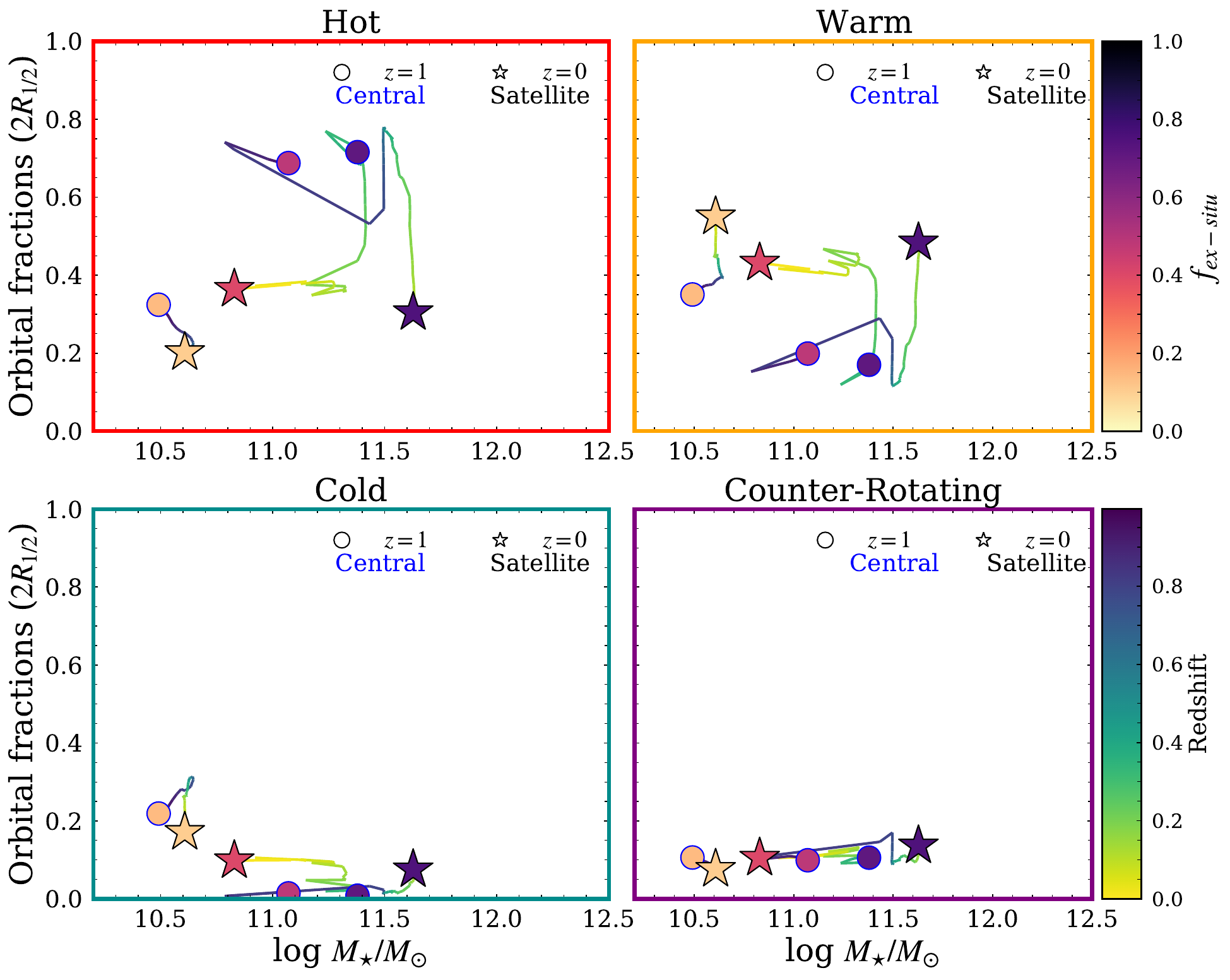}
    \caption{
    Evolution of individual galaxies in orbital-fraction–stellar-mass space from $z=1$ to $z=0$.
    Circles mark the location of each galaxy at $z=1$, while stars indicate their descendants at $z=0$; lines connect the two epochs and trace the evolutionary path of each system.
    Galaxies are color-coded by their ex situ stellar mass fraction, and marker edge colors distinguish central and satellite galaxies. The sample consists of three representative systems selected at $z=1$. The diversity of trajectories illustrates that galaxies with similar stellar masses and orbital fractions at $z=1$ can follow markedly different evolutionary paths toward $z=0$.
    }
    \label{fig:exsitu_orbital_tracks}
\end{figure*}

\subsection{Evolutionary trajectories in orbital fraction--stellar mass space}
\label{sec:exsitu_orbital_tracks}

We now investigate the statistical evolution of galaxies in orbital-fraction–stellar-mass space from $z=1$ to $z=0$. This enables us to identify population-level trends in the evolution of different orbital components and to connect these changes to the build-up of ex situ stellar mass through galaxy assembly processes.

In Fig.~\ref{fig:exsitu_orbital_tracks_arrow}, galaxies are binned by stellar mass and orbital fraction at $z=1$, with a minimum of five galaxies per bin. The median positions of each bin at $z=1$ and $z=0$ are connected by arrows, whose tails and heads mark the respective redshifts. Their length and direction indicate the magnitude and preferred pathway of orbital evolution. The arrow colour represents the change in the logit-transformed ex situ stellar mass fraction,
$
\Delta \logit(f_{\rm ex\text{-}situ}) =
\logit(f_{\rm ex\text{-}situ}^{z=0}) -
\logit(f_{\rm ex\text{-}situ}^{z=1}),
$
where $\logit(x)=\ln(x/(1-x))$.\footnote{The logit transformation maps $0<x<1$ onto the full real line, allowing relative changes at low and high ex situ fractions to be compared more directly.}

Figure~\ref{fig:exsitu_orbital_tracks_arrow} reveals clear preferred directions in the evolution of galaxies through orbital-fraction–stellar-mass space. In the hot component, galaxies generally evolve toward both higher stellar masses and higher hot-orbit fractions. The strongest upward evolution is observed for systems that initially possess relatively low hot fractions, and these bins also exhibit the largest increases in ex situ stellar mass fraction. This suggests that the build-up of hot stellar orbits is closely connected to ex situ mass assembly. A similar behavior is visible in the warm component, where galaxies with initially large warm fractions evolve toward lower warm-orbit fractions while simultaneously experiencing strong ex situ growth. Part of this evolution between hot and warm components may reflect the imposed hard separation between these orbital classes, since stars can transition between warm and hot orbital categories during merger-driven dynamical heating.
The cold component shows a more differentiated evolutionary behavior. Galaxies with already low cold fractions at $z=1$ tend to evolve primarily horizontally in the diagram, indicating substantial stellar mass growth but comparatively little change in their cold-orbit content. In contrast, galaxies with initially high cold fractions show a much stronger downward evolution in orbital fraction, implying a significant depletion of cold disk-like orbits over time. These systems are again associated with larger increases in ex situ fraction, supporting a scenario in which accretion and merger activity contribute to the progressive destruction or heating of dynamically cold stellar structures. The CR component displays predominantly horizontal evolution which shows that its fractional contribution remains approximately constant across most of the stellar mass range despite continued stellar mass growth. Only at the highest stellar masses do the arrows show a modest upward tilt.

While the previous sections focused on median trends and population-level correlations, we now test the evolutionary trajectories of individual galaxies in orbital-fraction–stellar-mass space. This enables a direct comparison between individual-galaxy evolution and statistical trends, allowing us to assess whether the observed population-level relations are driven by coherent evolutionary pathways or by a mixture of diverse histories.

We show this by selecting three random galaxies, as shown in Fig. \ref{fig:exsitu_orbital_tracks}. Each galaxy is shown at $z=1$ (circles) and at $z=0$ (stars), with symbols connected by lines to trace their evolutionary trajectories. Galaxies are color-coded by their ex situ stellar mass fraction, while the marker edge color denotes their central or satellite status. We observe a strong diversity in the evolutionary paths: the selected galaxies do not follow a common direction of motion in the stellar mass–orbital fraction space. Instead, their trajectories spread out substantially by $z=0$, with some systems evolving toward higher hot fractions and larger stellar masses, while others exhibit comparatively modest changes.

Moreover, this behavior is not unique to the particular example shown in the figure. Repeating the same exercise multiple times with different randomly selected galaxies consistently reveals a similarly broad diversity of evolutionary tracks. Furthermore, we repeated the analysis by selecting galaxies within narrow bins of stellar mass and orbital fraction at $z=1$. Even for galaxies that begin with nearly identical initial conditions in this parameter space, we do not observe a unified evolutionary direction toward $z=0$. Some systems evolve toward hotter orbital structures, others remain relatively unchanged, and some even evolve in opposite directions. This scatter highlights the stochastic and highly individual nature of galaxy evolution, demonstrating that galaxies with similar initial masses and orbital compositions can nonetheless experience markedly different evolutionary outcomes. Taken together, these figures demonstrate that while individual galaxies exhibit highly diverse and non-monotonic evolutionary paths, a statistical treatment reveals well-defined and physically interpretable trends.

\section{Discussion}
\label{sec:discussion}

\begin{figure*}[t]
    \centering
    \includegraphics[width=1\textwidth]{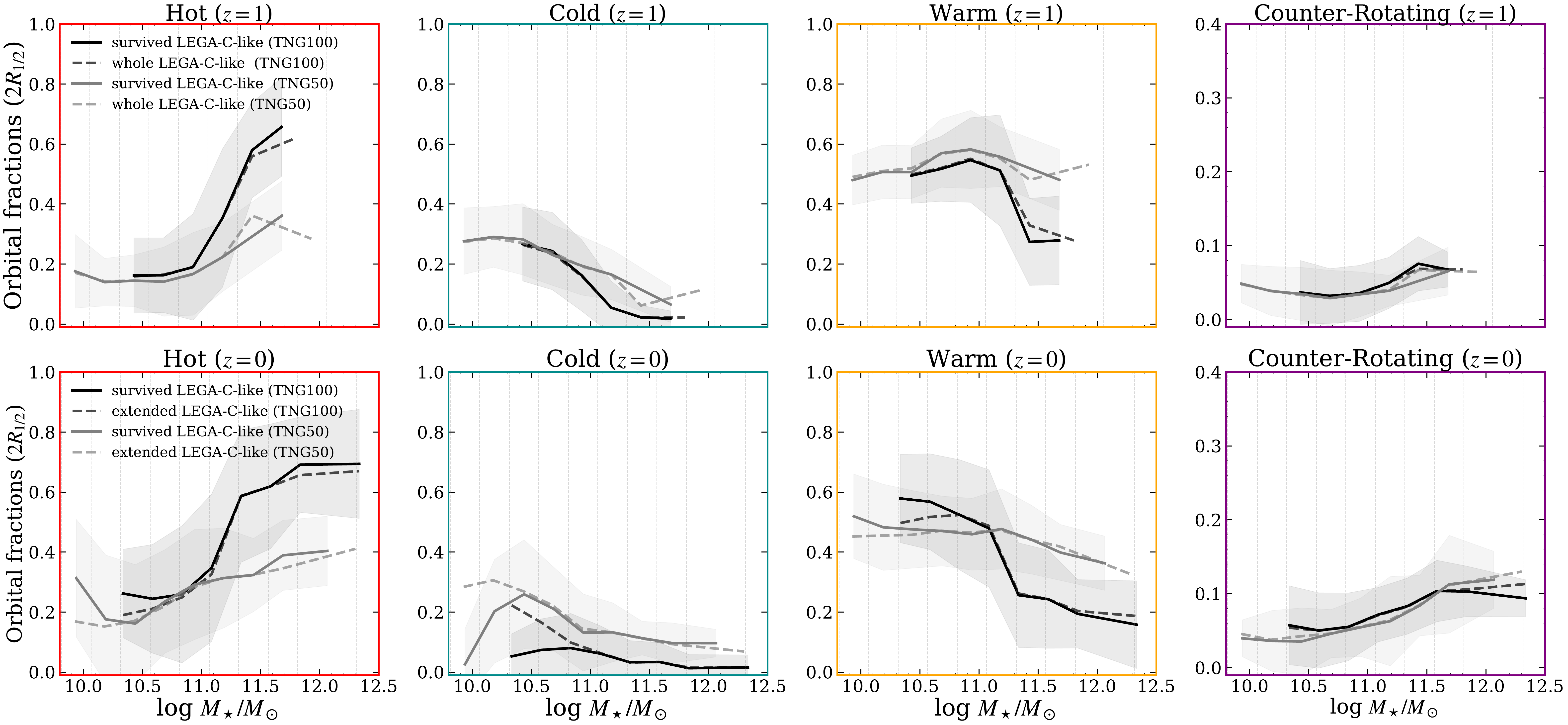}
    \caption{Comparison of median orbital fractions between \texttt{TNG100} and \texttt{TNG50} simulations as a function of stellar mass at $z=1$ (top row) and $z=0$ (bottom row). From left to right: Hot, cold, warm, and CR orbital components. Solid lines represent the \texttt{survived LEGA-C-like} sample, while dashed lines show the \texttt{whole LEGA-C-like} (\texttt{extended LEGA-C-like}) sample at $z=1$ ($z=0$) for each simulation. \texttt{TNG100} is same as shown in  Fig. 
    \ref{fig:fraction_plot_z1_z0}. Both simulations exhibit consistent qualitative trends across all orbital components and redshifts. At $z=0$, however, \texttt{TNG50} shows systematically higher cold and warm fractions at fixed stellar mass, reflecting its improved ability to resolve thin, rotationally supported discs. Divergence at $\log(M_\star/M_\odot) \gtrsim 11.2$ may be attributed to small-number statistics in the \texttt{TNG50} volume.}
    \label{figOrbtalFract_TNG100_TNG50}
\end{figure*}

\subsection{Resolution effects: Comparison with TNG50}
\label{sec:TNG50}

The primary analysis in this work is based on \texttt{TNG100}, which provides the statistical power needed to characterize orbital structures of massive galaxies across a representative cosmological volume. However, orbital decompositions can be sensitive to numerical resolution, particularly for dynamically cold substructures (thin discs, bars, and cold streams) that require both high mass resolution and small gravitational softening. We therefore repeat the full orbital-decomposition pipeline on galaxies drawn from the higher-resolution \texttt{TNG50} simulation and compare the resulting orbital fractions to \texttt{TNG100} to assess the robustness of our conclusions against resolution effects.

The \texttt{TNG50} and \texttt{TNG100} runs share the same galaxy-formation model, but \texttt{TNG50} improves the baryonic mass resolution and gravitational softening by roughly an order of magnitude \citep{Pillepich2018b}, enabling a more faithful representation of kinematically cold structures. The trade-off is a smaller volume and, consequently, limited sampling of the most massive systems. For \texttt{TNG50}, we independently select galaxies at $z=1$ in the stellar-mass range $10^{9.8} < M_\star/M_\odot < 10^{11.5}$, yielding a \texttt{whole sample} of 1159 galaxies, of which 918 survive to $z=0$ and comprise our \texttt{survived LEGA-C-like} sample. This sample is not a one-to-one cross-match with the \texttt{TNG100} galaxies, but rather a separately selected population subject to the same selection criteria. For a consistent comparison between simulations, we apply the same dynamical decomposition based on the snapshot-averaged circularity $\overline{\lambda_z}$ (Sect.~\ref{sec:orbitaldecomposition}) and measure orbital fractions within $2R_{1/2}$.

At $z=1$, the correspondence between \texttt{TNG100} and \texttt{TNG50} is particularly close. Both runs exhibit warm-dominated orbital structures over most of the mass range, with only modest cold fractions at the low-mass end and a strong rise of the hot component toward higher stellar mass (Fig.~\ref{figOrbtalFract_TNG100_TNG50}). Counter-rotating orbits remain a minor component in both simulations. The near-identical mass dependence at $z=1$ suggests that the emergence of warm/thick discs and the onset of hot, dispersion-dominated systems around the massive end are robust predictions of the \texttt{IllustrisTNG} model. At the highest stellar masses (roughly $\log(M_\star/M_\odot)\gtrsim 11.2$), the comparison becomes noisier, and apparent discrepancies between \texttt{TNG50} and \texttt{TNG100} should not be over-interpreted. In this regime, the \texttt{TNG50} volume contains few galaxies per mass bin, so median trends are more sensitive to small-number statistics and to individual assembly histories. We therefore treat the massive end as primarily constrained by \texttt{TNG100}, while using \texttt{TNG50} to validate the existence of the mass-dependent trends in the cold/warm/hot partition.

By $z=0$, both simulations again recover the same qualitative evolution relative to $z=1$: the hot component increases while the warm component decreases, and cold orbits remain strongly suppressed in massive galaxies. Nevertheless, \texttt{TNG50} shows a systematic quantitative offset at fixed stellar mass, with somewhat higher cold and warm fractions and correspondingly lower hot fractions compared to \texttt{TNG100}$~$(Fig.~\ref{figOrbtalFract_TNG100_TNG50}). To test whether this offset is driven by differences in the central/satellite mix, we repeated the comparison using only central galaxies and found that the discrepancy persists. This suggests that the offset is not solely caused by environmental differences associated with satellite populations. Previous comparisons between \texttt{TNG100} and \texttt{TNG50} reported closer agreement for central galaxies \citep{Zhang2025}. One important difference is that those studies typically measured orbital properties within $1R_{\rm e}$, excluded galaxies undergoing mergers, and used phase-space-averaged circularity definitions, whereas our analysis measures orbital fractions within $2R_{1/2}$, includes galaxies irrespective of merger state, and adopts a snapshot-averaged circularity definition. Since orbital fractions can vary with radius, mergers can significantly perturb the orbital distribution, and different circularity definitions can systematically affect orbit classification; these methodological differences may contribute to the observed offset. Furthermore, the discrepancy becomes more pronounced at the massive end, where the limited volume of \texttt{TNG50} results in only a small number of galaxies per mass bin, making the median trends increasingly sensitive to individual systems and recent assembly histories.

\subsection{Comparison with previous studies}
\label{sec:comparison_literature}

Our dynamical decomposition follows the orbit-based framework developed for large integral-field surveys, where galaxies are separated into cold, warm, hot, and CR components using the stellar circularity distribution \citep[e.g.,][]{Zhu2018b, Zhu2018MNRAS2, Jin2020MNRAS, Santucci2022ApJ}. At $z=0$, the stellar-mass dependence we recover is consistent with the trend observed from Schwarzschild modeling of \texttt{CALIFA} galaxies \citep{Zhu2018b, Zhu2018MNRAS2} and with analogous orbit-based decompositions from MaNGA and \texttt{SAMI} \citep[e.g.,][]{Jin2020MNRAS, Santucci2022ApJ, Santucci2023}. In particular, observational studies consistently find that the most massive systems are predominantly dispersion-supported, while disk-dominated galaxies occupy the lower-to-intermediate mass regime and show substantial cold+warm components \citep{Zhu2018b, Jin2020MNRAS}. Our \texttt{TNG}-based results, therefore, support the interpretation that the observed mass–orbit-fraction trends primarily reflect systematic changes in assembly histories with stellar mass.

On the simulation side, our results are closely aligned with the \texttt{IllustrisTNG} analysis of \citet{Xu2019MNRAS}, who constructed a $z=0$ \texttt{TNG100} sample matched to \texttt{CALIFA} and showed that \texttt{TNG} broadly reproduces the observed mass trends in orbital fractions, with some offsets in the absolute cold/CR fractions at fixed mass. A more recent comparison between \texttt{CALIFA} and \texttt{TNG}, focusing on the physical origins of dynamically hot components, also reports overall consistency in the relative prominence of cold, warm, and hot orbits in \texttt{TNG50} galaxies \citep{Zhang2025}. Beyond \texttt{IllustrisTNG}, \citet{Santucci2024MNRAS} performed an orbital decomposition for $z\simeq 0$ galaxies in the \texttt{EAGLE} simulations and likewise found a strong stellar-mass dependence, including signatures of merger-driven heating in systems with more active accretion histories. This qualitatively similar behavior emerges across \texttt{IllustrisTNG} and \texttt{EAGLE}, despite differences in subgrid physics, feedback implementations, and numerical resolution, suggests that the dominant trends are set by hierarchical growth and the associated redistribution of stellar angular momentum.

Differences in the orbital fractions between studies are expected and they can largely be traced to methodological and selection choices. First, sample definitions vary: our primary analysis includes galaxies irrespective of morphology or whether they are in a dynamically disturbed phase, whereas some \texttt{TNG–CALIFA} matching efforts explicitly exclude systems undergoing an on-going merger \citep[e.g.,][]{Zhang2025}. Such cuts can shift medians toward different values of orbital fractions at fixed mass. Second, studies adopt different apertures ($1R_{\rm e}$, $2R_{1/2}$ as used here, or total), which changes the relative weight of inner bulge-dominated regions versus the outer disk and halo, and thus affects the inferred component fractions. Third, the definition of circularity differs across the literature. Observational orbit distributions are typically derived from Schwarzschild models and reported in terms of orbit families in a best-fitting potential \citep{Zhu2018b, Zhu2018MNRAS2}, while simulation studies may use instantaneous circularity, phase-space averaging, or time/snapshot averaging. Our use of snapshot-averaged circularity reduces misclassifications of stars on eccentric or boxy orbits that can momentarily exhibit high $L_z$, and is designed to more directly trace the underlying orbit class. As a consequence, small systematic shifts in the cold–warm boundary are possible.

The key advance of our work relative to previous orbit-based observational studies is the explicit evolutionary perspective. Existing observational results are largely restricted to the Local Universe ($z\sim0$; e.g., \citealt{Zhu2018b, Jin2020MNRAS, Santucci2022ApJ}), whereas here we show for the first time that the overall mass-dependent ordering of orbital fractions is already largely established by $z=1$. In other words, when considering the galaxy population as a whole, the relative dominance of cold, warm, hot, and CR components across stellar mass is remarkably similar at both epochs.

At the same time, this apparent population-level stability does not imply passive or uniform evolution for individual systems. On the contrary, individual galaxies can undergo substantial orbital transformations between $z=1$ and $z=0$, as illustrated in Fig.\ref{fig:exsitu_orbital_tracks}. In particular, mergers and ex situ mass growth can strongly suppress dynamically cold orbits while enhancing hot and, to a lesser extent, CR components. Other galaxies, especially those that continue forming stars without experiencing major mergers, are able to preserve or even rebuild rotationally supported structures. Our results therefore reveal an important dual picture: while the statistical orbital distribution of the galaxy population is already broadly established by $z=1$, the subsequent evolutionary pathways of individual galaxies remain highly diverse and are strongly shaped by their merger and assembly histories.

\subsection{Implications of merger-driven orbital evolution for
\texttt{LEGA-C} and local observations}
\label{sec:observational_implications}

Our results provide a physical framework for interpreting the orbital
fractions of galaxies observed at $z\sim1$ and
those measured in the Local Universe. The strong stellar-mass
dependence of orbital structure is already established by $z=1$.
Lower-mass galaxies retain substantial cold and warm components,
whereas galaxies above
$\log(M_\star/M_\odot)\sim11$--$11.2$ are increasingly dominated by
hot, dispersion-supported orbits. The large hot-orbit fractions of
massive galaxies at this epoch therefore suggest that a substantial
part of their dynamical transformation occurred before they entered
the \texttt{LEGA-C} redshift range. Their orbital distributions should
thus not be interpreted solely as a consequence of their instantaneous
star formation activity or morphology at $z\sim1$, but rather as the
cumulative outcome of their earlier assembly histories.

The connection between orbital structure and merger history from
$z=1$ to $z=0$ is, however, not determined by the occurrence of
mergers alone. Our results show that the mass ratio, gas content, and
timing of mergers jointly determine the orbital configuration of the
remnant. Major mergers are particularly efficient at suppressing the
cold and warm components and enhancing the contribution of hot orbits.
These results are consistent with the established theoretical framework
of merger-driven morphological transformation, in which major mergers
strongly perturb the stellar potential and can completely disrupt the
progenitor discs, leading to the formation of pressure-supported
elliptical remnants \citep{Naab2006MNRAS,Hopkins2009MNRA}. Minor
mergers, on the other hand, act as a cumulative heating mechanism.
Although each individual event is less disruptive, repeated minor
mergers can progressively thicken stellar discs and increase the
contribution of dynamically hot components over time
\citep{Dekel2009ApJ,Fiacconi2015MNRAS,Welker2017MNRAS}. We note,
however, that mergers are not the only processes capable of altering
galaxy orbital structure. Secular evolution, disk instabilities, and
environmental effects may also contribute to the build-up of
dynamically hot components.

Moreover, the dynamical outcome of a merger also depends strongly on the
available gas reservoir. Dry mergers predominantly increase the hot and
CR fractions because the existing stellar components are
redistributed through largely collisionless processes. Wet mergers, by
contrast, can preserve part of the rotationally supported structure or
supply gas from which a new stellar disk can subsequently form
\citep{Robertson2006A,Athanassoula2016ApJ,Sparre2017MNRAS,
Zhu2022A&A}. Continued star formation can therefore help preserve or
rebuild rotationally supported components after a merger, whereas
quenching suppresses this recovery by preventing the formation of new
stars on cold and warm orbits. In this sense, mergers provide a
mechanism for dynamical heating, while star formation and quenching
regulate whether its imprint persists over time.

Consequently, a large hot-orbit fraction at fixed stellar mass can be
interpreted statistically as evidence for a comparatively merger-rich,
ex situ-dominated, or gas-poor assembly history. This interpretation is
supported by the positive partial correlation between the ex situ
stellar mass fraction and the hot component, together with the negative
correlations between the ex situ fraction and the cold and warm
components.

The converse interpretation, however, requires greater caution.
Substantial cold or warm fractions do not necessarily imply an absence
of mergers. Galaxies whose last major merger occurred at an early epoch
have had several gigayears during which gas accretion and in situ star
formation could rebuild rotationally supported structures. Moreover,
the major mergers occurring at the early epochs are predominantly wet
in our sample. Their remnants can therefore recover substantial warm
and, to a lesser extent, cold components by $z=0$. Thus, a present-day galaxy
with a significant rotationally supported component might have either experienced a relatively quiet merger history or undergone an early gas-rich merger followed by prolonged disk regrowth. Orbital
fractions can therefore constrain assembly histories statistically,
but they do not uniquely determine the merger history of an individual
galaxy.

These ambiguities become particularly important when interpreting
observed evolutionary trends. A fundamental challenge in observational
studies of galaxy evolution is that observations provide only a
snapshot of the galaxy population at a given redshift. Generally, individual
galaxies cannot be followed directly across cosmic time, and
evolutionary trends must therefore be inferred by comparing
statistically selected populations at different epochs. Such
comparisons reveal how the properties of the population as a whole
change, but they do not necessarily show how the same individual
galaxies evolve.

When samples at two epochs are compared at fixed stellar mass, it is
often implicitly assumed that the high-redshift population represents
the progenitors of the low-redshift population. In reality, the two
samples are not composed of the same evolutionary population. Galaxies
continuously enter a given stellar-mass range through star formation
and mass growth, while others leave it by merging into more massive
systems or undergoing strong dynamical transformations that alter their
properties. Differences between galaxy populations at different epochs
therefore reflect not only the physical evolution of individual
galaxies, but also changes in the demographic composition of the
selected samples. This effect, known as progenitor bias
\citep{Dokkum1996,Dokkum2001}, can substantially distort inferences
about evolutionary trends when individual galaxies are not tracked
directly.

Our analysis provides a direct and quantitative demonstration of this
effect. At $z=0$, the orbital structure of the
\texttt{extended LEGA-C-like} sample differs from that of the
\texttt{survived LEGA-C-like} sample, particularly at
$\log(M_\star/M_\odot)\lesssim11$. While the survived population
exhibits strongly suppressed cold-orbit fractions and enhanced hot
components relative to $z=1$, the extended sample recovers orbital
fractions more similar to those observed at $z=1$, including larger
cold fractions and more substantial rotational support
(Fig.~\ref{fig:fraction_plot_z1_z0}). This divergence arises because
galaxies that enter the stellar-mass selection after $z=1$ are, on
average, more disk-like and rotationally supported than galaxies that
were already included in the sample at $z=1$. Their inclusion therefore
changes the orbital composition of the complete mass-selected
population at $z=0$.

The consequences for observational comparisons are significant. A
direct comparison between a mass-selected sample at $z\sim1$, such as
\texttt{LEGA-C}, and a mass-selected low-redshift sample, such as
\texttt{CALIFA} or \texttt{MaNGA}, would not necessarily connect the
same evolutionary population. Without accounting for the absence of a
direct progenitor--descendant relationship, the inferred amount of
orbital heating between the two epochs could be
systematically underestimated as a result.

As part of our future work, we plan to perform Schwarzschild-based orbital decompositions of \texttt{LEGA-C} galaxies at $z\sim1$ and compare them with analogous measurements at $z=0$. This will allow us to test these predictions and constrain how merger-driven heating, continued star formation, and quenching contribute to the observed evolution of galaxy orbital structure.

\section{Summary and conclusions}
\label{sec:Summary_Conclusions}

We present the first systematic study of the orbital decomposition and dynamical evolution of galaxies similar in mass and redshift to the LEGA-C survey \citep{van20161, van2021,Straatman2018}, \texttt{LEGA-C}–like galaxies in the \texttt{IllustrisTNG} cosmological hydrodynamical simulations, tracing individual galaxies from $z=1$ to the present day. Using the \texttt{LEGA-C} analogs constructed by \citet{Wu2021} from \texttt{TNG100}, we followed 2,281 galaxies along their main descendant branches across merger trees, enabling a direct, galaxy-by-galaxy connection between intermediate-redshift orbital structure and present-day dynamical state. Stellar orbits were decomposed into cold ($0.8 \leq \overline{\lambda_z} \leq 1$), warm ($0.25 < \overline{\lambda_z} < 0.8$), hot ($-0.25 \leq \overline{\lambda_z} \leq 0.25$), and CR ($-1 \leq \overline{\lambda_z} < -0.25$) components using the snapshot-averaged circularity $\overline{\lambda_z}$, measured within $2R_{1/2}$. This physically motivated, orbit-based framework provides a more robust and interpretable classification of internal galaxy structure than traditional photometric bulge–disk decompositions and is directly comparable to the Schwarzschild orbit-superposition models applied to large integral-field spectroscopic surveys in the Local Universe. Our principal findings are as follows:

\begin{enumerate}

    \item Orbital structure at $z=1$ is already strongly mass-dependent: Across the full stellar-mass range probed ($10.3 \lesssim \log(M_\star/M_\odot)\lesssim 11.8$), warm and hot orbits collectively dominate the stellar budget at $z=1$ (see Fig. \ref{fig:fraction_plot_z1_z0}). At the low-mass end, rotationally supported structures are still present, with cold orbits contributing $\sim 25$--$30\%$ and warm orbits $\sim 50\%$ of the total stellar mass. With increasing stellar mass, the warm fraction peaks near $\log(M_\star/M_\odot) \sim 11$ before declining sharply, while the hot component rises steeply and dominates above $\log(M_\star/M_\odot) \gtrsim 11.2$, where dispersion-supported orbits account for $\sim 55\%$ of the stellar mass. We recover the same mass dependent behavior at $z=0$ as well.
    
    \item Galaxies undergo orbital heating from $z=1$ to $z=0$: We find that the hot orbital fraction increases across nearly all stellar-mass bins between $z=1$ and $z=0$, while the cold fraction declines monotonically and the warm fraction decreases particularly in massive systems (see Fig.  \ref{fig:orbital_evolution}). Lower-mass galaxies ($\log(M_\star/M_\odot) \lesssim 11.2$ at $z=0$) attain their maximum cold-orbit fractions at $z=1$ and experience a gradual, progressive decline thereafter, accompanied by a redistribution of stellar orbits toward warmer and hotter dynamical components. In contrast, massive galaxies ($\log(M_\star/M_\odot) \gtrsim 11.2$ at $z=0$) already host large hot components at $z=1$, reaching $\sim 60\%$, and evolve into strongly dispersion-dominated systems by $z=0$ through the steady accumulation of hot orbits.
        
    \item Star formation activity retains the cold orbits: At fixed stellar mass, star-forming galaxies host larger cold and warm fractions than their quiescent counterparts at both $z=1$ and $z=0$, while quiescent galaxies are characterized by enhanced hot fractions (see again Fig. \ref{fig:fraction_plot_z1_z0}). Quiescent galaxies exhibit a monotonic and near-total depletion of cold orbits by $z=0$, whereas star-forming galaxies retain significant rotationally supported components, sustained by ongoing gas accretion and in situ star formation that continuously replenishes the dynamically cold stellar disk. This shows that quenching does not merely halt disk growth, it removes the mechanism by which dynamically cold structures are regenerated, leaving the existing cold disk exposed to secular heating without replenishment. The strong suppression of cold orbits in the overall \texttt{survived LEGA-C-like} population at $z=0$ can therefore be traced, in large part, to the growth of the quiescent fraction over cosmic time. The central–satellite classification, in contrast, produces only minor differences in orbital fractions at both redshifts.

    \item Major mergers are the primary drivers of dynamical transformation: When comparing galaxies that experienced at least one merger with mass ratio $\mu > 1/10$ since $z=1$ (\texttt{merger-rich}) to those without such events (\texttt{merger-poor}), we find a contrast in their orbital compositions. \texttt{Merger-rich} galaxies, particularly at high stellar mass, are strongly dominated by hot and CR orbits, whereas \texttt{merger-poor} galaxies retain predominantly warm structures across the stellar-mass range, with only a modest and gradual rise of the hot component at the massive end (see Fig. \ref{fig:Combined_mergers_comparison}). When separating major ($\mu \geq 1/4$) from minor ($1/10 \leq \mu < 1/4$) mergers, we find that major mergers are more efficient per event (see Fig. \ref{fig:Num_major_minor_mergers_z0}). Even a single major merger is sufficient to significantly suppress warm and cold components and strongly enhance the hot fraction. The timing of the last major merger further effects the orbital fraction: galaxies with recent major mergers exhibit hot-orbit fractions exceeding $\sim 60\%$, while galaxies whose last major merger occurred at early epochs ($\gtrsim 9$--$10$~Gyr ago) more substantial contribution of warm and cold components (see Fig. \ref{fig:LastMajorMerger}).

    \item The gas content of mergers strongly modulates their dynamical impact: Wet (gas-rich) mergers, defined here by a merger gas fraction $f_{\rm gas,\,merger} \ge 0.2$, are associated with enhanced warm and cold orbital fractions in the remnant, while dry (gas-poor) mergers drive efficient orbital heating and the build-up of dispersion-dominated structures (see Fig. \ref{fig:WetToDryMerger}). These trends are present across both stellar-mass bins ($10.2 \leq \log M_\star/M_\odot < 11$ and $11 \leq \log M_\star/M_\odot < 12.5$) studied and are consistent with the established theoretical picture in which dissipation during gas-rich mergers promotes disk survival and regrowth, whereas collisionless relaxation in gas-poor mergers efficiently randomizes stellar orbits. At fixed wet-to-total merger ratio, $f_{\rm wet}$, massive galaxies consistently exhibit higher hot fractions and lower cold and warm fractions compared to lower-mass systems, reflecting their predominantly dry assembly histories.

    \item Orbital fractions correlate directly with ex situ stellar mass fraction, independently of stellar mass: We find that, at fixed $M_\star$, galaxies with larger ex situ fractions host significantly enhanced hot and CR components and strongly suppressed warm and cold components (see Fig. \ref{fig:exsitu_correlation}). These results shows that the present-day orbital structure of galaxies carries a statistically significant imprint of their integrated accretion history, beyond what is encoded in stellar mass alone.

\end{enumerate}

While this work provides a detailed theoretical framework for the orbital evolution of galaxies from $z = 1$ to the present day, a critical observational gap remains. To date, orbit-based dynamical decompositions have been almost exclusively limited to the Local Universe. Large integral-field spectroscopic surveys such as \texttt{CALIFA, MaNGA, and SAMI} have enabled Schwarzschild orbit-superposition modeling for thousands of nearby galaxies, revealing systematic trends in orbital structure with stellar mass, morphology, and environment. However, no comparable orbit-based constraints currently exist at intermediate or high redshift, leaving the temporal evolution of stellar orbital structures largely unconstrained by observations.

Our results demonstrate that substantial and systematic orbital transformation already occurs between $z = 1$ and $z = 0$, with galaxies undergoing significant dynamical heating that is strongly mass- and merger-dependent. In particular, massive galaxies are predicted to be dominated by hot, dispersion-supported orbits already at $z \sim 1$, while lower-mass systems retain substantial warm and cold components. Although the population-wide orbital fractions evolve only moderately over this period, our results predict clear mass-dependent trends in the internal dynamical structure of galaxies. Our results also show that the orbital evolution of an individual galaxy cannot be predicted uniquely from its properties at a single epoch because galaxies with similar initial orbital structures may follow different evolutionary pathways depending on their subsequent mergers, gas accretion, star formation, and quenching histories. The \texttt{LEGA-C} survey provides a unique opportunity to test whether these trends are also present in the observed Universe at $z\sim1$. In particular, planned future work of orbit-based dynamical modeling of \texttt{LEGA-C} galaxies will enable direct comparisons between the predicted orbital distributions of simulated galaxies and the observed dynamical structures of massive galaxies during this critical epoch of galaxy assembly.

This will help us interpret the observations within the evolutionary context provided by simulations, while also revealing any discrepancies, given that simulations are typically calibrated using galaxies at $z=0$. More generally, it will be possible to empirically trace the build-up of galaxy structure across cosmic time and, more specifically, to address the questions of whether massive galaxies at $z\sim1$ that already exhibit the dominance of hot orbits predicted by simulations. The next question is related to whether the transition from rotation-supported to dispersion-supported systems is as strongly mass-dependent as suggested in this work. Finally, we consider whether the signatures of recent mergers could be directly linked to changes in orbital composition at earlier cosmic times.

Future extensions of this approach to higher-redshift samples with upcoming facilities such as \texttt{JWST} will further enable direct dynamical studies of galaxies during the peak of cosmic star formation. Altogether, these efforts promise a unified, orbit-based view of galaxy evolution that links present-day dynamical structures to their origins at $z \gtrsim 1$.

\begin{acknowledgements}

This work made use of the following software packages: \texttt{astropy } \citep{astropy:2013,astropy:2018,astropy:2022} \texttt{Jupyter } \citep{2007CSE.....9c..21P,kluyver2016jupyter}, \texttt{matplotlib } \citep{Hunter:2007}, \texttt{numpy } \citep{numpy} and \texttt{pandas } in \texttt{python} language \citep{python}.  Software citation information aggregated using \texttt{\href{https://www.tomwagg.com/software-citation-station/}{The Software Citation Station}} \citep{software-citation-station-paper,software-citation-station-zenodo}. Moreover, the computational results have been achieved using the Austrian Scientific Computing (ASC) infrastructure.

\end{acknowledgements}

\bibliographystyle{aa}
\bibliography{ref}

@ARTICLE{wellons2015formation,
       author = {{Wellons}, Sarah and {Torrey}, Paul and {Ma}, Chung-Pei and {Rodriguez-Gomez}, Vicente and {Vogelsberger}, Mark and {Kriek}, Mariska and {van Dokkum}, Pieter and {Nelson}, Erica and {Genel}, Shy and {Pillepich}, Annalisa and {Springel}, Volker and {Sijacki}, Debora and {Snyder}, Gregory and {Nelson}, Dylan and {Sales}, Laura and {Hernquist}, Lars},
        title = "{The formation of massive, compact galaxies at z = 2 in the Illustris simulation}",
      journal = {\mnras},
         year = 2015,
        month = may,
       volume = {449},
       number = {1},
        pages = {361-372},
          doi = {10.1093/mnras/stv303},
archivePrefix = {arXiv},
       eprint = {1411.0667},
 primaryClass = {astro-ph.GA},
       adsurl = {https://ui.adsabs.harvard.edu/abs/2015MNRAS.449..361W}
}

@ARTICLE{Lackner2012,
       author = {{Lackner}, C.~N. and {Cen}, R. and {Ostriker}, J.~P. and {Joung}, M.~R.},
        title = "{Building galaxies by accretion and in situ star formation}",
      journal = {\mnras},
         year = 2012,
        month = sep,
       volume = {425},
       number = {1},
        pages = {641-656},
          doi = {10.1111/j.1365-2966.2012.21525.x},
archivePrefix = {arXiv},
       eprint = {1206.0295},
 primaryClass = {astro-ph.CO},
       adsurl = {https://ui.adsabs.harvard.edu/abs/2012MNRAS.425..641L}
}

@ARTICLE{Schweizer1982,
       author = {{Schweizer}, F.},
        title = "{Colliding and merging galaxies. I. Evidence for the recent merging of two disk galaxies in NGC 7252.}",
      journal = {\apj},
         year = 1982,
        month = jan,
       volume = {252},
        pages = {455-460},
          doi = {10.1086/159573},
       adsurl = {https://ui.adsabs.harvard.edu/abs/1982ApJ...252..455S}
}

@ARTICLE{Mihos1996,
       author = {{Mihos}, J. Christopher and {Hernquist}, Lars},
        title = "{Gasdynamics and Starbursts in Major Mergers}",
      journal = {\apj},
         year = 1996,
        month = jun,
       volume = {464},
        pages = {641},
          doi = {10.1086/177353},
archivePrefix = {arXiv},
       eprint = {astro-ph/9512099},
 primaryClass = {astro-ph},
       adsurl = {https://ui.adsabs.harvard.edu/abs/1996ApJ...464..641M}
}

@ARTICLE{Barnes1992,
       author = {{Barnes}, Joshua E.},
        title = "{Transformations of Galaxies. I. Mergers of Equal-Mass Stellar Disks}",
      journal = {\apj},
         year = 1992,
        month = jul,
       volume = {393},
        pages = {484},
          doi = {10.1086/171522},
       adsurl = {https://ui.adsabs.harvard.edu/abs/1992ApJ...393..484B}
}

@ARTICLE{Naab2009,
       author = {{Naab}, Thorsten and {Johansson}, Peter H. and {Ostriker}, Jeremiah P.},
        title = "{Minor Mergers and the Size Evolution of Elliptical Galaxies}",
      journal = {\apjl},
         year = 2009,
        month = jul,
       volume = {699},
       number = {2},
        pages = {L178-L182},
          doi = {10.1088/0004-637X/699/2/L178},
archivePrefix = {arXiv},
       eprint = {0903.1636},
 primaryClass = {astro-ph.CO},
       adsurl = {https://ui.adsabs.harvard.edu/abs/2009ApJ...699L.178N}
}

@ARTICLE{Dokkum2010,
       author = {{van Dokkum}, Pieter G. and {Whitaker}, Katherine E. and {Brammer}, Gabriel and {Franx}, Marijn and {Kriek}, Mariska and {Labb{\'e}}, Ivo and {Marchesini}, Danilo and {Quadri}, Ryan and {Bezanson}, Rachel and {Illingworth}, Garth D. and {Muzzin}, Adam and {Rudnick}, Gregory and {Tal}, Tomer and {Wake}, David},
        title = "{The Growth of Massive Galaxies Since z = 2}",
      journal = {\apj},
         year = 2010,
        month = feb,
       volume = {709},
       number = {2},
        pages = {1018-1041},
          doi = {10.1088/0004-637X/709/2/1018},
archivePrefix = {arXiv},
       eprint = {0912.0514},
 primaryClass = {astro-ph.CO},
       adsurl = {https://ui.adsabs.harvard.edu/abs/2010ApJ...709.1018V}
}

@ARTICLE{Newman2012,
       author = {{Newman}, Andrew B. and {Ellis}, Richard S. and {Bundy}, Kevin and {Treu}, Tommaso},
        title = "{Can Minor Merging Account for the Size Growth of Quiescent Galaxies? New Results from the CANDELS Survey}",
      journal = {\apj},
         year = 2012,
        month = feb,
       volume = {746},
       number = {2},
          eid = {162},
        pages = {162},
          doi = {10.1088/0004-637X/746/2/162},
archivePrefix = {arXiv},
       eprint = {1110.1637},
 primaryClass = {astro-ph.CO},
       adsurl = {https://ui.adsabs.harvard.edu/abs/2012ApJ...746..162N}
}

@ARTICLE{Ferreras2014,
       author = {{Ferreras}, I. and {Trujillo}, I. and {M{\'a}rmol-Queralt{\'o}}, E. and {P{\'e}rez-Gonz{\'a}lez}, P.~G. and {Cava}, A. and {Barro}, G. and {Cenarro}, J. and {Hern{\'a}n-Caballero}, A. and {Cardiel}, N. and {Rodr{\'\i}guez-Zaur{\'\i}n}, J. and {Cebri{\'a}n}, M.},
        title = "{Constraints on the merging channel of massive galaxies since z {\ensuremath{\sim}} 1}",
      journal = {\mnras},
         year = 2014,
        month = oct,
       volume = {444},
       number = {1},
        pages = {906-918},
          doi = {10.1093/mnras/stu1425},
archivePrefix = {arXiv},
       eprint = {1312.5317},
 primaryClass = {astro-ph.CO},
       adsurl = {https://ui.adsabs.harvard.edu/abs/2014MNRAS.444..906F}
}

@ARTICLE{Treister2012,
       author = {{Treister}, E. and {Schawinski}, K. and {Urry}, C.~M. and {Simmons}, B.~D.},
        title = "{Major Galaxy Mergers Only Trigger the Most Luminous Active Galactic Nuclei}",
      journal = {\apjl},
         year = 2012,
        month = oct,
       volume = {758},
       number = {2},
          eid = {L39},
        pages = {L39},
          doi = {10.1088/2041-8205/758/2/L39},
archivePrefix = {arXiv},
       eprint = {1209.5393},
 primaryClass = {astro-ph.CO},
       adsurl = {https://ui.adsabs.harvard.edu/abs/2012ApJ...758L..39T}
}

@ARTICLE{Ellison2019,
       author = {{Ellison}, Sara L. and {Viswanathan}, Akshara and {Patton}, David R. and {Bottrell}, Connor and {McConnachie}, Alan W. and {Gwyn}, Stephen and {Cuillandre}, Jean-Charles},
        title = "{A definitive merger-AGN connection at z {\ensuremath{\sim}} 0 with CFIS: mergers have an excess of AGN and AGN hosts are more frequently disturbed}",
      journal = {\mnras},
         year = 2019,
        month = aug,
       volume = {487},
       number = {2},
        pages = {2491-2504},
          doi = {10.1093/mnras/stz1431},
archivePrefix = {arXiv},
       eprint = {1905.08830},
 primaryClass = {astro-ph.GA},
       adsurl = {https://ui.adsabs.harvard.edu/abs/2019MNRAS.487.2491E}
}

@ARTICLE{Torrey2012,
       author = {{Torrey}, Paul and {Cox}, T.~J. and {Kewley}, Lisa and {Hernquist}, Lars},
        title = "{The Metallicity Evolution of Interacting Galaxies}",
      journal = {\apj},
         year = 2012,
        month = feb,
       volume = {746},
       number = {1},
          eid = {108},
        pages = {108},
          doi = {10.1088/0004-637X/746/1/108},
archivePrefix = {arXiv},
       eprint = {1107.0001},
 primaryClass = {astro-ph.GA},
       adsurl = {https://ui.adsabs.harvard.edu/abs/2012ApJ...746..108T}
}

@ARTICLE{Barrera2015,
       author = {{Barrera-Ballesteros}, J.~K. and {S{\'a}nchez}, S.~F. and {Garc{\'\i}a-Lorenzo}, B. and {Falc{\'o}n-Barroso}, J. and {Mast}, D. and {Garc{\'\i}a-Benito}, R. and {Husemann}, B. and {van de Ven}, G. and {Iglesias-P{\'a}ramo}, J. and {Rosales-Ortega}, F.~F. and {P{\'e}rez-Torres}, M.~A. and {M{\'a}rquez}, I. and {Kehrig}, C. and {Marino}, R.~A. and {Vilchez}, J.~M. and {Galbany}, L. and {L{\'o}pez-S{\'a}nchez}, {\'A}. R. and {Walcher}, C.~J. and {CALIFA Collaboration}},
        title = "{Central star formation and metallicity in CALIFA interacting galaxies}",
      journal = {\aap},
         year = 2015,
        month = jul,
       volume = {579},
          eid = {A45},
        pages = {A45},
          doi = {10.1051/0004-6361/201425397},
archivePrefix = {arXiv},
       eprint = {1505.03153},
 primaryClass = {astro-ph.GA},
       adsurl = {https://ui.adsabs.harvard.edu/abs/2015A&A...579A..45B}
}

@ARTICLE{Bustamante2018,
       author = {{Bustamante}, Sebasti{\'a}n and {Sparre}, Martin and {Springel}, Volker and {Grand}, Robert J.~J.},
        title = "{Merger-induced metallicity dilution in cosmological galaxy formation simulations}",
      journal = {\mnras},
         year = 2018,
        month = sep,
       volume = {479},
       number = {3},
        pages = {3381-3392},
          doi = {10.1093/mnras/sty1692},
archivePrefix = {arXiv},
       eprint = {1712.03250},
 primaryClass = {astro-ph.GA},
       adsurl = {https://ui.adsabs.harvard.edu/abs/2018MNRAS.479.3381B}
}

@ARTICLE{Buck2023,
       author = {{Buck}, Tobias and {Obreja}, Aura and {Ratcliffe}, Bridget and {Lu}, Yuxi(Lucy) and {Minchev}, Ivan and {Macci{\`o}}, Andrea V.},
        title = "{The impact of early massive mergers on the chemical evolution of Milky Way-like galaxies: insights from NIHAO-UHD simulations}",
      journal = {\mnras},
         year = 2023,
        month = jul,
       volume = {523},
       number = {1},
        pages = {1565-1576},
          doi = {10.1093/mnras/stad1503},
archivePrefix = {arXiv},
       eprint = {2305.13759},
 primaryClass = {astro-ph.GA},
       adsurl = {https://ui.adsabs.harvard.edu/abs/2023MNRAS.523.1565B}
}

@ARTICLE{Pan2025,
       author = {{Pan}, Hsi-An and {Lin}, Lihwai and {S{\'a}nchez}, Sebasti{\'a}n F. and {Barrera-Ballesteros}, Jorge K. and {Hsieh}, Bau-Ching},
        title = "{SDSS-IV MaNGA: Spatial Evolution of Gas-phase Metallicity Changes Induced by Galaxy Interactions}",
      journal = {\apj},
         year = 2025,
        month = apr,
       volume = {982},
       number = {2},
          eid = {130},
        pages = {130},
          doi = {10.3847/1538-4357/adbbd2},
archivePrefix = {arXiv},
       eprint = {2503.02184},
 primaryClass = {astro-ph.GA},
       adsurl = {https://ui.adsabs.harvard.edu/abs/2025ApJ...982..130P}
}

@ARTICLE{Fall1980,
       author = {{Fall}, S.~M. and {Efstathiou}, G.},
        title = "{Formation and rotation of disc galaxies with haloes.}",
      journal = {\mnras},
         year = 1980,
        month = oct,
       volume = {193},
        pages = {189-206},
          doi = {10.1093/mnras/193.2.189},
       adsurl = {https://ui.adsabs.harvard.edu/abs/1980MNRAS.193..189F}
}

@ARTICLE{Lagos2017,
       author = {{Lagos}, Claudia del P. and {Theuns}, Tom and {Stevens}, Adam R.~H. and {Cortese}, Luca and {Padilla}, Nelson D. and {Davis}, Timothy A. and {Contreras}, Sergio and {Croton}, Darren},
        title = "{Angular momentum evolution of galaxies in EAGLE}",
      journal = {\mnras},
         year = 2017,
        month = feb,
       volume = {464},
       number = {4},
        pages = {3850-3870},
          doi = {10.1093/mnras/stw2610},
archivePrefix = {arXiv},
       eprint = {1609.01739},
 primaryClass = {astro-ph.GA},
       adsurl = {https://ui.adsabs.harvard.edu/abs/2017MNRAS.464.3850L}
}

@ARTICLE{White1978,
       author = {{White}, S.~D.~M. and {Rees}, M.~J.},
        title = "{Core condensation in heavy halos: a two-stage theory for galaxy formation and clustering.}",
      journal = {\mnras},
         year = 1978,
        month = may,
       volume = {183},
        pages = {341-358},
          doi = {10.1093/mnras/183.3.341},
       adsurl = {https://ui.adsabs.harvard.edu/abs/1978MNRAS.183..341W}
}

@ARTICLE{Leaman2017,
       author = {{Leaman}, Ryan and {Mendel}, J. Trevor and {Wisnioski}, Emily and {Brooks}, Alyson M. and {Beasley}, Michael A. and {Starkenburg}, Else and {Martig}, Marie and {Battaglia}, Giuseppina and {Christensen}, Charlotte and {Cole}, Andrew A. and {de Boer}, T.~J.~L. and {Wills}, Drew},
        title = "{A unified model for age-velocity dispersion relations in Local Group galaxies: disentangling ISM turbulence and latent dynamical heating}",
      journal = {\mnras},
         year = 2017,
        month = dec,
       volume = {472},
       number = {2},
        pages = {1879-1896},
          doi = {10.1093/mnras/stx2014},
archivePrefix = {arXiv},
       eprint = {1710.01317},
 primaryClass = {astro-ph.GA},
       adsurl = {https://ui.adsabs.harvard.edu/abs/2017MNRAS.472.1879L}
}

@ARTICLE{Marinacci2018a,
       author = {{Marinacci}, Federico and {Vogelsberger}, Mark and {Pakmor}, R{\"u}diger and {Torrey}, Paul and {Springel}, Volker and {Hernquist}, Lars and {Nelson}, Dylan and {Weinberger}, Rainer and {Pillepich}, Annalisa and {Naiman}, Jill and {Genel}, Shy},
        title = "{First results from the IllustrisTNG simulations: radio haloes and magnetic fields}",
      journal = {\mnras},
         year = 2018,
        month = nov,
       volume = {480},
       number = {4},
        pages = {5113-5139},
          doi = {10.1093/mnras/sty2206},
archivePrefix = {arXiv},
       eprint = {1707.03396},
 primaryClass = {astro-ph.CO},
       adsurl = {https://ui.adsabs.harvard.edu/abs/2018MNRAS.480.5113M}
}

@ARTICLE{Naiman2018a,
       author = {{Naiman}, Jill P. and {Pillepich}, Annalisa and {Springel}, Volker and {Ramirez-Ruiz}, Enrico and {Torrey}, Paul and {Vogelsberger}, Mark and {Pakmor}, R{\"u}diger and {Nelson}, Dylan and {Marinacci}, Federico and {Hernquist}, Lars and {Weinberger}, Rainer and {Genel}, Shy},
        title = "{First results from the IllustrisTNG simulations: a tale of two elements - chemical evolution of magnesium and europium}",
      journal = {\mnras},
         year = 2018,
        month = jun,
       volume = {477},
       number = {1},
        pages = {1206-1224},
          doi = {10.1093/mnras/sty618},
archivePrefix = {arXiv},
       eprint = {1707.03401},
 primaryClass = {astro-ph.GA},
       adsurl = {https://ui.adsabs.harvard.edu/abs/2018MNRAS.477.1206N}
}

@ARTICLE{Nelson2018a,
       author = {{Nelson}, Dylan and {Pillepich}, Annalisa and {Springel}, Volker and {Weinberger}, Rainer and {Hernquist}, Lars and {Pakmor}, R{\"u}diger and {Genel}, Shy and {Torrey}, Paul and {Vogelsberger}, Mark and {Kauffmann}, Guinevere and {Marinacci}, Federico and {Naiman}, Jill},
        title = "{First results from the IllustrisTNG simulations: the galaxy colour bimodality}",
      journal = {\mnras},
         year = 2018,
        month = mar,
       volume = {475},
       number = {1},
        pages = {624-647},
          doi = {10.1093/mnras/stx3040},
archivePrefix = {arXiv},
       eprint = {1707.03395},
 primaryClass = {astro-ph.GA},
       adsurl = {https://ui.adsabs.harvard.edu/abs/2018MNRAS.475..624N}
}

@ARTICLE{Pillepich2018a,
       author = {{Pillepich}, Annalisa and {Nelson}, Dylan and {Hernquist}, Lars and {Springel}, Volker and {Pakmor}, R{\"u}diger and {Torrey}, Paul and {Weinberger}, Rainer and {Genel}, Shy and {Naiman}, Jill P. and {Marinacci}, Federico and {Vogelsberger}, Mark},
        title = "{First results from the IllustrisTNG simulations: the stellar mass content of groups and clusters of galaxies}",
      journal = {\mnras},
         year = 2018,
        month = mar,
       volume = {475},
       number = {1},
        pages = {648-675},
          doi = {10.1093/mnras/stx3112},
archivePrefix = {arXiv},
       eprint = {1707.03406},
 primaryClass = {astro-ph.GA},
       adsurl = {https://ui.adsabs.harvard.edu/abs/2018MNRAS.475..648P}
}

@ARTICLE{Springel2018a,
       author = {{Springel}, Volker and {Pakmor}, R{\"u}diger and {Pillepich}, Annalisa and {Weinberger}, Rainer and {Nelson}, Dylan and {Hernquist}, Lars and {Vogelsberger}, Mark and {Genel}, Shy and {Torrey}, Paul and {Marinacci}, Federico and {Naiman}, Jill},
        title = "{First results from the IllustrisTNG simulations: matter and galaxy clustering}",
      journal = {\mnras},
         year = 2018,
        month = mar,
       volume = {475},
       number = {1},
        pages = {676-698},
          doi = {10.1093/mnras/stx3304},
archivePrefix = {arXiv},
       eprint = {1707.03397},
 primaryClass = {astro-ph.GA},
       adsurl = {https://ui.adsabs.harvard.edu/abs/2018MNRAS.475..676S}
}

@ARTICLE{Springel2010a,
       author = {{Springel}, Volker},
        title = "{E pur si muove: Galilean-invariant cosmological hydrodynamical simulations on a moving mesh}",
      journal = {\mnras},
         year = 2010,
        month = jan,
       volume = {401},
       number = {2},
        pages = {791-851},
          doi = {10.1111/j.1365-2966.2009.15715.x},
archivePrefix = {arXiv},
       eprint = {0901.4107},
 primaryClass = {astro-ph.CO},
       adsurl = {https://ui.adsabs.harvard.edu/abs/2010MNRAS.401..791S}
}

@ARTICLE{Pakmor2011a,
       author = {{Pakmor}, Ruediger and {Bauer}, Andreas and {Springel}, Volker},
        title = "{Magnetohydrodynamics on an unstructured moving grid}",
      journal = {\mnras},
         year = 2011,
        month = dec,
       volume = {418},
       number = {2},
        pages = {1392-1401},
          doi = {10.1111/j.1365-2966.2011.19591.x},
archivePrefix = {arXiv},
       eprint = {1108.1792},
 primaryClass = {astro-ph.IM},
       adsurl = {https://ui.adsabs.harvard.edu/abs/2011MNRAS.418.1392P}
}

@ARTICLE{Pakmor2013a,
       author = {{Pakmor}, R{\"u}diger and {Springel}, Volker},
        title = "{Simulations of magnetic fields in isolated disc galaxies}",
      journal = {\mnras},
         year = 2013,
        month = jun,
       volume = {432},
       number = {1},
        pages = {176-193},
          doi = {10.1093/mnras/stt428},
archivePrefix = {arXiv},
       eprint = {1212.1452},
 primaryClass = {astro-ph.CO},
       adsurl = {https://ui.adsabs.harvard.edu/abs/2013MNRAS.432..176P}
}

@ARTICLE{Pakmor2016a,
       author = {{Pakmor}, R{\"u}diger and {Springel}, Volker and {Bauer}, Andreas and {Mocz}, Philip and {Munoz}, Diego J. and {Ohlmann}, Sebastian T. and {Schaal}, Kevin and {Zhu}, Chenchong},
        title = "{Improving the convergence properties of the moving-mesh code AREPO}",
      journal = {\mnras},
         year = 2016,
        month = jan,
       volume = {455},
       number = {1},
        pages = {1134-1143},
          doi = {10.1093/mnras/stv2380},
archivePrefix = {arXiv},
       eprint = {1503.00562},
 primaryClass = {astro-ph.GA},
       adsurl = {https://ui.adsabs.harvard.edu/abs/2016MNRAS.455.1134P}
}

@ARTICLE{Vogelsberger2013b,
       author = {{Vogelsberger}, Mark and {Genel}, Shy and {Sijacki}, Debora and {Torrey}, Paul and {Springel}, Volker and {Hernquist}, Lars},
        title = "{A model for cosmological simulations of galaxy formation physics}",
      journal = {\mnras},
         year = 2013,
        month = dec,
       volume = {436},
       number = {4},
        pages = {3031-3067},
          doi = {10.1093/mnras/stt1789},
archivePrefix = {arXiv},
       eprint = {1305.2913},
 primaryClass = {astro-ph.CO},
       adsurl = {https://ui.adsabs.harvard.edu/abs/2013MNRAS.436.3031V}
}

@ARTICLE{Vogelsberger2014a,
       author = {{Vogelsberger}, M. and {Genel}, S. and {Springel}, V. and {Torrey}, P. and {Sijacki}, D. and {Xu}, D. and {Snyder}, G. and {Bird}, S. and {Nelson}, D. and {Hernquist}, L.},
        title = "{Properties of galaxies reproduced by a hydrodynamic simulation}",
      journal = {\nat},
         year = 2014,
        month = may,
       volume = {509},
       number = {7499},
        pages = {177-182},
          doi = {10.1038/nature13316},
archivePrefix = {arXiv},
       eprint = {1405.1418},
 primaryClass = {astro-ph.CO},
       adsurl = {https://ui.adsabs.harvard.edu/abs/2014Natur.509..177V}
}

@ARTICLE{Vogelsberger2014b,
       author = {{Vogelsberger}, Mark and {Genel}, Shy and {Springel}, Volker and {Torrey}, Paul and {Sijacki}, Debora and {Xu}, Dandan and {Snyder}, Greg and {Nelson}, Dylan and {Hernquist}, Lars},
        title = "{Introducing the Illustris Project: simulating the coevolution of dark and visible matter in the Universe}",
      journal = {\mnras},
         year = 2014,
        month = oct,
       volume = {444},
       number = {2},
        pages = {1518-1547},
          doi = {10.1093/mnras/stu1536},
archivePrefix = {arXiv},
       eprint = {1405.2921},
 primaryClass = {astro-ph.CO},
       adsurl = {https://ui.adsabs.harvard.edu/abs/2014MNRAS.444.1518V}
}

@ARTICLE{Genel2014a,
       author = {{Genel}, Shy and {Vogelsberger}, Mark and {Springel}, Volker and {Sijacki}, Debora and {Nelson}, Dylan and {Snyder}, Greg and {Rodriguez-Gomez}, Vicente and {Torrey}, Paul and {Hernquist}, Lars},
        title = "{Introducing the Illustris project: the evolution of galaxy populations across cosmic time}",
      journal = {\mnras},
         year = 2014,
        month = nov,
       volume = {445},
       number = {1},
        pages = {175-200},
          doi = {10.1093/mnras/stu1654},
archivePrefix = {arXiv},
       eprint = {1405.3749},
 primaryClass = {astro-ph.CO},
       adsurl = {https://ui.adsabs.harvard.edu/abs/2014MNRAS.445..175G}
}

@ARTICLE{Nelson2015A&C,
       author = {{Nelson}, D. and {Pillepich}, A. and {Genel}, S. and {Vogelsberger}, M. and {Springel}, V. and {Torrey}, P. and {Rodriguez-Gomez}, V. and {Sijacki}, D. and {Snyder}, G.~F. and {Griffen}, B. and {Marinacci}, F. and {Blecha}, L. and {Sales}, L. and {Xu}, D. and {Hernquist}, L.},
        title = "{The illustris simulation: Public data release}",
      journal = {Astronomy and Computing},
         year = 2015,
        month = nov,
       volume = {13},
        pages = {12-37},
          doi = {10.1016/j.ascom.2015.09.003},
archivePrefix = {arXiv},
       eprint = {1504.00362},
 primaryClass = {astro-ph.CO},
       adsurl = {https://ui.adsabs.harvard.edu/abs/2015A&C....13...12N}
}

@ARTICLE{Sijacki2015a,
       author = {{Sijacki}, Debora and {Vogelsberger}, Mark and {Genel}, Shy and {Springel}, Volker and {Torrey}, Paul and {Snyder}, Gregory F. and {Nelson}, Dylan and {Hernquist}, Lars},
        title = "{The Illustris simulation: the evolving population of black holes across cosmic time}",
      journal = {\mnras},
         year = 2015,
        month = sep,
       volume = {452},
       number = {1},
        pages = {575-596},
          doi = {10.1093/mnras/stv1340},
archivePrefix = {arXiv},
       eprint = {1408.6842},
 primaryClass = {astro-ph.GA},
       adsurl = {https://ui.adsabs.harvard.edu/abs/2015MNRAS.452..575S}
}

@ARTICLE{Pillepich2018b,
       author = {{Pillepich}, Annalisa and {Nelson}, Dylan and {Hernquist}, Lars and {Springel}, Volker and {Pakmor}, R{\"u}diger and {Torrey}, Paul and {Weinberger}, Rainer and {Genel}, Shy and {Naiman}, Jill P. and {Marinacci}, Federico and {Vogelsberger}, Mark},
        title = "{First results from the IllustrisTNG simulations: the stellar mass content of groups and clusters of galaxies}",
      journal = {\mnras},
         year = 2018,
        month = mar,
       volume = {475},
       number = {1},
        pages = {648-675},
          doi = {10.1093/mnras/stx3112},
archivePrefix = {arXiv},
       eprint = {1707.03406},
 primaryClass = {astro-ph.GA},
       adsurl = {https://ui.adsabs.harvard.edu/abs/2018MNRAS.475..648P}
}

@ARTICLE{Donnari2019,
       author = {{Donnari}, Martina and {Pillepich}, Annalisa and {Nelson}, Dylan and {Vogelsberger}, Mark and {Genel}, Shy and {Weinberger}, Rainer and {Marinacci}, Federico and {Springel}, Volker and {Hernquist}, Lars},
        title = "{The star formation activity of IllustrisTNG galaxies: main sequence, UVJ diagram, quenched fractions, and systematics}",
      journal = {\mnras},
         year = 2019,
        month = jun,
       volume = {485},
       number = {4},
        pages = {4817-4840},
          doi = {10.1093/mnras/stz712},
archivePrefix = {arXiv},
       eprint = {1812.07584},
 primaryClass = {astro-ph.GA},
       adsurl = {https://ui.adsabs.harvard.edu/abs/2019MNRAS.485.4817D}
}

@ARTICLE{Donnari2021,
       author = {{Donnari}, Martina and {Pillepich}, Annalisa and {Joshi}, Gandhali D. and {Nelson}, Dylan and {Genel}, Shy and {Marinacci}, Federico and {Rodriguez-Gomez}, Vicente and {Pakmor}, R{\"u}diger and {Torrey}, Paul and {Vogelsberger}, Mark and {Hernquist}, Lars},
        title = "{Quenched fractions in the IllustrisTNG simulations: the roles of AGN feedback, environment, and pre-processing}",
      journal = {\mnras},
         year = 2021,
        month = jan,
       volume = {500},
       number = {3},
        pages = {4004-4024},
          doi = {10.1093/mnras/staa3006},
archivePrefix = {arXiv},
       eprint = {2008.00005},
 primaryClass = {astro-ph.GA},
       adsurl = {https://ui.adsabs.harvard.edu/abs/2021MNRAS.500.4004D}
}

@ARTICLE{Genel2018,
       author = {{Genel}, Shy and {Nelson}, Dylan and {Pillepich}, Annalisa and {Springel}, Volker and {Pakmor}, R{\"u}diger and {Weinberger}, Rainer and {Hernquist}, Lars and {Naiman}, Jill and {Vogelsberger}, Mark and {Marinacci}, Federico and {Torrey}, Paul},
        title = "{The size evolution of star-forming and quenched galaxies in the IllustrisTNG simulation}",
      journal = {\mnras},
         year = 2018,
        month = mar,
       volume = {474},
       number = {3},
        pages = {3976-3996},
          doi = {10.1093/mnras/stx3078},
archivePrefix = {arXiv},
       eprint = {1707.05327},
 primaryClass = {astro-ph.GA},
       adsurl = {https://ui.adsabs.harvard.edu/abs/2018MNRAS.474.3976G}
}

@ARTICLE{Huertas2019a,
       author = {{Huertas-Company}, Marc and {Rodriguez-Gomez}, Vicente and {Nelson}, Dylan and {Pillepich}, Annalisa and {Bottrell}, Connor and {Bernardi}, Mariangela and {Dom{\'\i}nguez-S{\'a}nchez}, Helena and {Genel}, Shy and {Pakmor}, Ruediger and {Snyder}, Gregory F. and {Vogelsberger}, Mark},
        title = "{The Hubble Sequence at z {\ensuremath{\sim}} 0 in the IllustrisTNG simulation with deep learning}",
      journal = {\mnras},
         year = 2019,
        month = oct,
       volume = {489},
       number = {2},
        pages = {1859-1879},
          doi = {10.1093/mnras/stz2191},
archivePrefix = {arXiv},
       eprint = {1903.07625},
 primaryClass = {astro-ph.GA},
       adsurl = {https://ui.adsabs.harvard.edu/abs/2019MNRAS.489.1859H}
}

@ARTICLE{Rodriguez2019a,
       author = {{Rodriguez-Gomez}, Vicente and {Snyder}, Gregory F. and {Lotz}, Jennifer M. and {Nelson}, Dylan and {Pillepich}, Annalisa and {Springel}, Volker and {Genel}, Shy and {Weinberger}, Rainer and {Tacchella}, Sandro and {Pakmor}, R{\"u}diger and {Torrey}, Paul and {Marinacci}, Federico and {Vogelsberger}, Mark and {Hernquist}, Lars and {Thilker}, David A.},
        title = "{The optical morphologies of galaxies in the IllustrisTNG simulation: a comparison to Pan-STARRS observations}",
      journal = {\mnras},
         year = 2019,
        month = mar,
       volume = {483},
       number = {3},
        pages = {4140-4159},
          doi = {10.1093/mnras/sty3345},
archivePrefix = {arXiv},
       eprint = {1809.08239},
 primaryClass = {astro-ph.GA},
       adsurl = {https://ui.adsabs.harvard.edu/abs/2019MNRAS.483.4140R}
}

@ARTICLE{Nelson2021a,
       author = {{Nelson}, Erica J. and {Tacchella}, Sandro and {Diemer}, Benedikt and {Leja}, Joel and {Hernquist}, Lars and {Whitaker}, Katherine E. and {Weinberger}, Rainer and {Pillepich}, Annalisa and {Nelson}, Dylan and {Terrazas}, Bryan A. and {Nevin}, Rebecca and {Brammer}, Gabriel B. and {Burkhart}, Blakesley and {Cochrane}, Rachel K. and {van Dokkum}, Pieter and {Johnson}, Benjamin D. and {Marinacci}, Federico and {Mowla}, Lamiya and {Pakmor}, R{\"u}diger and {Skelton}, Rosalind E. and {Speagle}, Joshua and {Springel}, Volker and {Torrey}, Paul and {Vogelsberger}, Mark and {Wuyts}, Stijn},
        title = "{Spatially resolved star formation and inside-out quenching in the TNG50 simulation and 3D-HST observations}",
      journal = {\mnras},
         year = 2021,
        month = nov,
       volume = {508},
       number = {1},
        pages = {219-235},
          doi = {10.1093/mnras/stab2131},
archivePrefix = {arXiv},
       eprint = {2101.12212},
 primaryClass = {astro-ph.GA},
       adsurl = {https://ui.adsabs.harvard.edu/abs/2021MNRAS.508..219N}
}

@ARTICLE{Weinberger2017a,
       author = {{Weinberger}, Rainer and {Springel}, Volker and {Hernquist}, Lars and {Pillepich}, Annalisa and {Marinacci}, Federico and {Pakmor}, R{\"u}diger and {Nelson}, Dylan and {Genel}, Shy and {Vogelsberger}, Mark and {Naiman}, Jill and {Torrey}, Paul},
        title = "{Simulating galaxy formation with black hole driven thermal and kinetic feedback}",
      journal = {\mnras},
         year = 2017,
        month = mar,
       volume = {465},
       number = {3},
        pages = {3291-3308},
          doi = {10.1093/mnras/stw2944},
archivePrefix = {arXiv},
       eprint = {1607.03486},
 primaryClass = {astro-ph.GA},
       adsurl = {https://ui.adsabs.harvard.edu/abs/2017MNRAS.465.3291W}
}

@ARTICLE{Davis1985,
       author = {{Davis}, M. and {Efstathiou}, G. and {Frenk}, C.~S. and {White}, S.~D.~M.},
        title = "{The evolution of large-scale structure in a universe dominated by cold dark matter}",
      journal = {\apj},
         year = 1985,
        month = may,
       volume = {292},
        pages = {371-394},
          doi = {10.1086/163168},
       adsurl = {https://ui.adsabs.harvard.edu/abs/1985ApJ...292..371D}
}

@ARTICLE{Springel2001a,
       author = {{Springel}, Volker and {White}, Simon D.~M. and {Tormen}, Giuseppe and {Kauffmann}, Guinevere},
        title = "{Populating a cluster of galaxies - I. Results at z=0}",
      journal = {\mnras},
         year = 2001,
        month = dec,
       volume = {328},
       number = {3},
        pages = {726-750},
          doi = {10.1046/j.1365-8711.2001.04912.x},
archivePrefix = {arXiv},
       eprint = {astro-ph/0012055},
 primaryClass = {astro-ph},
       adsurl = {https://ui.adsabs.harvard.edu/abs/2001MNRAS.328..726S}
}

@ARTICLE{Rodriguez2015,
       author = {{Rodriguez-Gomez}, Vicente and {Genel}, Shy and {Vogelsberger}, Mark and {Sijacki}, Debora and {Pillepich}, Annalisa and {Sales}, Laura V. and {Torrey}, Paul and {Snyder}, Greg and {Nelson}, Dylan and {Springel}, Volker and {Ma}, Chung-Pei and {Hernquist}, Lars},
        title = "{The merger rate of galaxies in the Illustris simulation: a comparison with observations and semi-empirical models}",
      journal = {\mnras},
         year = 2015,
        month = may,
       volume = {449},
       number = {1},
        pages = {49-64},
          doi = {10.1093/mnras/stv264},
archivePrefix = {arXiv},
       eprint = {1502.01339},
 primaryClass = {astro-ph.GA},
       adsurl = {https://ui.adsabs.harvard.edu/abs/2015MNRAS.449...49R}
}

@ARTICLE{Dolag2009,
       author = {{Dolag}, K. and {Borgani}, S. and {Murante}, G. and {Springel}, V.},
        title = "{Substructures in hydrodynamical cluster simulations}",
      journal = {\mnras},
         year = 2009,
        month = oct,
       volume = {399},
       number = {2},
        pages = {497-514},
          doi = {10.1111/j.1365-2966.2009.15034.x},
archivePrefix = {arXiv},
       eprint = {0808.3401},
 primaryClass = {astro-ph},
       adsurl = {https://ui.adsabs.harvard.edu/abs/2009MNRAS.399..497D}
}

@ARTICLE{Lucia2007,
       author = {{De Lucia}, Gabriella and {Blaizot}, J{\'e}r{\'e}my},
        title = "{The hierarchical formation of the brightest cluster galaxies}",
      journal = {\mnras},
         year = 2007,
        month = feb,
       volume = {375},
       number = {1},
        pages = {2-14},
          doi = {10.1111/j.1365-2966.2006.11287.x},
archivePrefix = {arXiv},
       eprint = {astro-ph/0606519},
 primaryClass = {astro-ph},
       adsurl = {https://ui.adsabs.harvard.edu/abs/2007MNRAS.375....2D}
}

@ARTICLE{van20161,
       author = {{van der Wel}, A. and {Noeske}, K. and {Bezanson}, R. and {Pacifici}, C. and {Gallazzi}, A. and {Franx}, M. and {Mu{\~n}oz-Mateos}, J.~C. and {Bell}, E.~F. and {Brammer}, G. and {Charlot}, S. and {Chauk{\'e}}, P. and {Labb{\'e}}, I. and {Maseda}, M.~V. and {Muzzin}, A. and {Rix}, H. -W. and {Sobral}, D. and {van de Sande}, J. and {van Dokkum}, P.~G. and {Wild}, V. and {Wolf}, C.},
        title = "{The VLT LEGA-C Spectroscopic Survey: The Physics of Galaxies at a Lookback Time of 7 Gyr}",
      journal = {\apjs},
         year = 2016,
        month = apr,
       volume = {223},
       number = {2},
          eid = {29},
        pages = {29},
          doi = {10.3847/0067-0049/223/2/29},
archivePrefix = {arXiv},
       eprint = {1603.05479},
 primaryClass = {astro-ph.GA},
       adsurl = {https://ui.adsabs.harvard.edu/abs/2016ApJS..223...29V}
}

@ARTICLE{Straatman2018,
       author = {{Straatman}, Caroline M.~S. and {van der Wel}, Arjen and {Bezanson}, Rachel and {Pacifici}, Camilla and {Gallazzi}, Anna and {Wu}, Po-Feng and {Noeske}, Kai and {Bari{\v{s}}i{\'c}}, Ivana and {Bell}, Eric F. and {Brammer}, Gabriel B. and {Calhau}, Jo{\~a}o and {Chauke}, Priscilla and {Franx}, Marijn and {van Houdt}, Josha and {Labb{\'e}}, Ivo and {Maseda}, Michael V. and {Mu{\~n}oz-Mateos}, Juan C. and {Muzzin}, Adam and {van de Sande}, Jesse and {Sobral}, David and {Spilker}, Justin S.},
        title = "{The Large Early Galaxy Astrophysics Census (LEGA-C) Data Release 2: Dynamical and Stellar Population Properties of z {\ensuremath{\lesssim}} 1 Galaxies in the COSMOS Field}",
      journal = {\apjs},
         year = 2018,
        month = dec,
       volume = {239},
       number = {2},
          eid = {27},
        pages = {27},
          doi = {10.3847/1538-4365/aae37a},
archivePrefix = {arXiv},
       eprint = {1809.08236},
 primaryClass = {astro-ph.GA},
       adsurl = {https://ui.adsabs.harvard.edu/abs/2018ApJS..239...27S}
}

@ARTICLE{van2021,
       author = {{van der Wel}, Arjen and {Bezanson}, Rachel and {D'Eugenio}, Francesco and {Straatman}, Caroline and {Franx}, Marijn and {van Houdt}, Josha and {Maseda}, Michael V. and {Gallazzi}, Anna and {Wu}, Po-Feng and {Pacifici}, Camilla and {Barisic}, Ivana and {Brammer}, Gabriel B. and {Munoz-Mateos}, Juan Carlos and {Vervalcke}, Sarah and {Zibetti}, Stefano and {Sobral}, David and {de Graaff}, Anna and {Calhau}, Joao and {Kaushal}, Yasha and {Muzzin}, Adam and {Bell}, Eric F. and {van Dokkum}, Pieter G.},
        title = "{The Large Early Galaxy Astrophysics Census (LEGA-C) Data Release 3: 3000 High-quality Spectra of K$_{s}$-selected Galaxies at z > 0.6}",
      journal = {\apjs},
         year = 2021,
        month = oct,
       volume = {256},
       number = {2},
          eid = {44},
        pages = {44},
          doi = {10.3847/1538-4365/ac1356},
archivePrefix = {arXiv},
       eprint = {2108.00744},
 primaryClass = {astro-ph.GA},
       adsurl = {https://ui.adsabs.harvard.edu/abs/2021ApJS..256...44V}
}

@ARTICLE{Scoville2007,
       author = {{Scoville}, N. and {Abraham}, R.~G. and {Aussel}, H. and {Barnes}, J.~E. and {Benson}, A. and {Blain}, A.~W. and {Calzetti}, D. and {Comastri}, A. and {Capak}, P. and {Carilli}, C. and {Carlstrom}, J.~E. and {Carollo}, C.~M. and {Colbert}, J. and {Daddi}, E. and {Ellis}, R.~S. and {Elvis}, M. and {Ewald}, S.~P. and {Fall}, M. and {Franceschini}, A. and {Giavalisco}, M. and {Green}, W. and {Griffiths}, R.~E. and {Guzzo}, L. and {Hasinger}, G. and {Impey}, C. and {Kneib}, J. -P. and {Koda}, J. and {Koekemoer}, A. and {Lefevre}, O. and {Lilly}, S. and {Liu}, C.~T. and {McCracken}, H.~J. and {Massey}, R. and {Mellier}, Y. and {Miyazaki}, S. and {Mobasher}, B. and {Mould}, J. and {Norman}, C. and {Refregier}, A. and {Renzini}, A. and {Rhodes}, J. and {Rich}, M. and {Sanders}, D.~B. and {Schiminovich}, D. and {Schinnerer}, E. and {Scodeggio}, M. and {Sheth}, K. and {Shopbell}, P.~L. and {Taniguchi}, Y. and {Tyson}, N.~D. and {Urry}, C.~M. and {Van Waerbeke}, L. and {Vettolani}, P. and {White}, S.~D.~M. and {Yan}, L.},
        title = "{COSMOS: Hubble Space Telescope Observations}",
      journal = {\apjs},
         year = 2007,
        month = sep,
       volume = {172},
       number = {1},
        pages = {38-45},
          doi = {10.1086/516580},
archivePrefix = {arXiv},
       eprint = {astro-ph/0612306},
 primaryClass = {astro-ph},
       adsurl = {https://ui.adsabs.harvard.edu/abs/2007ApJS..172...38S}
}

@ARTICLE{Muzzin2013,
       author = {{Muzzin}, Adam and {Marchesini}, Danilo and {Stefanon}, Mauro and {Franx}, Marijn and {McCracken}, Henry J. and {Milvang-Jensen}, Bo and {Dunlop}, James S. and {Fynbo}, J.~P.~U. and {Brammer}, Gabriel and {Labb{\'e}}, Ivo and {van Dokkum}, Pieter G.},
        title = "{The Evolution of the Stellar Mass Functions of Star-forming and Quiescent Galaxies to z = 4 from the COSMOS/UltraVISTA Survey}",
      journal = {\apj},
         year = 2013,
        month = nov,
       volume = {777},
       number = {1},
          eid = {18},
        pages = {18},
          doi = {10.1088/0004-637X/777/1/18},
archivePrefix = {arXiv},
       eprint = {1303.4409},
 primaryClass = {astro-ph.CO},
       adsurl = {https://ui.adsabs.harvard.edu/abs/2013ApJ...777...18M}
}

@ARTICLE{Wu2021,
       author = {{Wu}, Po-Feng and {Nelson}, Dylan and {van der Wel}, Arjen and {Pillepich}, Annalisa and {Zibetti}, Stefano and {Bezanson}, Rachel and {DEugenio}, Francesco and {Gallazzi}, Anna and {Pacifici}, Camilla and {Straatman}, Caroline M.~S. and {Bari{\v{s}}i{\'c}}, Ivana and {Bell}, Eric F. and {Maseda}, Michael V. and {Muzzin}, Adam and {Sobral}, David and {Whitaker}, Katherine E.},
        title = "{Toward Precise Galaxy Evolution: A Comparison between Spectral Indices of z  1 Galaxies in the IllustrisTNG Simulation and the LEGA-C Survey}",
      journal = {\aj},
         year = 2021,
        month = nov,
       volume = {162},
       number = {5},
          eid = {201},
        pages = {201},
          doi = {10.3847/1538-3881/ac20d6},
archivePrefix = {arXiv},
       eprint = {2108.10455},
 primaryClass = {astro-ph.GA},
       adsurl = {https://ui.adsabs.harvard.edu/abs/2021AJ....162..201W}
}

@ARTICLE{Zhu2018b,
       author = {{Zhu}, Ling and {van de Ven}, Glenn and {van den Bosch}, Remco and {Rix}, Hans-Walter and {Lyubenova}, Mariya and {Falc{\'o}n-Barroso}, Jes{\'u}s and {Martig}, Marie and {Mao}, Shude and {Xu}, Dandan and {Jin}, Yunpeng and {Obreja}, Aura and {Grand}, Robert J.~J. and {Dutton}, Aaron A. and {Macci{\`o}}, Andrea V. and {G{\'o}mez}, Facundo A. and {Walcher}, Jakob C. and {Garc{\'\i}a-Benito}, Rub{\'e}n and {Zibetti}, Stefano and {S{\'a}nchez}, Sebastian F.},
        title = "{The stellar orbit distribution in present-day galaxies inferred from the CALIFA survey}",
      journal = {Nature Astronomy},
         year = 2018,
        month = jan,
       volume = {2},
        pages = {233-238},
          doi = {10.1038/s41550-017-0348-1},
archivePrefix = {arXiv},
       eprint = {1711.06728},
 primaryClass = {astro-ph.GA},
       adsurl = {https://ui.adsabs.harvard.edu/abs/2018NatAs...2..233Z}
}

@ARTICLE{Rodriguez2016,
       author = {{Rodriguez-Gomez}, Vicente and {Pillepich}, Annalisa and {Sales}, Laura V. and {Genel}, Shy and {Vogelsberger}, Mark and {Zhu}, Qirong and {Wellons}, Sarah and {Nelson}, Dylan and {Torrey}, Paul and {Springel}, Volker and {Ma}, Chung-Pei and {Hernquist}, Lars},
        title = "{The stellar mass assembly of galaxies in the Illustris simulation: growth by mergers and the spatial distribution of accreted stars}",
      journal = {\mnras},
         year = 2016,
        month = may,
       volume = {458},
       number = {3},
        pages = {2371-2390},
          doi = {10.1093/mnras/stw456},
archivePrefix = {arXiv},
       eprint = {1511.08804},
 primaryClass = {astro-ph.GA},
       adsurl = {https://ui.adsabs.harvard.edu/abs/2016MNRAS.458.2371R}
}

@ARTICLE{Oser2010,
       author = {{Oser}, Ludwig and {Ostriker}, Jeremiah P. and {Naab}, Thorsten and {Johansson}, Peter H. and {Burkert}, Andreas},
        title = "{The Two Phases of Galaxy Formation}",
      journal = {\apj},
         year = 2010,
        month = dec,
       volume = {725},
       number = {2},
        pages = {2312-2323},
          doi = {10.1088/0004-637X/725/2/2312},
archivePrefix = {arXiv},
       eprint = {1010.1381},
 primaryClass = {astro-ph.CO},
       adsurl = {https://ui.adsabs.harvard.edu/abs/2010ApJ...725.2312O}
}

@ARTICLE{Guo2008MNRAS,
       author = {{Guo}, Q. and {White}, S.~D.~M.},
        title = "{Galaxy growth in the concordance {\ensuremath{\Lambda}}CDM cosmology}",
      journal = {\mnras},
         year = 2008,
        month = feb,
       volume = {384},
       number = {1},
        pages = {2-10},
          doi = {10.1111/j.1365-2966.2007.12619.x},
archivePrefix = {arXiv},
       eprint = {0708.1814},
 primaryClass = {astro-ph},
       adsurl = {https://ui.adsabs.harvard.edu/abs/2008MNRAS.384....2G}
}

@ARTICLE{Cora2018MNRAS,
       author = {{Cora}, Sof{\'\i}a A. and {Vega-Mart{\'\i}nez}, Cristian A. and {Hough}, Tom{\'a}s and {Ruiz}, Andr{\'e}s N. and {Orsi}, {\'A}lvaro A. and {Mu{\~n}oz Arancibia}, Alejandra M. and {Gargiulo}, Ignacio D. and {Collacchioni}, Florencia and {Padilla}, Nelson D. and {Gottl{\"o}ber}, Stefan and {Yepes}, Gustavo},
        title = "{Semi-analytic galaxies - I. Synthesis of environmental and star-forming regulation mechanisms}",
      journal = {\mnras},
         year = 2018,
        month = sep,
       volume = {479},
       number = {1},
        pages = {2-24},
          doi = {10.1093/mnras/sty1131},
archivePrefix = {arXiv},
       eprint = {1801.03883},
 primaryClass = {astro-ph.GA},
       adsurl = {https://ui.adsabs.harvard.edu/abs/2018MNRAS.479....2C}
}

@ARTICLE{Lee2013ApJ,
       author = {{Lee}, Jaehyun and {Yi}, Sukyoung K.},
        title = "{On the Assembly History of Stellar Components in Massive Galaxies}",
      journal = {\apj},
         year = 2013,
        month = mar,
       volume = {766},
       number = {1},
          eid = {38},
        pages = {38},
          doi = {10.1088/0004-637X/766/1/38},
archivePrefix = {arXiv},
       eprint = {1302.1588},
 primaryClass = {astro-ph.CO},
       adsurl = {https://ui.adsabs.harvard.edu/abs/2013ApJ...766...38L}
}

@ARTICLE{Santucci2024MNRAS,
       author = {{Santucci}, Giulia and {Lagos}, Claudia Del P. and {Harborne}, Katherine E. and {Ludlow}, Aaron and {Proctor}, Katy L. and {Foster}, Caroline and {McDermid}, Richard and {Poci}, Adriano and {Thater}, Sabine and {van de Ven}, Glenn and {Zhu}, Ling and {Walo Mart{\'\i}n}, Daniel},
        title = "{The distribution of stellar orbits in EAGLE galaxies - the effect of mergers, gas accretion, and secular evolution}",
      journal = {\mnras},
         year = 2024,
        month = feb,
       volume = {528},
       number = {2},
        pages = {2326-2345},
          doi = {10.1093/mnras/stae113},
archivePrefix = {arXiv},
       eprint = {2310.06885},
 primaryClass = {astro-ph.GA},
       adsurl = {https://ui.adsabs.harvard.edu/abs/2024MNRAS.528.2326S}
}

@ARTICLE{Kamdar2019,
       author = {{Kamdar}, Harshil and {Conroy}, Charlie and {Ting}, Yuan-Sen and {Bonaca}, Ana and {Smith}, Martin C. and {Brown}, Anthony G.~A.},
        title = "{Stars that Move Together Were Born Together}",
      journal = {\apjl},
         year = 2019,
        month = oct,
       volume = {884},
       number = {2},
          eid = {L42},
        pages = {L42},
          doi = {10.3847/2041-8213/ab4997},
archivePrefix = {arXiv},
       eprint = {1904.02159},
 primaryClass = {astro-ph.GA},
       adsurl = {https://ui.adsabs.harvard.edu/abs/2019ApJ...884L..42K}
}

@ARTICLE{Bird2013,
       author = {{Bird}, Jonathan C. and {Kazantzidis}, Stelios and {Weinberg}, David H. and {Guedes}, Javiera and {Callegari}, Simone and {Mayer}, Lucio and {Madau}, Piero},
        title = "{Inside out and Upside down: Tracing the Assembly of a Simulated Disk Galaxy Using Mono-age Stellar Populations}",
      journal = {\apj},
         year = 2013,
        month = aug,
       volume = {773},
       number = {1},
          eid = {43},
        pages = {43},
          doi = {10.1088/0004-637X/773/1/43},
archivePrefix = {arXiv},
       eprint = {1301.0620},
 primaryClass = {astro-ph.GA},
       adsurl = {https://ui.adsabs.harvard.edu/abs/2013ApJ...773...43B}
}

@ARTICLE{Peng2002AJ,
       author = {{Peng}, Chien Y. and {Ho}, Luis C. and {Impey}, Chris D. and {Rix}, Hans-Walter},
        title = "{Detailed Structural Decomposition of Galaxy Images}",
      journal = {\aj},
         year = 2002,
        month = jul,
       volume = {124},
       number = {1},
        pages = {266-293},
          doi = {10.1086/340952},
archivePrefix = {arXiv},
       eprint = {astro-ph/0204182},
 primaryClass = {astro-ph},
       adsurl = {https://ui.adsabs.harvard.edu/abs/2002AJ....124..266P}
}

@ARTICLE{Simard2011ApJS,
       author = {{Simard}, Luc and {Mendel}, J. Trevor and {Patton}, David R. and {Ellison}, Sara L. and {McConnachie}, Alan W.},
        title = "{A Catalog of Bulge+disk Decompositions and Updated Photometry for 1.12 Million Galaxies in the Sloan Digital Sky Survey}",
      journal = {\apjs},
         year = 2011,
        month = sep,
       volume = {196},
       number = {1},
          eid = {11},
        pages = {11},
          doi = {10.1088/0067-0049/196/1/11},
archivePrefix = {arXiv},
       eprint = {1107.1518},
 primaryClass = {astro-ph.CO},
       adsurl = {https://ui.adsabs.harvard.edu/abs/2011ApJS..196...11S}
}

@ARTICLE{Meert2015MNRAS,
       author = {{Meert}, Alan and {Vikram}, Vinu and {Bernardi}, Mariangela},
        title = "{A catalogue of 2D photometric decompositions in the SDSS-DR7 spectroscopic main galaxy sample: preferred models and systematics}",
      journal = {\mnras},
         year = 2015,
        month = feb,
       volume = {446},
       number = {4},
        pages = {3943-3974},
          doi = {10.1093/mnras/stu2333},
archivePrefix = {arXiv},
       eprint = {1406.4179},
 primaryClass = {astro-ph.GA},
       adsurl = {https://ui.adsabs.harvard.edu/abs/2015MNRAS.446.3943M}
}

@BOOK{Sérsic1968,
  author    = {{Sérsic}, José Luis},
  title     = {Atlas de Galaxias Australes},
  year      = {1968},
  publisher = {Observatorio Astronómico, Universidad Nacional de Córdoba},
  address   = {Córdoba, Argentina},
  adsurl    = {https://ui.adsabs.harvard.edu/abs/1968adga.book.....S}
}

@ARTICLE{Freeman1970ApJ,
       author = {{Freeman}, K.~C.},
        title = "{On the Disks of Spiral and S0 Galaxies}",
      journal = {\apj},
         year = 1970,
        month = jun,
       volume = {160},
        pages = {811},
          doi = {10.1086/150474},
       adsurl = {https://ui.adsabs.harvard.edu/abs/1970ApJ...160..811F}
}

@ARTICLE{Teklu2015ApJ,
       author = {{Teklu}, Adelheid F. and {Remus}, Rhea-Silvia and {Dolag}, Klaus and {Beck}, Alexander M. and {Burkert}, Andreas and {Schmidt}, Andreas S. and {Schulze}, Felix and {Steinborn}, Lisa K.},
        title = "{Connecting Angular Momentum and Galactic Dynamics: The Complex Interplay between Spin, Mass, and Morphology}",
      journal = {\apj},
         year = 2015,
        month = oct,
       volume = {812},
       number = {1},
          eid = {29},
        pages = {29},
          doi = {10.1088/0004-637X/812/1/29},
archivePrefix = {arXiv},
       eprint = {1503.03501},
 primaryClass = {astro-ph.GA},
       adsurl = {https://ui.adsabs.harvard.edu/abs/2015ApJ...812...29T}
}

@ARTICLE{Obreja2016MNRAS,
       author = {{Obreja}, Aura and {Stinson}, Gregory S. and {Dutton}, Aaron A. and {Macci{\`o}}, Andrea V. and {Wang}, Liang and {Kang}, Xi},
        title = "{NIHAO VI. The hidden discs of simulated galaxies}",
      journal = {\mnras},
         year = 2016,
        month = jun,
       volume = {459},
       number = {1},
        pages = {467-486},
          doi = {10.1093/mnras/stw690},
archivePrefix = {arXiv},
       eprint = {1603.01703},
 primaryClass = {astro-ph.GA},
       adsurl = {https://ui.adsabs.harvard.edu/abs/2016MNRAS.459..467O}
}

@ARTICLE{Zhu2018MNRAS,
       author = {{Zhu}, Ling and {van de Ven}, Glenn and {M{\'e}ndez-Abreu}, Jairo and {Obreja}, Aura},
        title = "{Morphology and kinematics of orbital components in CALIFA galaxies across the Hubble sequence}",
      journal = {\mnras},
         year = 2018,
        month = sep,
       volume = {479},
       number = {1},
        pages = {945-960},
          doi = {10.1093/mnras/sty1521},
archivePrefix = {arXiv},
       eprint = {1806.02886},
 primaryClass = {astro-ph.GA},
       adsurl = {https://ui.adsabs.harvard.edu/abs/2018MNRAS.479..945Z}
}

@ARTICLE{Scannapieco2010MNRAS,
       author = {{Scannapieco}, Cecilia and {Gadotti}, Dimitri A. and {Jonsson}, Patrik and {White}, Simon D.~M.},
        title = "{An observer's view of simulated galaxies: disc-to-total ratios, bars and (pseudo-)bulges}",
      journal = {\mnras},
         year = 2010,
        month = sep,
       volume = {407},
       number = {1},
        pages = {L41-L45},
          doi = {10.1111/j.1745-3933.2010.00900.x},
archivePrefix = {arXiv},
       eprint = {1001.4890},
 primaryClass = {astro-ph.GA},
       adsurl = {https://ui.adsabs.harvard.edu/abs/2010MNRAS.407L..41S}
}

@ARTICLE{Martig2012ApJ,
       author = {{Martig}, Marie and {Bournaud}, Fr{\'e}d{\'e}ric and {Croton}, Darren J. and {Dekel}, Avishai and {Teyssier}, Romain},
        title = "{A Diversity of Progenitors and Histories for Isolated Spiral Galaxies}",
      journal = {\apj},
         year = 2012,
        month = sep,
       volume = {756},
       number = {1},
          eid = {26},
        pages = {26},
          doi = {10.1088/0004-637X/756/1/26},
archivePrefix = {arXiv},
       eprint = {1201.1079},
 primaryClass = {astro-ph.CO},
       adsurl = {https://ui.adsabs.harvard.edu/abs/2012ApJ...756...26M}
}

@ARTICLE{Schwarzschild1979ApJ,
       author = {{Schwarzschild}, M.},
        title = "{A numerical model for a triaxial stellar system in dynamical equilibrium.}",
      journal = {\apj},
         year = 1979,
        month = aug,
       volume = {232},
        pages = {236-247},
          doi = {10.1086/157282},
       adsurl = {https://ui.adsabs.harvard.edu/abs/1979ApJ...232..236S}
}

@ARTICLE{Marel1998ApJ,
       author = {{van der Marel}, Roeland P. and {Cretton}, N. and {de Zeeuw}, P. Tim and {Rix}, Hans-Walter},
        title = "{Improved Evidence for a Black Hole in M32 from HST/FOS Spectra. II. Axisymmetric Dynamical Models}",
      journal = {\apj},
         year = 1998,
        month = jan,
       volume = {493},
       number = {2},
        pages = {613-631},
          doi = {10.1086/305147},
archivePrefix = {arXiv},
       eprint = {astro-ph/9705081},
 primaryClass = {astro-ph},
       adsurl = {https://ui.adsabs.harvard.edu/abs/1998ApJ...493..613V}
}

@ARTICLE{Verolme2002MNRA,
       author = {{Verolme}, E.~K. and {Cappellari}, M. and {Copin}, Y. and {van der Marel}, R.~P. and {Bacon}, R. and {Bureau}, M. and {Davies}, R.~L. and {Miller}, B.~M. and {de Zeeuw}, P.~T.},
        title = "{A SAURON study of M32: measuring the intrinsic flattening and the central black hole mass}",
      journal = {\mnras},
         year = 2002,
        month = sep,
       volume = {335},
       number = {3},
        pages = {517-525},
          doi = {10.1046/j.1365-8711.2002.05664.x},
archivePrefix = {arXiv},
       eprint = {astro-ph/0201086},
 primaryClass = {astro-ph},
       adsurl = {https://ui.adsabs.harvard.edu/abs/2002MNRAS.335..517V}
}

@ARTICLE{Gebhardt2003ApJ,
       author = {{Gebhardt}, Karl and {Richstone}, Douglas and {Tremaine}, Scott and {Lauer}, Tod R. and {Bender}, Ralf and {Bower}, Gary and {Dressler}, Alan and {Faber}, S.~M. and {Filippenko}, Alexei V. and {Green}, Richard and {Grillmair}, Carl and {Ho}, Luis C. and {Kormendy}, John and {Magorrian}, John and {Pinkney}, Jason},
        title = "{Axisymmetric Dynamical Models of the Central Regions of Galaxies}",
      journal = {\apj},
         year = 2003,
        month = jan,
       volume = {583},
       number = {1},
        pages = {92-115},
          doi = {10.1086/345081},
archivePrefix = {arXiv},
       eprint = {astro-ph/0209483},
 primaryClass = {astro-ph},
       adsurl = {https://ui.adsabs.harvard.edu/abs/2003ApJ...583...92G}
}

@ARTICLE{Valluri2004ApJ,
       author = {{Valluri}, Monica and {Merritt}, David and {Emsellem}, Eric},
        title = "{Difficulties with Recovering the Masses of Supermassive Black Holes from Stellar Kinematical Data}",
      journal = {\apj},
         year = 2004,
        month = feb,
       volume = {602},
       number = {1},
        pages = {66-92},
          doi = {10.1086/380896},
archivePrefix = {arXiv},
       eprint = {astro-ph/0210379},
 primaryClass = {astro-ph},
       adsurl = {https://ui.adsabs.harvard.edu/abs/2004ApJ...602...66V}
}

@ARTICLE{Krajnovic2009MNRAS,
       author = {{Krajnovi{\'c}}, Davor and {McDermid}, Richard M. and {Cappellari}, Michele and {Davies}, Roger L.},
        title = "{Determination of masses of the central black holes in NGC 524 and 2549 using laser guide star adaptive optics}",
      journal = {\mnras},
         year = 2009,
        month = nov,
       volume = {399},
       number = {4},
        pages = {1839-1857},
          doi = {10.1111/j.1365-2966.2009.15415.x},
archivePrefix = {arXiv},
       eprint = {0907.3748},
 primaryClass = {astro-ph.GA},
       adsurl = {https://ui.adsabs.harvard.edu/abs/2009MNRAS.399.1839K}
}

@ARTICLE{Rusli2013AJ,
       author = {{Rusli}, S.~P. and {Thomas}, J. and {Saglia}, R.~P. and {Fabricius}, M. and {Erwin}, P. and {Bender}, R. and {Nowak}, N. and {Lee}, C.~H. and {Riffeser}, A. and {Sharp}, R.},
        title = "{The Influence of Dark Matter Halos on Dynamical Estimates of Black Hole Mass: 10 New Measurements for High-{\ensuremath{\sigma}} Early-type Galaxies}",
      journal = {\aj},
         year = 2013,
        month = sep,
       volume = {146},
       number = {3},
          eid = {45},
        pages = {45},
          doi = {10.1088/0004-6256/146/3/45},
archivePrefix = {arXiv},
       eprint = {1306.1124},
 primaryClass = {astro-ph.CO},
       adsurl = {https://ui.adsabs.harvard.edu/abs/2013AJ....146...45R}
}

@ARTICLE{Seth2014Natur,
       author = {{Seth}, Anil C. and {van den Bosch}, Remco and {Mieske}, Steffen and {Baumgardt}, Holger and {Brok}, Mark Den and {Strader}, Jay and {Neumayer}, Nadine and {Chilingarian}, Igor and {Hilker}, Michael and {McDermid}, Richard and {Spitler}, Lee and {Brodie}, Jean and {Frank}, Matthias J. and {Walsh}, Jonelle L.},
        title = "{A supermassive black hole in an ultra-compact dwarf galaxy}",
      journal = {\nat},
         year = 2014,
        month = sep,
       volume = {513},
       number = {7518},
        pages = {398-400},
          doi = {10.1038/nature13762},
archivePrefix = {arXiv},
       eprint = {1409.4769},
 primaryClass = {astro-ph.GA},
       adsurl = {https://ui.adsabs.harvard.edu/abs/2014Natur.513..398S}
}

@ARTICLE{Thater2017A&A,
       author = {{Thater}, S. and {Krajnovi{\'c}}, D. and {Bourne}, M.~A. and {Cappellari}, M. and {de Zeeuw}, T. and {Emsellem}, E. and {Magorrian}, J. and {McDermid}, R.~M. and {Sarzi}, M. and {van de Ven}, G.},
        title = "{A low upper mass limit for the central black hole in the late-type galaxy NGC 4414}",
      journal = {\aap},
         year = 2017,
        month = jan,
       volume = {597},
          eid = {A18},
        pages = {A18},
          doi = {10.1051/0004-6361/201629480},
archivePrefix = {arXiv},
       eprint = {1610.04269},
 primaryClass = {astro-ph.GA},
       adsurl = {https://ui.adsabs.harvard.edu/abs/2017A&A...597A..18T}
}

@ARTICLE{Thater2019A&A,
       author = {{Thater}, Sabine and {Krajnovi{\'c}}, Davor and {Cappellari}, Michele and {Davis}, Timothy A. and {de Zeeuw}, P. Tim and {McDermid}, Richard M. and {Sarzi}, Marc},
        title = "{Six new supermassive black hole mass determinations from adaptive-optics assisted SINFONI observations}",
      journal = {\aap},
         year = 2019,
        month = may,
       volume = {625},
          eid = {A62},
        pages = {A62},
          doi = {10.1051/0004-6361/201834808},
archivePrefix = {arXiv},
       eprint = {1902.10175},
 primaryClass = {astro-ph.GA},
       adsurl = {https://ui.adsabs.harvard.edu/abs/2019A&A...625A..62T}
}

@ARTICLE{Liepold2020ApJ,
       author = {{Liepold}, Emily R. and {Quenneville}, Matthew E. and {Ma}, Chung-Pei and {Walsh}, Jonelle L. and {McConnell}, Nicholas J. and {Greene}, Jenny E. and {Blakeslee}, John P.},
        title = "{The MASSIVE Survey. XV. A Stellar Dynamical Mass Measurement of the Supermassive Black Hole in Massive Elliptical Galaxy NGC 1453}",
      journal = {\apj},
         year = 2020,
        month = mar,
       volume = {891},
       number = {1},
          eid = {4},
        pages = {4},
          doi = {10.3847/1538-4357/ab6f71},
archivePrefix = {arXiv},
       eprint = {2001.08753},
 primaryClass = {astro-ph.GA},
       adsurl = {https://ui.adsabs.harvard.edu/abs/2020ApJ...891....4L}
}

@ARTICLE{Quenneville2021ApJS,
       author = {{Quenneville}, Matthew E. and {Liepold}, Emily R. and {Ma}, Chung-Pei},
        title = "{Dynamical Modeling of Galaxies and Supermassive Black Holes: Axisymmetry in Triaxial Schwarzschild Orbit Superposition Models}",
      journal = {\apjs},
         year = 2021,
        month = jun,
       volume = {254},
       number = {2},
          eid = {25},
        pages = {25},
          doi = {10.3847/1538-4365/abe6a0},
archivePrefix = {arXiv},
       eprint = {2005.00542},
 primaryClass = {astro-ph.GA},
       adsurl = {https://ui.adsabs.harvard.edu/abs/2021ApJS..254...25Q}
}

@ARTICLE{Cappellari2016ARA,
       author = {{Cappellari}, Michele},
        title = "{Structure and Kinematics of Early-Type Galaxies from Integral Field Spectroscopy}",
      journal = {\araa},
         year = 2016,
        month = sep,
       volume = {54},
        pages = {597-665},
          doi = {10.1146/annurev-astro-082214-122432},
archivePrefix = {arXiv},
       eprint = {1602.04267},
 primaryClass = {astro-ph.GA},
       adsurl = {https://ui.adsabs.harvard.edu/abs/2016ARA&A..54..597C}
}

@ARTICLE{Thomas2007MNRAS,
       author = {{Thomas}, J. and {Saglia}, R.~P. and {Bender}, R. and {Thomas}, D. and {Gebhardt}, K. and {Magorrian}, J. and {Corsini}, E.~M. and {Wegner}, G.},
        title = "{Dynamical modelling of luminous and dark matter in 17 Coma early-type galaxies}",
      journal = {\mnras},
         year = 2007,
        month = dec,
       volume = {382},
       number = {2},
        pages = {657-684},
          doi = {10.1111/j.1365-2966.2007.12434.x},
archivePrefix = {arXiv},
       eprint = {0709.0691},
 primaryClass = {astro-ph},
       adsurl = {https://ui.adsabs.harvard.edu/abs/2007MNRAS.382..657T}
}

@ARTICLE{van2008MNRAS,
       author = {{van de Ven}, G. and {de Zeeuw}, P.~T. and {van den Bosch}, R.~C.~E.},
        title = "{Recovery of the internal orbital structure of galaxies}",
      journal = {\mnras},
         year = 2008,
        month = apr,
       volume = {385},
       number = {2},
        pages = {614-646},
          doi = {10.1111/j.1365-2966.2008.12873.x},
archivePrefix = {arXiv},
       eprint = {0712.0309},
 primaryClass = {astro-ph},
       adsurl = {https://ui.adsabs.harvard.edu/abs/2008MNRAS.385..614V}
}

@ARTICLE{Thomas2014ApJ,
       author = {{Thomas}, J. and {Saglia}, R.~P. and {Bender}, R. and {Erwin}, P. and {Fabricius}, M.},
        title = "{The Dynamical Fingerprint of Core Scouring in Massive Elliptical Galaxies}",
      journal = {\apj},
         year = 2014,
        month = feb,
       volume = {782},
       number = {1},
          eid = {39},
        pages = {39},
          doi = {10.1088/0004-637X/782/1/39},
archivePrefix = {arXiv},
       eprint = {1311.3783},
 primaryClass = {astro-ph.GA},
       adsurl = {https://ui.adsabs.harvard.edu/abs/2014ApJ...782...39T}
}

@ARTICLE{Fahrion2019A,
       author = {{Fahrion}, Katja and {Lyubenova}, Mariya and {van de Ven}, Glenn and {Leaman}, Ryan and {Hilker}, Michael and {Mart{\'\i}n-Navarro}, Ignacio and {Zhu}, Ling and {Alfaro-Cuello}, Mayte and {Coccato}, Lodovico and {Corsini}, Enrico M. and {Falc{\'o}n-Barroso}, Jes{\'u}s and {Iodice}, Enrichetta and {McDermid}, Richard M. and {Sarzi}, Marc and {de Zeeuw}, Tim},
        title = "{Constraining nuclear star cluster formation using MUSE-AO observations of the early-type galaxy FCC 47}",
      journal = {\aap},
         year = 2019,
        month = aug,
       volume = {628},
          eid = {A92},
        pages = {A92},
          doi = {10.1051/0004-6361/201935832},
archivePrefix = {arXiv},
       eprint = {1907.01007},
 primaryClass = {astro-ph.GA},
       adsurl = {https://ui.adsabs.harvard.edu/abs/2019A&A...628A..92F}
}

@ARTICLE{Poci2019MNRAS,
       author = {{Poci}, Adriano and {McDermid}, Richard M. and {Zhu}, Ling and {van de Ven}, Glenn},
        title = "{Combining stellar populations with orbit-superposition dynamical modelling: the formation history of the lenticular galaxy NGC 3115}",
      journal = {\mnras},
         year = 2019,
        month = aug,
       volume = {487},
       number = {3},
        pages = {3776-3796},
          doi = {10.1093/mnras/stz1154},
archivePrefix = {arXiv},
       eprint = {1904.11605},
 primaryClass = {astro-ph.GA},
       adsurl = {https://ui.adsabs.harvard.edu/abs/2019MNRAS.487.3776P}
}

@ARTICLE{Thater2022MNRAS,
       author = {{Thater}, Sabine and {Krajnovi{\'c}}, Davor and {Weilbacher}, Peter M. and {Nguyen}, Dieu D. and {Bureau}, Martin and {Cappellari}, Michele and {Davis}, Timothy A. and {Iguchi}, Satoru and {McDermid}, Richard and {Onishi}, Kyoko and {Sarzi}, Marc and {van de Ven}, Glenn},
        title = "{Cross-checking SMBH mass estimates in NGC 6958 - I. Stellar dynamics from adaptive optics-assisted MUSE observations}",
      journal = {\mnras},
         year = 2022,
        month = feb,
       volume = {509},
       number = {4},
        pages = {5416-5436},
          doi = {10.1093/mnras/stab3210},
archivePrefix = {arXiv},
       eprint = {2111.01620},
 primaryClass = {astro-ph.GA},
       adsurl = {https://ui.adsabs.harvard.edu/abs/2022MNRAS.509.5416T}
}

@ARTICLE{Santucci2022ApJ,
       author = {{Santucci}, Giulia and {Brough}, Sarah and {van de Sande}, Jesse and {McDermid}, Richard M. and {van de Ven}, Glenn and {Zhu}, Ling and {D'Eugenio}, Francesco and {Bland-Hawthorn}, Joss and {Barsanti}, Stefania and {Bryant}, Julia J. and {Croom}, Scott M. and {Davies}, Roger L. and {Green}, Andrew W. and {Lawrence}, Jon S. and {Lorente}, Nuria P.~F. and {Owers}, Matt S. and {Poci}, Adriano and {Richards}, Samuel N. and {Thater}, Sabine and {Yi}, Sukyoung},
        title = "{The SAMI Galaxy Survey: The Internal Orbital Structure and Mass Distribution of Passive Galaxies from Triaxial Orbit-superposition Schwarzschild Models}",
      journal = {\apj},
         year = 2022,
        month = may,
       volume = {930},
       number = {2},
          eid = {153},
        pages = {153},
          doi = {10.3847/1538-4357/ac5bd5},
archivePrefix = {arXiv},
       eprint = {2203.03648},
 primaryClass = {astro-ph.GA},
       adsurl = {https://ui.adsabs.harvard.edu/abs/2022ApJ...930..153S}
}

@ARTICLE{vande2006A,
       author = {{van de Ven}, G. and {van den Bosch}, R.~C.~E. and {Verolme}, E.~K. and {de Zeeuw}, P.~T.},
        title = "{The dynamical distance and intrinsic structure of the globular cluster {\ensuremath{\omega}} Centauri}",
      journal = {\aap},
         year = 2006,
        month = jan,
       volume = {445},
       number = {2},
        pages = {513-543},
          doi = {10.1051/0004-6361:20053061},
archivePrefix = {arXiv},
       eprint = {astro-ph/0509228},
 primaryClass = {astro-ph},
       adsurl = {https://ui.adsabs.harvard.edu/abs/2006A&A...445..513V}
}

@ARTICLE{Feldmeier2017MNRAS,
       author = {{Feldmeier-Krause}, A. and {Zhu}, L. and {Neumayer}, N. and {van de Ven}, G. and {de Zeeuw}, P.~T. and {Sch{\"o}del}, R.},
        title = "{Triaxial orbit-based modelling of the Milky Way Nuclear Star Cluster}",
      journal = {\mnras},
         year = 2017,
        month = apr,
       volume = {466},
       number = {4},
        pages = {4040-4052},
          doi = {10.1093/mnras/stw3377},
archivePrefix = {arXiv},
       eprint = {1701.01583},
 primaryClass = {astro-ph.GA},
       adsurl = {https://ui.adsabs.harvard.edu/abs/2017MNRAS.466.4040F}
}

@ARTICLE{Zocchi2012A&A,
       author = {{Zocchi}, A. and {Bertin}, G. and {Varri}, A.~L.},
        title = "{A dynamical study of Galactic globular clusters under different relaxation conditions}",
      journal = {\aap},
         year = 2012,
        month = mar,
       volume = {539},
          eid = {A65},
        pages = {A65},
          doi = {10.1051/0004-6361/201117977},
archivePrefix = {arXiv},
       eprint = {1201.1466},
 primaryClass = {astro-ph.GA},
       adsurl = {https://ui.adsabs.harvard.edu/abs/2012A&A...539A..65Z}
}

@ARTICLE{deZeeuw2002MNRAS,
       author = {{de Zeeuw}, P.~T. and {Bureau}, M. and {Emsellem}, Eric and {Bacon}, R. and {Carollo}, C.~M. and {Copin}, Y. and {Davies}, Roger L. and {Kuntschner}, Harald and {Miller}, Bryan W. and {Monnet}, G. and {Peletier}, Reynier F. and {Verolme}, E.~K.},
        title = "{The SAURON project - II. Sample and early results}",
      journal = {\mnras},
         year = 2002,
        month = jan,
       volume = {329},
       number = {3},
        pages = {513-530},
          doi = {10.1046/j.1365-8711.2002.05059.x},
archivePrefix = {arXiv},
       eprint = {astro-ph/0109511},
 primaryClass = {astro-ph},
       adsurl = {https://ui.adsabs.harvard.edu/abs/2002MNRAS.329..513D}
}

@ARTICLE{Cappellari2011MNRAS,
       author = {{Cappellari}, Michele and {Emsellem}, Eric and {Krajnovi{\'c}}, Davor and {McDermid}, Richard M. and {Scott}, Nicholas and {Verdoes Kleijn}, G.~A. and {Young}, Lisa M. and {Alatalo}, Katherine and {Bacon}, R. and {Blitz}, Leo and {Bois}, Maxime and {Bournaud}, Fr{\'e}d{\'e}ric and {Bureau}, M. and {Davies}, Roger L. and {Davis}, Timothy A. and {de Zeeuw}, P.~T. and {Duc}, Pierre-Alain and {Khochfar}, Sadegh and {Kuntschner}, Harald and {Lablanche}, Pierre-Yves and {Morganti}, Raffaella and {Naab}, Thorsten and {Oosterloo}, Tom and {Sarzi}, Marc and {Serra}, Paolo and {Weijmans}, Anne-Marie},
        title = "{The ATLAS$^{3D}$ project - I. A volume-limited sample of 260 nearby early-type galaxies: science goals and selection criteria}",
      journal = {\mnras},
         year = 2011,
        month = may,
       volume = {413},
       number = {2},
        pages = {813-836},
          doi = {10.1111/j.1365-2966.2010.18174.x},
archivePrefix = {arXiv},
       eprint = {1012.1551},
 primaryClass = {astro-ph.CO},
       adsurl = {https://ui.adsabs.harvard.edu/abs/2011MNRAS.413..813C}
}

@ARTICLE{Sanchez2023MNRAS,
       author = {{S{\'a}nchez}, S.~F. and {Galbany}, L. and {Walcher}, C.~J. and {Garc{\'\i}a-Benito}, R. and {Barrera-Ballesteros}, J.~K.},
        title = "{The Calar Alto Legacy Integral Field Area survey: extended and remastered data release}",
      journal = {\mnras},
         year = 2023,
        month = dec,
       volume = {526},
       number = {4},
        pages = {5555-5589},
          doi = {10.1093/mnras/stad3119},
archivePrefix = {arXiv},
       eprint = {2304.13022},
 primaryClass = {astro-ph.GA},
       adsurl = {https://ui.adsabs.harvard.edu/abs/2023MNRAS.526.5555S}
}

@ARTICLE{Croom2012MNRAS,
       author = {{Croom}, Scott M. and {Lawrence}, Jon S. and {Bland-Hawthorn}, Joss and {Bryant}, Julia J. and {Fogarty}, Lisa and {Richards}, Samuel and {Goodwin}, Michael and {Farrell}, Tony and {Miziarski}, Stan and {Heald}, Ron and {Jones}, D. Heath and {Lee}, Steve and {Colless}, Matthew and {Brough}, Sarah and {Hopkins}, Andrew M. and {Bauer}, Amanda E. and {Birchall}, Michael N. and {Ellis}, Simon and {Horton}, Anthony and {Leon-Saval}, Sergio and {Lewis}, Geraint and {L{\'o}pez-S{\'a}nchez}, {\'A}. R. and {Min}, Seong-Sik and {Trinh}, Christopher and {Trowland}, Holly},
        title = "{The Sydney-AAO Multi-object Integral field spectrograph}",
      journal = {\mnras},
         year = 2012,
        month = mar,
       volume = {421},
       number = {1},
        pages = {872-893},
          doi = {10.1111/j.1365-2966.2011.20365.x},
archivePrefix = {arXiv},
       eprint = {1112.3367},
 primaryClass = {astro-ph.CO},
       adsurl = {https://ui.adsabs.harvard.edu/abs/2012MNRAS.421..872C}
}

@ARTICLE{Bryant2015MNRAS,
       author = {{Bryant}, J.~J. and {Owers}, M.~S. and {Robotham}, A.~S.~G. and {Croom}, S.~M. and {Driver}, S.~P. and {Drinkwater}, M.~J. and {Lorente}, N.~P.~F. and {Cortese}, L. and {Scott}, N. and {Colless}, M. and {Schaefer}, A. and {Taylor}, E.~N. and {Konstantopoulos}, I.~S. and {Allen}, J.~T. and {Baldry}, I. and {Barnes}, L. and {Bauer}, A.~E. and {Bland-Hawthorn}, J. and {Bloom}, J.~V. and {Brooks}, A.~M. and {Brough}, S. and {Cecil}, G. and {Couch}, W. and {Croton}, D. and {Davies}, R. and {Ellis}, S. and {Fogarty}, L.~M.~R. and {Foster}, C. and {Glazebrook}, K. and {Goodwin}, M. and {Green}, A. and {Gunawardhana}, M.~L. and {Hampton}, E. and {Ho}, I.-T. and {Hopkins}, A.~M. and {Kewley}, L. and {Lawrence}, J.~S. and {Leon-Saval}, S.~G. and {Leslie}, S. and {McElroy}, R. and {Lewis}, G. and {Liske}, J. and {L{\'o}pez-S{\'a}nchez}, {\'A}. R. and {Mahajan}, S. and {Medling}, A.~M. and {Metcalfe}, N. and {Meyer}, M. and {Mould}, J. and {Obreschkow}, D. and {O'Toole}, S. and {Pracy}, M. and {Richards}, S.~N. and {Shanks}, T. and {Sharp}, R. and {Sweet}, S.~M. and {Thomas}, A.~D. and {Tonini}, C. and {Walcher}, C.~J.},
        title = "{The SAMI Galaxy Survey: instrument specification and target selection}",
      journal = {\mnras},
         year = 2015,
        month = mar,
       volume = {447},
       number = {3},
        pages = {2857-2879},
          doi = {10.1093/mnras/stu2635},
archivePrefix = {arXiv},
       eprint = {1407.7335},
 primaryClass = {astro-ph.GA},
       adsurl = {https://ui.adsabs.harvard.edu/abs/2015MNRAS.447.2857B}
}

@ARTICLE{Croom2021MNRAS,
       author = {{Croom}, Scott M. and {Owers}, Matt S. and {Scott}, Nicholas and {Poetrodjojo}, Henry and {Groves}, Brent and {van de Sande}, Jesse and {Barone}, Tania M. and {Cortese}, Luca and {D'Eugenio}, Francesco and {Bland-Hawthorn}, Joss and {Bryant}, Julia and {Oh}, Sree and {Brough}, Sarah and {Agostino}, James and {Casura}, Sarah and {Catinella}, Barbara and {Colless}, Matthew and {Cecil}, Gerald and {Davies}, Roger L. and {Drinkwater}, Michael J. and {Driver}, Simon P. and {Ferreras}, Ignacio and {Foster}, Caroline and {Fraser-McKelvie}, Amelia and {Lawrence}, Jon and {Leslie}, Sarah K. and {Liske}, Jochen and {L{\'o}pez-S{\'a}nchez}, {\'A}ngel R. and {Lorente}, Nuria P.~F. and {McElroy}, Rebecca and {Medling}, Anne M. and {Obreschkow}, Danail and {Richards}, Samuel N. and {Sharp}, Rob and {Sweet}, Sarah M. and {Taranu}, Dan S. and {Taylor}, Edward N. and {Tescari}, Edoardo and {Thomas}, Adam D. and {Tocknell}, James and {Vaughan}, Sam P.},
        title = "{The SAMI Galaxy Survey: the third and final data release}",
      journal = {\mnras},
         year = 2021,
        month = jul,
       volume = {505},
       number = {1},
        pages = {991-1016},
          doi = {10.1093/mnras/stab229},
archivePrefix = {arXiv},
       eprint = {2101.12224},
 primaryClass = {astro-ph.GA},
       adsurl = {https://ui.adsabs.harvard.edu/abs/2021MNRAS.505..991C}
}

@ARTICLE{Bundy2015ApJ,
       author = {{Bundy}, Kevin and {Bershady}, Matthew A. and {Law}, David R. and {Yan}, Renbin and {Drory}, Niv and {MacDonald}, Nicholas and {Wake}, David A. and {Cherinka}, Brian and {S{\'a}nchez-Gallego}, Jos{\'e} R. and {Weijmans}, Anne-Marie and {Thomas}, Daniel and {Tremonti}, Christy and {Masters}, Karen and {Coccato}, Lodovico and {Diamond-Stanic}, Aleksandar M. and {Arag{\'o}n-Salamanca}, Alfonso and {Avila-Reese}, Vladimir and {Badenes}, Carles and {Falc{\'o}n-Barroso}, J{\'e}sus and {Belfiore}, Francesco and {Bizyaev}, Dmitry and {Blanc}, Guillermo A. and {Bland-Hawthorn}, Joss and {Blanton}, Michael R. and {Brownstein}, Joel R. and {Byler}, Nell and {Cappellari}, Michele and {Conroy}, Charlie and {Dutton}, Aaron A. and {Emsellem}, Eric and {Etherington}, James and {Frinchaboy}, Peter M. and {Fu}, Hai and {Gunn}, James E. and {Harding}, Paul and {Johnston}, Evelyn J. and {Kauffmann}, Guinevere and {Kinemuchi}, Karen and {Klaene}, Mark A. and {Knapen}, Johan H. and {Leauthaud}, Alexie and {Li}, Cheng and {Lin}, Lihwai and {Maiolino}, Roberto and {Malanushenko}, Viktor and {Malanushenko}, Elena and {Mao}, Shude and {Maraston}, Claudia and {McDermid}, Richard M. and {Merrifield}, Michael R. and {Nichol}, Robert C. and {Oravetz}, Daniel and {Pan}, Kaike and {Parejko}, John K. and {Sanchez}, Sebastian F. and {Schlegel}, David and {Simmons}, Audrey and {Steele}, Oliver and {Steinmetz}, Matthias and {Thanjavur}, Karun and {Thompson}, Benjamin A. and {Tinker}, Jeremy L. and {van den Bosch}, Remco C.~E. and {Westfall}, Kyle B. and {Wilkinson}, David and {Wright}, Shelley and {Xiao}, Ting and {Zhang}, Kai},
        title = "{Overview of the SDSS-IV MaNGA Survey: Mapping nearby Galaxies at Apache Point Observatory}",
      journal = {\apj},
         year = 2015,
        month = jan,
       volume = {798},
       number = {1},
          eid = {7},
        pages = {7},
          doi = {10.1088/0004-637X/798/1/7},
archivePrefix = {arXiv},
       eprint = {1412.1482},
 primaryClass = {astro-ph.GA},
       adsurl = {https://ui.adsabs.harvard.edu/abs/2015ApJ...798....7B}
}

@ARTICLE{Sarzi2018,
       author = {{Sarzi}, M. and {Iodice}, E. and {Coccato}, L. and {Corsini}, E.~M. and {de Zeeuw}, P.~T. and {Falc{\'o}n-Barroso}, J. and {Gadotti}, D.~A. and {Lyubenova}, M. and {McDermid}, R.~M. and {van de Ven}, G. and {Fahrion}, K. and {Pizzella}, A. and {Zhu}, L.},
        title = "{Fornax3D project: Overall goals, galaxy sample, MUSE data analysis, and initial results}",
      journal = {\aap},
         year = 2018,
        month = aug,
       volume = {616},
          eid = {A121},
        pages = {A121},
          doi = {10.1051/0004-6361/201833137},
archivePrefix = {arXiv},
       eprint = {1804.06795},
 primaryClass = {astro-ph.GA},
       adsurl = {https://ui.adsabs.harvard.edu/abs/2018A&A...616A.121S}
}

@ARTICLE{Jin2020MNRAS,
       author = {{Jin}, Yunpeng and {Zhu}, Ling and {Long}, R.~J. and {Mao}, Shude and {Wang}, Lan and {van de Ven}, Glenn},
        title = "{SDSS-IV MaNGA: Internal mass distributions and orbital structures of early-type galaxies and their dependence on environment}",
      journal = {\mnras},
         year = 2020,
        month = jan,
       volume = {491},
       number = {2},
        pages = {1690-1708},
          doi = {10.1093/mnras/stz3072},
archivePrefix = {arXiv},
       eprint = {1911.00777},
 primaryClass = {astro-ph.GA},
       adsurl = {https://ui.adsabs.harvard.edu/abs/2020MNRAS.491.1690J}
}

@ARTICLE{Xu2019MNRAS,
       author = {{Xu}, Dandan and {Zhu}, Ling and {Grand}, Robert and {Springel}, Volker and {Mao}, Shude and {van de Ven}, Glenn and {Lu}, Shengdong and {Wang}, Yougang and {Pillepich}, Annalisa and {Genel}, Shy and {Nelson}, Dylan and {Rodriguez-Gomez}, Vicente and {Pakmor}, R{\"u}diger and {Weinberger}, Rainer and {Marinacci}, Federico and {Vogelsberger}, Mark and {Torrey}, Paul and {Naiman}, Jill and {Hernquist}, Lars},
        title = "{A study of stellar orbit fractions: simulated IllustrisTNG galaxies compared to CALIFA observations}",
      journal = {\mnras},
         year = 2019,
        month = oct,
       volume = {489},
       number = {1},
        pages = {842-854},
          doi = {10.1093/mnras/stz2164},
archivePrefix = {arXiv},
       eprint = {1905.11415},
 primaryClass = {astro-ph.GA},
       adsurl = {https://ui.adsabs.harvard.edu/abs/2019MNRAS.489..842X}
}

@ARTICLE{McAlpine2016,
       author = {{McAlpine}, S. and {Helly}, J.~C. and {Schaller}, M. and {Trayford}, J.~W. and {Qu}, Y. and {Furlong}, M. and {Bower}, R.~G. and {Crain}, R.~A. and {Schaye}, J. and {Theuns}, T. and {Dalla Vecchia}, C. and {Frenk}, C.~S. and {McCarthy}, I.~G. and {Jenkins}, A. and {Rosas-Guevara}, Y. and {White}, S.~D.~M. and {Baes}, M. and {Camps}, P. and {Lemson}, G.},
        title = "{The EAGLE simulations of galaxy formation: Public release of halo and galaxy catalogues}",
      journal = {Astronomy and Computing},
         year = 2016,
        month = apr,
       volume = {15},
        pages = {72-89},
          doi = {10.1016/j.ascom.2016.02.004},
archivePrefix = {arXiv},
       eprint = {1510.01320},
 primaryClass = {astro-ph.GA},
       adsurl = {https://ui.adsabs.harvard.edu/abs/2016A&C....15...72M}
}

@ARTICLE{Walcher2014A,
       author = {{Walcher}, C.~J. and {Wisotzki}, L. and {Bekerait{\'e}}, S. and {Husemann}, B. and {Iglesias-P{\'a}ramo}, J. and {Backsmann}, N. and {Barrera Ballesteros}, J. and {Catal{\'a}n-Torrecilla}, C. and {Cortijo}, C. and {del Olmo}, A. and {Garcia Lorenzo}, B. and {Falc{\'o}n-Barroso}, J. and {Jilkova}, L. and {Kalinova}, V. and {Mast}, D. and {Marino}, R.~A. and {M{\'e}ndez-Abreu}, J. and {Pasquali}, A. and {S{\'a}nchez}, S.~F. and {Trager}, S. and {Zibetti}, S. and {Aguerri}, J.~A.~L. and {Alves}, J. and {Bland-Hawthorn}, J. and {Boselli}, A. and {Castillo Morales}, A. and {Cid Fernandes}, R. and {Flores}, H. and {Galbany}, L. and {Gallazzi}, A. and {Garc{\'\i}a-Benito}, R. and {Gil de Paz}, A. and {Gonz{\'a}lez-Delgado}, R.~M. and {Jahnke}, K. and {Jungwiert}, B. and {Kehrig}, C. and {Lyubenova}, M. and {M{\'a}rquez Perez}, I. and {Masegosa}, J. and {Monreal Ibero}, A. and {P{\'e}rez}, E. and {Quirrenbach}, A. and {Rosales-Ortega}, F.~F. and {Roth}, M.~M. and {Sanchez-Blazquez}, P. and {Spekkens}, K. and {Tundo}, E. and {van de Ven}, G. and {Verheijen}, M.~A.~W. and {Vilchez}, J.~V. and {Ziegler}, B.},
        title = "{CALIFA: a diameter-selected sample for an integral field spectroscopy galaxy survey}",
      journal = {\aap},
         year = 2014,
        month = sep,
       volume = {569},
          eid = {A1},
        pages = {A1},
          doi = {10.1051/0004-6361/201424198},
archivePrefix = {arXiv},
       eprint = {1407.2939},
 primaryClass = {astro-ph.GA},
       adsurl = {https://ui.adsabs.harvard.edu/abs/2014A&A...569A...1W}
}

@ARTICLE{Obreschkow2009,
       author = {{Obreschkow}, D. and {Rawlings}, S.},
        title = "{Compactness of cold gas in high-redshift galaxies}",
      journal = {\mnras},
         year = 2009,
        month = dec,
       volume = {400},
       number = {2},
        pages = {665-669},
          doi = {10.1111/j.1365-2966.2009.15500.x},
archivePrefix = {arXiv},
       eprint = {0908.0724},
 primaryClass = {astro-ph.CO},
       adsurl = {https://ui.adsabs.harvard.edu/abs/2009MNRAS.400..665O}
}

@ARTICLE{Lagos2011MNRAS,
       author = {{Lagos}, Claudia Del P. and {Baugh}, Carlton M. and {Lacey}, Cedric G. and {Benson}, Andrew J. and {Kim}, Han-Seek and {Power}, Chris},
        title = "{Cosmic evolution of the atomic and molecular gas contents of galaxies}",
      journal = {\mnras},
         year = 2011,
        month = dec,
       volume = {418},
       number = {3},
        pages = {1649-1667},
          doi = {10.1111/j.1365-2966.2011.19583.x},
archivePrefix = {arXiv},
       eprint = {1105.2294},
 primaryClass = {astro-ph.CO},
       adsurl = {https://ui.adsabs.harvard.edu/abs/2011MNRAS.418.1649L}
}

@ARTICLE{Conselice2009MNRAS,
       author = {{Conselice}, Christopher J. and {Arnold}, Jessica},
        title = "{The structures of distant galaxies - II. Diverse galaxy structures and local environments at z = 4-6 implications for early galaxy assembly}",
      journal = {\mnras},
         year = 2009,
        month = jul,
       volume = {397},
       number = {1},
        pages = {208-231},
          doi = {10.1111/j.1365-2966.2009.14959.x},
archivePrefix = {arXiv},
       eprint = {0904.4250},
 primaryClass = {astro-ph.CO},
       adsurl = {https://ui.adsabs.harvard.edu/abs/2009MNRAS.397..208C}
}

@ARTICLE{Madau1996MNRAS,
       author = {{Madau}, Piero and {Ferguson}, Henry C. and {Dickinson}, Mark E. and {Giavalisco}, Mauro and {Steidel}, Charles C. and {Fruchter}, Andrew},
        title = "{High-redshift galaxies in the Hubble Deep Field: colour selection and star formation history to z\raisebox{-0.5ex}\textasciitilde4}",
      journal = {\mnras},
         year = 1996,
        month = dec,
       volume = {283},
       number = {4},
        pages = {1388-1404},
          doi = {10.1093/mnras/283.4.1388},
archivePrefix = {arXiv},
       eprint = {astro-ph/9607172},
 primaryClass = {astro-ph},
       adsurl = {https://ui.adsabs.harvard.edu/abs/1996MNRAS.283.1388M}
}

@ARTICLE{Planck2016,
       author = {{Planck Collaboration} and {Ade}, P.~A.~R. and {Aghanim}, N. and {Arnaud}, M. and {Ashdown}, M. and {Aumont}, J. and {Baccigalupi}, C. and {Banday}, A.~J. and {Barreiro}, R.~B. and {Bartlett}, J.~G. and {Bartolo}, N. and {Battaner}, E. and {Battye}, R. and {Benabed}, K. and {Beno{\^\i}t}, A. and {Benoit-L{\'e}vy}, A. and {Bernard}, J.-P. and {Bersanelli}, M. and {Bielewicz}, P. and {Bock}, J.~J. and {Bonaldi}, A. and {Bonavera}, L. and {Bond}, J.~R. and {Borrill}, J. and {Bouchet}, F.~R. and {Boulanger}, F. and {Bucher}, M. and {Burigana}, C. and {Butler}, R.~C. and {Calabrese}, E. and {Cardoso}, J.-F. and {Catalano}, A. and {Challinor}, A. and {Chamballu}, A. and {Chary}, R.-R. and {Chiang}, H.~C. and {Chluba}, J. and {Christensen}, P.~R. and {Church}, S. and {Clements}, D.~L. and {Colombi}, S. and {Colombo}, L.~P.~L. and {Combet}, C. and {Coulais}, A. and {Crill}, B.~P. and {Curto}, A. and {Cuttaia}, F. and {Danese}, L. and {Davies}, R.~D. and {Davis}, R.~J. and {de Bernardis}, P. and {de Rosa}, A. and {de Zotti}, G. and {Delabrouille}, J. and {D{\'e}sert}, F.-X. and {Di Valentino}, E. and {Dickinson}, C. and {Diego}, J.~M. and {Dolag}, K. and {Dole}, H. and {Donzelli}, S. and {Dor{\'e}}, O. and {Douspis}, M. and {Ducout}, A. and {Dunkley}, J. and {Dupac}, X. and {Efstathiou}, G. and {Elsner}, F. and {En{\ss}lin}, T.~A. and {Eriksen}, H.~K. and {Farhang}, M. and {Fergusson}, J. and {Finelli}, F. and {Forni}, O. and {Frailis}, M. and {Fraisse}, A.~A. and {Franceschi}, E. and {Frejsel}, A. and {Galeotta}, S. and {Galli}, S. and {Ganga}, K. and {Gauthier}, C. and {Gerbino}, M. and {Ghosh}, T. and {Giard}, M. and {Giraud-H{\'e}raud}, Y. and {Giusarma}, E. and {Gjerl{\o}w}, E. and {Gonz{\'a}lez-Nuevo}, J. and {G{\'o}rski}, K.~M. and {Gratton}, S. and {Gregorio}, A. and {Gruppuso}, A. and {Gudmundsson}, J.~E. and {Hamann}, J. and {Hansen}, F.~K. and {Hanson}, D. and {Harrison}, D.~L. and {Helou}, G. and {Henrot-Versill{\'e}}, S. and {Hern{\'a}ndez-Monteagudo}, C. and {Herranz}, D. and {Hildebrandt}, S.~R. and {Hivon}, E. and {Hobson}, M. and {Holmes}, W.~A. and {Hornstrup}, A. and {Hovest}, W. and {Huang}, Z. and {Huffenberger}, K.~M. and {Hurier}, G. and {Jaffe}, A.~H. and {Jaffe}, T.~R. and {Jones}, W.~C. and {Juvela}, M. and {Keih{\"a}nen}, E. and {Keskitalo}, R. and {Kisner}, T.~S. and {Kneissl}, R. and {Knoche}, J. and {Knox}, L. and {Kunz}, M. and {Kurki-Suonio}, H. and {Lagache}, G. and {L{\"a}hteenm{\"a}ki}, A. and {Lamarre}, J.-M. and {Lasenby}, A. and {Lattanzi}, M. and {Lawrence}, C.~R. and {Leahy}, J.~P. and {Leonardi}, R. and {Lesgourgues}, J. and {Levrier}, F. and {Lewis}, A. and {Liguori}, M. and {Lilje}, P.~B. and {Linden-V{\o}rnle}, M. and {L{\'o}pez-Caniego}, M. and {Lubin}, P.~M. and {Mac{\'\i}as-P{\'e}rez}, J.~F. and {Maggio}, G. and {Maino}, D. and {Mandolesi}, N. and {Mangilli}, A. and {Marchini}, A. and {Maris}, M. and {Martin}, P.~G. and {Martinelli}, M. and {Mart{\'\i}nez-Gonz{\'a}lez}, E. and {Masi}, S. and {Matarrese}, S. and {McGehee}, P. and {Meinhold}, P.~R. and {Melchiorri}, A. and {Melin}, J.-B. and {Mendes}, L. and {Mennella}, A. and {Migliaccio}, M. and {Millea}, M. and {Mitra}, S. and {Miville-Desch{\^e}nes}, M.-A. and {Moneti}, A. and {Montier}, L. and {Morgante}, G. and {Mortlock}, D. and {Moss}, A. and {Munshi}, D. and {Murphy}, J.~A. and {Naselsky}, P. and {Nati}, F. and {Natoli}, P. and {Netterfield}, C.~B. and {N{\o}rgaard-Nielsen}, H.~U. and {Noviello}, F. and {Novikov}, D. and {Novikov}, I. and {Oxborrow}, C.~A. and {Paci}, F. and {Pagano}, L. and {Pajot}, F. and {Paladini}, R. and {Paoletti}, D. and {Partridge}, B. and {Pasian}, F. and {Patanchon}, G. and {Pearson}, T.~J. and {Perdereau}, O. and {Perotto}, L. and {Perrotta}, F. and {Pettorino}, V. and {Piacentini}, F. and {Piat}, M. and {Pierpaoli}, E. and {Pietrobon}, D. and {Plaszczynski}, S. and {Pointecouteau}, E. and {Polenta}, G. and {Popa}, L. and {Pratt}, G.~W. and {Pr{\'e}zeau}, G.},
        title = "{Planck 2015 results. XIII. Cosmological parameters}",
      journal = {\aap},
         year = 2016,
        month = sep,
       volume = {594},
          eid = {A13},
        pages = {A13},
          doi = {10.1051/0004-6361/201525830},
archivePrefix = {arXiv},
       eprint = {1502.01589},
 primaryClass = {astro-ph.CO},
       adsurl = {https://ui.adsabs.harvard.edu/abs/2016A&A...594A..13P}
}

@ARTICLE{Zhu2018MNRAS2,
       author = {{Zhu}, Ling and {van den Bosch}, Remco and {van de Ven}, Glenn and {Lyubenova}, Mariya and {Falc{\'o}n-Barroso}, Jes{\'u}s and {Meidt}, Sharon E. and {Martig}, Marie and {Shen}, Juntai and {Li}, Zhao-Yu and {Yildirim}, Akin and {Walcher}, C. Jakob and {Sanchez}, Sebastian F.},
        title = "{Orbital decomposition of CALIFA spiral galaxies}",
      journal = {\mnras},
         year = 2018,
        month = jan,
       volume = {473},
       number = {3},
        pages = {3000-3018},
          doi = {10.1093/mnras/stx2409},
archivePrefix = {arXiv},
       eprint = {1709.06649},
 primaryClass = {astro-ph.GA},
       adsurl = {https://ui.adsabs.harvard.edu/abs/2018MNRAS.473.3000Z}
    }

@ARTICLE{Jin2019MNRAS,
       author = {{Jin}, Yunpeng and {Zhu}, Ling and {Long}, R.~J. and {Mao}, Shude and {Xu}, Dandan and {Li}, Hongyu and {van de Ven}, Glenn},
        title = "{Evaluating the ability of triaxial Schwarzschild modelling to estimate properties of galaxies from the Illustris simulation}",
      journal = {\mnras},
         year = 2019,
        month = jul,
       volume = {486},
       number = {4},
        pages = {4753-4772},
          doi = {10.1093/mnras/stz1170},
archivePrefix = {arXiv},
       eprint = {1904.12942},
 primaryClass = {astro-ph.GA},
       adsurl = {https://ui.adsabs.harvard.edu/abs/2019MNRAS.486.4753J}
}

@ARTICLE{Hoffman2009,
       author = {{Hoffman}, Loren and {Cox}, Thomas J. and {Dutta}, Suvendra and {Hernquist}, Lars},
        title = "{The Imprint of Dissipation on the Shapes of Merger Remnant LOSVDs}",
      journal = {\apj},
         year = 2009,
        month = nov,
       volume = {705},
       number = {1},
        pages = {920-925},
          doi = {10.1088/0004-637X/705/1/920},
archivePrefix = {arXiv},
       eprint = {0903.3064},
 primaryClass = {astro-ph.CO},
       adsurl = {https://ui.adsabs.harvard.edu/abs/2009ApJ...705..920H}
}

@ARTICLE{Robertson2006A,
       author = {{Robertson}, Brant and {Bullock}, James S. and {Cox}, Thomas J. and {Di Matteo}, Tiziana and {Hernquist}, Lars and {Springel}, Volker and {Yoshida}, Naoki},
        title = "{A Merger-driven Scenario for Cosmological Disk Galaxy Formation}",
      journal = {\apj},
         year = 2006,
        month = jul,
       volume = {645},
       number = {2},
        pages = {986-1000},
          doi = {10.1086/504412},
archivePrefix = {arXiv},
       eprint = {astro-ph/0503369},
 primaryClass = {astro-ph},
       adsurl = {https://ui.adsabs.harvard.edu/abs/2006ApJ...645..986R}
}

@ARTICLE{Barnes1996ApJ,
       author = {{Barnes}, Joshua E. and {Hernquist}, Lars},
        title = "{Transformations of Galaxies. II. Gasdynamics in Merging Disk Galaxies}",
      journal = {\apj},
         year = 1996,
        month = nov,
       volume = {471},
        pages = {115},
          doi = {10.1086/177957},
       adsurl = {https://ui.adsabs.harvard.edu/abs/1996ApJ...471..115B}
}

@ARTICLE{Springel2003,
       author = {{Springel}, Volker and {Hernquist}, Lars},
        title = "{Cosmological smoothed particle hydrodynamics simulations: a hybrid multiphase model for star formation}",
      journal = {\mnras},
         year = 2003,
        month = feb,
       volume = {339},
       number = {2},
        pages = {289-311},
          doi = {10.1046/j.1365-8711.2003.06206.x},
archivePrefix = {arXiv},
       eprint = {astro-ph/0206393},
 primaryClass = {astro-ph},
       adsurl = {https://ui.adsabs.harvard.edu/abs/2003MNRAS.339..289S}
}

@ARTICLE{Naab2006MNRAS,
       author = {{Naab}, Thorsten and {Jesseit}, Roland and {Burkert}, Andreas},
        title = "{The influence of gas on the structure of merger remnants}",
      journal = {\mnras},
         year = 2006,
        month = oct,
       volume = {372},
       number = {2},
        pages = {839-852},
          doi = {10.1111/j.1365-2966.2006.10902.x},
archivePrefix = {arXiv},
       eprint = {astro-ph/0605155},
 primaryClass = {astro-ph},
       adsurl = {https://ui.adsabs.harvard.edu/abs/2006MNRAS.372..839N}
}

@ARTICLE{Hopkins2009MNRA,
       author = {{Hopkins}, Philip F. and {Somerville}, Rachel S. and {Cox}, Thomas J. and {Hernquist}, Lars and {Jogee}, Shardha and {Kere{\v{s}}}, Dusan and {Ma}, Chung-Pei and {Robertson}, Brant and {Stewart}, Kyle},
        title = "{The effects of gas on morphological transformation in mergers: implications for bulge and disc demographics}",
      journal = {\mnras},
         year = 2009,
        month = aug,
       volume = {397},
       number = {2},
        pages = {802-814},
          doi = {10.1111/j.1365-2966.2009.14983.x},
archivePrefix = {arXiv},
       eprint = {0901.4111},
 primaryClass = {astro-ph.CO},
       adsurl = {https://ui.adsabs.harvard.edu/abs/2009MNRAS.397..802H}
}

@ARTICLE{Lagos2018MNRAS,
       author = {{Lagos}, Claudia del P. and {Stevens}, Adam R.~H. and {Bower}, Richard G. and {Davis}, Timothy A. and {Contreras}, Sergio and {Padilla}, Nelson D. and {Obreschkow}, Danail and {Croton}, Darren and {Trayford}, James W. and {Welker}, Charlotte and {Theuns}, Tom},
        title = "{Quantifying the impact of mergers on the angular momentum of simulated galaxies}",
      journal = {\mnras},
         year = 2018,
        month = feb,
       volume = {473},
       number = {4},
        pages = {4956-4974},
          doi = {10.1093/mnras/stx2667},
archivePrefix = {arXiv},
       eprint = {1701.04407},
 primaryClass = {astro-ph.GA},
       adsurl = {https://ui.adsabs.harvard.edu/abs/2018MNRAS.473.4956L}
}

@ARTICLE{Bundy2009ApJ,
       author = {{Bundy}, Kevin and {Fukugita}, Masataka and {Ellis}, Richard S. and {Targett}, Thomas A. and {Belli}, Sirio and {Kodama}, Tadayuki},
        title = "{The Greater Impact of Mergers on the Growth of Massive Galaxies: Implications for Mass Assembly and Evolution since z sime 1}",
      journal = {\apj},
         year = 2009,
        month = jun,
       volume = {697},
       number = {2},
        pages = {1369-1383},
          doi = {10.1088/0004-637X/697/2/1369},
archivePrefix = {arXiv},
       eprint = {0902.1188},
 primaryClass = {astro-ph.CO},
       adsurl = {https://ui.adsabs.harvard.edu/abs/2009ApJ...697.1369B}
}

@ARTICLE{Pillepich2015ApJ,
       author = {{Pillepich}, Annalisa and {Madau}, Piero and {Mayer}, Lucio},
        title = "{Building Late-type Spiral Galaxies by In-situ and Ex-situ Star Formation}",
      journal = {\apj},
         year = 2015,
        month = feb,
       volume = {799},
       number = {2},
          eid = {184},
        pages = {184},
          doi = {10.1088/0004-637X/799/2/184},
archivePrefix = {arXiv},
       eprint = {1407.7855},
 primaryClass = {astro-ph.GA},
       adsurl = {https://ui.adsabs.harvard.edu/abs/2015ApJ...799..184P}
}

@ARTICLE{Zhu2022A&A,
       author = {{Zhu}, Ling and {Pillepich}, Annalisa and {van de Ven}, Glenn and {Leaman}, Ryan and {Hernquist}, Lars and {Nelson}, Dylan and {Pakmor}, Ruediger and {Vogelsberger}, Mark and {Zhang}, Le},
        title = "{Mass of the dynamically hot inner stellar halo predicts the ancient accreted stellar mass}",
      journal = {\aap},
         year = 2022,
        month = apr,
       volume = {660},
          eid = {A20},
        pages = {A20},
          doi = {10.1051/0004-6361/202142496},
archivePrefix = {arXiv},
       eprint = {2110.13172},
 primaryClass = {astro-ph.GA},
       adsurl = {https://ui.adsabs.harvard.edu/abs/2022A&A...660A..20Z}
}

@ARTICLE{Wu2025,
       author = {{Wu}, Xinyi and {Zhu}, Ling and {Chang}, Jiang and {Zeng}, Guangquan and {Lei}, Yu},
        title = "{The effects of the orbital configurations of mergers on reshaping galaxy structures}",
      journal = {\aap},
         year = 2025,
        month = jul,
       volume = {699},
          eid = {A374},
        pages = {A374},
          doi = {10.1051/0004-6361/202554638},
archivePrefix = {arXiv},
       eprint = {2507.02051},
 primaryClass = {astro-ph.GA},
       adsurl = {https://ui.adsabs.harvard.edu/abs/2025A&A...699A.374W}
}

@ARTICLE{Rodriguez2017MNRAS,
       author = {{Rodriguez-Gomez}, Vicente and {Sales}, Laura V. and {Genel}, Shy and {Pillepich}, Annalisa and {Zjupa}, Jolanta and {Nelson}, Dylan and {Griffen}, Brendan and {Torrey}, Paul and {Snyder}, Gregory F. and {Vogelsberger}, Mark and {Springel}, Volker and {Ma}, Chung-Pei and {Hernquist}, Lars},
        title = "{The role of mergers and halo spin in shaping galaxy morphology}",
      journal = {\mnras},
         year = 2017,
        month = may,
       volume = {467},
       number = {3},
        pages = {3083-3098},
          doi = {10.1093/mnras/stx305},
archivePrefix = {arXiv},
       eprint = {1609.09498},
 primaryClass = {astro-ph.GA},
       adsurl = {https://ui.adsabs.harvard.edu/abs/2017MNRAS.467.3083R}
}

@ARTICLE{Zhang2025,
       author = {{Zhang}, Le and {Zhu}, Ling and {Pillepich}, Annalisa and {Du}, Min and {Jiang}, Fangzhou and {Falc{\'o}n-Barroso}, Jes{\'u}s},
        title = "{The diverse physical origins of stars in the dynamically hot bulge: CALIFA versus IllustrisTNG}",
      journal = {\aap},
         year = 2025,
        month = jul,
       volume = {699},
          eid = {A320},
        pages = {A320},
          doi = {10.1051/0004-6361/202451292},
archivePrefix = {arXiv},
       eprint = {2501.07151},
 primaryClass = {astro-ph.GA},
       adsurl = {https://ui.adsabs.harvard.edu/abs/2025A&A...699A.320Z}
}

@ARTICLE{Santucci2023,
       author = {{Santucci}, Giulia and {Brough}, Sarah and {van de Sande}, Jesse and {McDermid}, Richard and {Barsanti}, Stefania and {Bland-Hawthorn}, Joss and {Bryant}, Julia J. and {Croom}, Scott M. and {Lagos}, Claudia and {Lawrence}, Jon S. and {Owers}, Matt S. and {van de Ven}, Glenn and {Vaughan}, Sam P. and {Yi}, Sukyoung K.},
        title = "{The SAMI Galaxy Survey: Environmental analysis of the orbital structures of passive galaxies}",
      journal = {\mnras},
         year = 2023,
        month = may,
       volume = {521},
       number = {2},
        pages = {2671-2691},
          doi = {10.1093/mnras/stad713},
archivePrefix = {arXiv},
       eprint = {2303.04161},
 primaryClass = {astro-ph.GA},
       adsurl = {https://ui.adsabs.harvard.edu/abs/2023MNRAS.521.2671S}
}

@ARTICLE{Dekel2009ApJ,
       author = {{Dekel}, Avishai and {Sari}, Re'em and {Ceverino}, Daniel},
        title = "{Formation of Massive Galaxies at High Redshift: Cold Streams, Clumpy Disks, and Compact Spheroids}",
      journal = {\apj},
         year = 2009,
        month = sep,
       volume = {703},
       number = {1},
        pages = {785-801},
          doi = {10.1088/0004-637X/703/1/785},
archivePrefix = {arXiv},
       eprint = {0901.2458},
 primaryClass = {astro-ph.GA},
       adsurl = {https://ui.adsabs.harvard.edu/abs/2009ApJ...703..785D}
}

@ARTICLE{Fiacconi2015MNRAS,
       author = {{Fiacconi}, Davide and {Feldmann}, Robert and {Mayer}, Lucio},
        title = "{The Argo simulation - II. The early build-up of the Hubble sequence}",
      journal = {\mnras},
         year = 2015,
        month = jan,
       volume = {446},
       number = {2},
        pages = {1957-1972},
          doi = {10.1093/mnras/stu2228},
archivePrefix = {arXiv},
       eprint = {1410.6818},
 primaryClass = {astro-ph.GA},
       adsurl = {https://ui.adsabs.harvard.edu/abs/2015MNRAS.446.1957F}
}

@ARTICLE{Welker2017MNRAS,
       author = {{Welker}, C. and {Dubois}, Y. and {Devriendt}, J. and {Pichon}, C. and {Kaviraj}, S. and {Peirani}, S.},
        title = "{The rise and fall of stellar discs across the peak of cosmic star formation history: effects of mergers versus diffuse stellar mass acquisition}",
      journal = {\mnras},
         year = 2017,
        month = feb,
       volume = {465},
       number = {1},
        pages = {1241-1258},
          doi = {10.1093/mnras/stw2778},
archivePrefix = {arXiv},
       eprint = {1502.05053},
 primaryClass = {astro-ph.GA},
       adsurl = {https://ui.adsabs.harvard.edu/abs/2017MNRAS.465.1241W}
}

@ARTICLE{Abadi2003ApJ,
       author = {{Abadi}, Mario G. and {Navarro}, Julio F. and {Steinmetz}, Matthias and {Eke}, Vincent R.},
        title = "{Simulations of Galaxy Formation in a {\ensuremath{\Lambda}} Cold Dark Matter Universe. I. Dynamical and Photometric Properties of a Simulated Disk Galaxy}",
      journal = {\apj},
         year = 2003,
        month = jul,
       volume = {591},
       number = {2},
        pages = {499-514},
          doi = {10.1086/375512},
archivePrefix = {arXiv},
       eprint = {astro-ph/0211331},
 primaryClass = {astro-ph},
       adsurl = {https://ui.adsabs.harvard.edu/abs/2003ApJ...591..499A}
}

@ARTICLE{Sellwood2014R,
       author = {{Sellwood}, J.~A.},
        title = "{Secular evolution in disk galaxies}",
      journal = {Reviews of Modern Physics},
         year = 2014,
        month = jan,
       volume = {86},
       number = {1},
        pages = {1-46},
          doi = {10.1103/RevModPhys.86.1},
archivePrefix = {arXiv},
       eprint = {1310.0403},
 primaryClass = {astro-ph.GA},
       adsurl = {https://ui.adsabs.harvard.edu/abs/2014RvMP...86....1S}
}

@ARTICLE{Read2008MNRAS,
       author = {{Read}, J.~I. and {Lake}, G. and {Agertz}, O. and {Debattista}, Victor P.},
        title = "{Thin, thick and dark discs in {\ensuremath{\Lambda}}CDM}",
      journal = {\mnras},
         year = 2008,
        month = sep,
       volume = {389},
       number = {3},
        pages = {1041-1057},
          doi = {10.1111/j.1365-2966.2008.13643.x},
archivePrefix = {arXiv},
       eprint = {0803.2714},
 primaryClass = {astro-ph},
       adsurl = {https://ui.adsabs.harvard.edu/abs/2008MNRAS.389.1041R}
}

@ARTICLE{Sparre2017MNRAS,
       author = {{Sparre}, Martin and {Springel}, Volker},
        title = "{The unorthodox evolution of major merger remnants into star-forming spiral galaxies}",
      journal = {\mnras},
         year = 2017,
        month = oct,
       volume = {470},
       number = {4},
        pages = {3946-3958},
          doi = {10.1093/mnras/stx1516},
archivePrefix = {arXiv},
       eprint = {1610.03850},
 primaryClass = {astro-ph.GA},
       adsurl = {https://ui.adsabs.harvard.edu/abs/2017MNRAS.470.3946S}
}

@ARTICLE{Nelson2019MNRAS,
       author = {{Nelson}, Dylan and {Pillepich}, Annalisa and {Springel}, Volker and {Pakmor}, R{\"u}diger and {Weinberger}, Rainer and {Genel}, Shy and {Torrey}, Paul and {Vogelsberger}, Mark and {Marinacci}, Federico and {Hernquist}, Lars},
        title = "{First results from the TNG50 simulation: galactic outflows driven by supernovae and black hole feedback}",
      journal = {\mnras},
         year = 2019,
        month = dec,
       volume = {490},
       number = {3},
        pages = {3234-3261},
          doi = {10.1093/mnras/stz2306},
archivePrefix = {arXiv},
       eprint = {1902.05554},
 primaryClass = {astro-ph.GA},
       adsurl = {https://ui.adsabs.harvard.edu/abs/2019MNRAS.490.3234N}
}

@ARTICLE{Somerville2015ARA&A,
       author = {{Somerville}, Rachel S. and {Dav{\'e}}, Romeel},
        title = "{Physical Models of Galaxy Formation in a Cosmological Framework}",
      journal = {\araa},
         year = 2015,
        month = aug,
       volume = {53},
        pages = {51-113},
          doi = {10.1146/annurev-astro-082812-140951},
archivePrefix = {arXiv},
       eprint = {1412.2712},
 primaryClass = {astro-ph.GA},
       adsurl = {https://ui.adsabs.harvard.edu/abs/2015ARA&A..53...51S}
}

@ARTICLE{Springel2005MNRAS,
       author = {{Springel}, Volker and {Di Matteo}, Tiziana and {Hernquist}, Lars},
        title = "{Modelling feedback from stars and black holes in galaxy mergers}",
      journal = {\mnras},
         year = 2005,
        month = aug,
       volume = {361},
       number = {3},
        pages = {776-794},
          doi = {10.1111/j.1365-2966.2005.09238.x},
archivePrefix = {arXiv},
       eprint = {astro-ph/0411108},
 primaryClass = {astro-ph},
       adsurl = {https://ui.adsabs.harvard.edu/abs/2005MNRAS.361..776S}
}

@ARTICLE{Bower2006MNRAS,
       author = {{Bower}, R.~G. and {Benson}, A.~J. and {Malbon}, R. and {Helly}, J.~C. and {Frenk}, C.~S. and {Baugh}, C.~M. and {Cole}, S. and {Lacey}, C.~G.},
        title = "{Breaking the hierarchy of galaxy formation}",
      journal = {\mnras},
         year = 2006,
        month = aug,
       volume = {370},
       number = {2},
        pages = {645-655},
          doi = {10.1111/j.1365-2966.2006.10519.x},
archivePrefix = {arXiv},
       eprint = {astro-ph/0511338},
 primaryClass = {astro-ph},
       adsurl = {https://ui.adsabs.harvard.edu/abs/2006MNRAS.370..645B}
}

@ARTICLE{Croton2006MNRAS,
       author = {{Croton}, Darren J. and {Springel}, Volker and {White}, Simon D.~M. and {De Lucia}, G. and {Frenk}, C.~S. and {Gao}, L. and {Jenkins}, A. and {Kauffmann}, G. and {Navarro}, J.~F. and {Yoshida}, N.},
        title = "{The many lives of active galactic nuclei: cooling flows, black holes and the luminosities and colours of galaxies}",
      journal = {\mnras},
         year = 2006,
        month = jan,
       volume = {365},
       number = {1},
        pages = {11-28},
          doi = {10.1111/j.1365-2966.2005.09675.x},
archivePrefix = {arXiv},
       eprint = {astro-ph/0508046},
 primaryClass = {astro-ph},
       adsurl = {https://ui.adsabs.harvard.edu/abs/2006MNRAS.365...11C}
}

@ARTICLE{Wright2022MNRAS,
       author = {{Wright}, Ruby J. and {Lagos}, Claudia del P. and {Power}, Chris and {Stevens}, Adam R.~H. and {Cortese}, Luca and {Poulton}, Rhys J.~J.},
        title = "{An orbital perspective on the starvation, stripping, and quenching of satellite galaxies in the EAGLE simulations}",
      journal = {\mnras},
         year = 2022,
        month = oct,
       volume = {516},
       number = {2},
        pages = {2891-2912},
          doi = {10.1093/mnras/stac2042},
archivePrefix = {arXiv},
       eprint = {2205.08414},
 primaryClass = {astro-ph.GA},
       adsurl = {https://ui.adsabs.harvard.edu/abs/2022MNRAS.516.2891W}
}

@ARTICLE{Wetzel2012MNRAS,
       author = {{Wetzel}, Andrew R. and {Tinker}, Jeremy L. and {Conroy}, Charlie},
        title = "{Galaxy evolution in groups and clusters: star formation rates, red sequence fractions and the persistent bimodality}",
      journal = {\mnras},
         year = 2012,
        month = jul,
       volume = {424},
       number = {1},
        pages = {232-243},
          doi = {10.1111/j.1365-2966.2012.21188.x},
archivePrefix = {arXiv},
       eprint = {1107.5311},
 primaryClass = {astro-ph.CO},
       adsurl = {https://ui.adsabs.harvard.edu/abs/2012MNRAS.424..232W}
}

@ARTICLE{Larson1980ApJ,
       author = {{Larson}, R.~B. and {Tinsley}, B.~M. and {Caldwell}, C.~N.},
        title = "{The evolution of disk galaxies and the origin of S0 galaxies}",
      journal = {\apj},
         year = 1980,
        month = may,
       volume = {237},
        pages = {692-707},
          doi = {10.1086/157917},
       adsurl = {https://ui.adsabs.harvard.edu/abs/1980ApJ...237..692L}
}

@ARTICLE{Merritt1983ApJ,
       author = {{Merritt}, D.},
        title = "{Relaxation and tidal stripping in rich clusters of galaxies. I. Evolution of the mass distribution.}",
      journal = {\apj},
         year = 1983,
        month = jan,
       volume = {264},
        pages = {24-48},
          doi = {10.1086/160571},
       adsurl = {https://ui.adsabs.harvard.edu/abs/1983ApJ...264...24M}
}

@ARTICLE{Gunn1972ApJ,
       author = {{Gunn}, James E. and {Gott}, III, J. Richard},
        title = "{On the Infall of Matter Into Clusters of Galaxies and Some Effects on Their Evolution}",
      journal = {\apj},
         year = 1972,
        month = aug,
       volume = {176},
        pages = {1},
          doi = {10.1086/151605},
       adsurl = {https://ui.adsabs.harvard.edu/abs/1972ApJ...176....1G}
}

@ARTICLE{Xie2025,
       author = {{Xie}, Lizhi and {De Lucia}, Gabriella and {Fossati}, Matteo and {Fontanot}, Fabio and {Hirschmann}, Michaela},
        title = "{The impact of ram pressure on cluster galaxies, insights from GAEA and TNG}",
      journal = {\aap},
         year = 2025,
        month = jun,
       volume = {698},
          eid = {A73},
        pages = {A73},
          doi = {10.1051/0004-6361/202553915},
archivePrefix = {arXiv},
       eprint = {2504.12863},
 primaryClass = {astro-ph.GA},
       adsurl = {https://ui.adsabs.harvard.edu/abs/2025A&A...698A..73X}
}

@ARTICLE{Boselli2022,
       author = {{Boselli}, Alessandro and {Fossati}, Matteo and {Sun}, Ming},
        title = "{Ram pressure stripping in high-density environments}",
      journal = {\aapr},
         year = 2022,
        month = dec,
       volume = {30},
       number = {1},
          eid = {3},
        pages = {3},
          doi = {10.1007/s00159-022-00140-3},
archivePrefix = {arXiv},
       eprint = {2109.13614},
 primaryClass = {astro-ph.GA},
       adsurl = {https://ui.adsabs.harvard.edu/abs/2022A&ARv..30....3B}
}

@ARTICLE{Rodrguez2025A&A,
       author = {{Rodr{\'\i}guez-Cardoso}, Ram{\'o}n and {Roca-F{\`a}brega}, Santi and {Jung}, Minyong and {Nguyễn}, Thịnh H. and {Kim}, Ji-Hoon and {Primack}, Joel and {Agertz}, Oscar and {Barrow}, Kirk S.~S. and {Gallego}, Jesus and {Nagamine}, Kentaro and {Powell}, Johnny W. and {Revaz}, Yves and {Vel{\'a}zquez}, Hector and {Genina}, Anna and {Kim}, Hyeonyong and {Lupi}, Alessandro and {Abel}, Tom and {Cen}, Renyue and {Ceverino}, Daniel and {Dekel}, Avishai and {Oh}, Boon Kiat and {Quinn}, Thomas R. and {The Agora Collaboration}},
        title = "{The AGORA High-Resolution Galaxy Simulations Comparison Project: VII. Satellite quenching in zoom-in simulation of a Milky Way-mass halo}",
      journal = {\aap},
         year = 2025,
        month = jun,
       volume = {698},
          eid = {A303},
        pages = {A303},
          doi = {10.1051/0004-6361/202453639},
archivePrefix = {arXiv},
       eprint = {2505.05844},
 primaryClass = {astro-ph.GA},
       adsurl = {https://ui.adsabs.harvard.edu/abs/2025A&A...698A.303R}
}

@ARTICLE{Daddi2007ApJ,
       author = {{Daddi}, E. and {Dickinson}, M. and {Morrison}, G. and {Chary}, R. and {Cimatti}, A. and {Elbaz}, D. and {Frayer}, D. and {Renzini}, A. and {Pope}, A. and {Alexander}, D.~M. and {Bauer}, F.~E. and {Giavalisco}, M. and {Huynh}, M. and {Kurk}, J. and {Mignoli}, M.},
        title = "{Multiwavelength Study of Massive Galaxies at z\raisebox{-0.5ex}\textasciitilde2. I. Star Formation and Galaxy Growth}",
      journal = {\apj},
         year = 2007,
        month = nov,
       volume = {670},
       number = {1},
        pages = {156-172},
          doi = {10.1086/521818},
archivePrefix = {arXiv},
       eprint = {0705.2831},
 primaryClass = {astro-ph},
       adsurl = {https://ui.adsabs.harvard.edu/abs/2007ApJ...670..156D}
}

@ARTICLE{Noeske2007ApJ,
       author = {{Noeske}, K.~G. and {Weiner}, B.~J. and {Faber}, S.~M. and {Papovich}, C. and {Koo}, D.~C. and {Somerville}, R.~S. and {Bundy}, K. and {Conselice}, C.~J. and {Newman}, J.~A. and {Schiminovich}, D. and {Le Floc'h}, E. and {Coil}, A.~L. and {Rieke}, G.~H. and {Lotz}, J.~M. and {Primack}, J.~R. and {Barmby}, P. and {Cooper}, M.~C. and {Davis}, M. and {Ellis}, R.~S. and {Fazio}, G.~G. and {Guhathakurta}, P. and {Huang}, J. and {Kassin}, S.~A. and {Martin}, D.~C. and {Phillips}, A.~C. and {Rich}, R.~M. and {Small}, T.~A. and {Willmer}, C.~N.~A. and {Wilson}, G.},
        title = "{Star Formation in AEGIS Field Galaxies since z=1.1: The Dominance of Gradually Declining Star Formation, and the Main Sequence of Star-forming Galaxies}",
      journal = {\apjl},
         year = 2007,
        month = may,
       volume = {660},
       number = {1},
        pages = {L43-L46},
          doi = {10.1086/517926},
archivePrefix = {arXiv},
       eprint = {astro-ph/0701924},
 primaryClass = {astro-ph},
       adsurl = {https://ui.adsabs.harvard.edu/abs/2007ApJ...660L..43N}
}

@ARTICLE{Whitaker2012ApJ,
       author = {{Whitaker}, Katherine E. and {van Dokkum}, Pieter G. and {Brammer}, Gabriel and {Franx}, Marijn},
        title = "{The Star Formation Mass Sequence Out to z = 2.5}",
      journal = {\apjl},
         year = 2012,
        month = aug,
       volume = {754},
       number = {2},
          eid = {L29},
        pages = {L29},
          doi = {10.1088/2041-8205/754/2/L29},
archivePrefix = {arXiv},
       eprint = {1205.0547},
 primaryClass = {astro-ph.CO},
       adsurl = {https://ui.adsabs.harvard.edu/abs/2012ApJ...754L..29W}
}

@ARTICLE{Tomczak2016ApJ,
       author = {{Tomczak}, Adam R. and {Quadri}, Ryan F. and {Tran}, Kim-Vy H. and {Labb{\'e}}, Ivo and {Straatman}, Caroline M.~S. and {Papovich}, Casey and {Glazebrook}, Karl and {Allen}, Rebecca and {Brammer}, Gabreil B. and {Cowley}, Michael and {Dickinson}, Mark and {Elbaz}, David and {Inami}, Hanae and {Kacprzak}, Glenn G. and {Morrison}, Glenn E. and {Nanayakkara}, Themiya and {Persson}, S. Eric and {Rees}, Glen A. and {Salmon}, Brett and {Schreiber}, Corentin and {Spitler}, Lee R. and {Whitaker}, Katherine E.},
        title = "{The SFR-M* Relation and Empirical Star-Formation Histories from ZFOURGE* at 0.5 < z < 4}",
      journal = {\apj},
         year = 2016,
        month = feb,
       volume = {817},
       number = {2},
          eid = {118},
        pages = {118},
          doi = {10.3847/0004-637X/817/2/118},
archivePrefix = {arXiv},
       eprint = {1510.06072},
 primaryClass = {astro-ph.GA},
       adsurl = {https://ui.adsabs.harvard.edu/abs/2016ApJ...817..118T}
}

@ARTICLE{Popesso2023MNRAS,
       author = {{Popesso}, P. and {Concas}, A. and {Cresci}, G. and {Belli}, S. and {Rodighiero}, G. and {Inami}, H. and {Dickinson}, M. and {Ilbert}, O. and {Pannella}, M. and {Elbaz}, D.},
        title = "{The main sequence of star-forming galaxies across cosmic times}",
      journal = {\mnras},
         year = 2023,
        month = feb,
       volume = {519},
       number = {1},
        pages = {1526-1544},
          doi = {10.1093/mnras/stac3214},
archivePrefix = {arXiv},
       eprint = {2203.10487},
 primaryClass = {astro-ph.GA},
       adsurl = {https://ui.adsabs.harvard.edu/abs/2023MNRAS.519.1526P}
}

@ARTICLE{Pillepich2019MNRAS,
       author = {{Pillepich}, Annalisa and {Nelson}, Dylan and {Springel}, Volker and {Pakmor}, R{\"u}diger and {Torrey}, Paul and {Weinberger}, Rainer and {Vogelsberger}, Mark and {Marinacci}, Federico and {Genel}, Shy and {van der Wel}, Arjen and {Hernquist}, Lars},
        title = "{First results from the TNG50 simulation: the evolution of stellar and gaseous discs across cosmic time}",
      journal = {\mnras},
         year = 2019,
        month = dec,
       volume = {490},
       number = {3},
        pages = {3196-3233},
          doi = {10.1093/mnras/stz2338},
archivePrefix = {arXiv},
       eprint = {1902.05553},
 primaryClass = {astro-ph.GA},
       adsurl = {https://ui.adsabs.harvard.edu/abs/2019MNRAS.490.3196P}
}

@ARTICLE{Bait2017MNRAS,
       author = {{Bait}, Omkar and {Barway}, Sudhanshu and {Wadadekar}, Yogesh},
        title = "{On the interdependence of galaxy morphology, star formation and environment in massive galaxies in the nearby Universe}",
      journal = {\mnras},
         year = 2017,
        month = nov,
       volume = {471},
       number = {3},
        pages = {2687-2702},
          doi = {10.1093/mnras/stx1688},
archivePrefix = {arXiv},
       eprint = {1707.00568},
 primaryClass = {astro-ph.GA},
       adsurl = {https://ui.adsabs.harvard.edu/abs/2017MNRAS.471.2687B}
}

@ARTICLE{Bluck2020MNRAS,
       author = {{Bluck}, Asa F.~L. and {Maiolino}, Roberto and {S{\'a}nchez}, Sebastian F. and {Ellison}, Sara L. and {Thorp}, Mallory D. and {Piotrowska}, Joanna M. and {Teimoorinia}, Hossen and {Bundy}, Kevin A.},
        title = "{Are galactic star formation and quenching governed by local, global, or environmental phenomena?}",
      journal = {\mnras},
         year = 2020,
        month = feb,
       volume = {492},
       number = {1},
        pages = {96-139},
          doi = {10.1093/mnras/stz3264},
archivePrefix = {arXiv},
       eprint = {1911.08857},
 primaryClass = {astro-ph.GA},
       adsurl = {https://ui.adsabs.harvard.edu/abs/2020MNRAS.492...96B}
}

@ARTICLE{Bluck2020MNRASB,
       author = {{Bluck}, Asa F.~L. and {Maiolino}, Roberto and {Piotrowska}, Joanna M. and {Trussler}, James and {Ellison}, Sara L. and {S{\'a}nchez}, Sebastian F. and {Thorp}, Mallory D. and {Teimoorinia}, Hossen and {Moreno}, Jorge and {Conselice}, Christopher J.},
        title = "{How do central and satellite galaxies quench? - Insights from spatially resolved spectroscopy in the MaNGA survey}",
      journal = {\mnras},
         year = 2020,
        month = nov,
       volume = {499},
       number = {1},
        pages = {230-268},
          doi = {10.1093/mnras/staa2806},
archivePrefix = {arXiv},
       eprint = {2009.05341},
 primaryClass = {astro-ph.GA},
       adsurl = {https://ui.adsabs.harvard.edu/abs/2020MNRAS.499..230B}
}

@ARTICLE{Baker2022MNRAS,
       author = {{Baker}, William M. and {Maiolino}, Roberto and {Bluck}, Asa F.~L. and {Lin}, Lihwai and {Ellison}, Sara L. and {Belfiore}, Francesco and {Pan}, Hsi-An and {Thorp}, Mallory},
        title = "{The ALMaQUEST survey IX: the nature of the resolved star forming main sequence}",
      journal = {\mnras},
         year = 2022,
        month = mar,
       volume = {510},
       number = {3},
        pages = {3622-3628},
          doi = {10.1093/mnras/stab3672},
archivePrefix = {arXiv},
       eprint = {2201.03592},
 primaryClass = {astro-ph.GA},
       adsurl = {https://ui.adsabs.harvard.edu/abs/2022MNRAS.510.3622B}
}

@ARTICLE{Athanassoula2016ApJ,
       author = {{Athanassoula}, E. and {Rodionov}, S.~A. and {Peschken}, N. and {Lambert}, J.~C.},
        title = "{Forming Disk Galaxies in Wet Major Mergers. I. Three Fiducial Examples}",
      journal = {\apj},
         year = 2016,
        month = apr,
       volume = {821},
       number = {2},
          eid = {90},
        pages = {90},
          doi = {10.3847/0004-637X/821/2/90},
archivePrefix = {arXiv},
       eprint = {1602.03189},
 primaryClass = {astro-ph.GA},
       adsurl = {https://ui.adsabs.harvard.edu/abs/2016ApJ...821...90A}
}

@INPROCEEDINGS{Thater2025IAUS,
       author = {{Thater}, S. and {Jethwa}, P. and {Lilley}, E.~J. and {Zocchi}, A. and {Santucci}, G. and {van de Ven}, G.},
        title = "{Dynamical modelling of ATLAS$^{3D}$ galaxies}",
    booktitle = {Dynamical Masses of Local Group Galaxies},
         year = 2025,
       editor = {{Bonifacio}, Piercarlo and {Cioni}, Maria-Rosa and {Hammer}, Fran{\c{c}}ois and {Pawloswki}, Marcel S. and {Taibi}, Salvatore},
       series = {IAU Symposium},
       volume = {379},
        month = jan,
        pages = {314-318},
          doi = {10.1017/S1743921323001850},
archivePrefix = {arXiv},
       eprint = {2305.09344},
 primaryClass = {astro-ph.GA},
       adsurl = {https://ui.adsabs.harvard.edu/abs/2025IAUS..379..314T}
}

@article{astropy:2013,
  adsurl        = {http://adsabs.harvard.edu/abs/2013A\%26A...558A..33A},
  archiveprefix = {arXiv},
  author        = {{Astropy Collaboration} and {Robitaille}, T.~P. and {Tollerud}, E.~J. and {Greenfield}, P. and {Droettboom}, M. and {Bray}, E. and {Aldcroft}, T. and {Davis}, M. and {Ginsburg}, A. and {Price-Whelan}, A.~M. and {Kerzendorf}, W.~E. and {Conley}, A. and {Crighton}, N. and {Barbary}, K. and {Muna}, D. and {Ferguson}, H. and {Grollier}, F. and {Parikh}, M.~M. and {Nair}, P.~H. and {Unther}, H.~M. and {Deil}, C. and {Woillez}, J. and {Conseil}, S. and {Kramer}, R. and {Turner}, J.~E.~H. and {Singer}, L. and {Fox}, R. and {Weaver}, B.~A. and {Zabalza}, V. and {Edwards}, Z.~I. and {Azalee Bostroem}, K. and {Burke}, D.~J. and {Casey}, A.~R. and {Crawford}, S.~M. and {Dencheva}, N. and {Ely}, J. and {Jenness}, T. and {Labrie}, K. and {Lim}, P.~L. and {Pierfederici}, F. and {Pontzen}, A. and {Ptak}, A. and {Refsdal}, B. and {Servillat}, M. and {Streicher}, O.},
  doi           = {10.1051/0004-6361/201322068},
  eid           = {A33},
  eprint        = {1307.6212},
  journal       = {\aap},
  month         = oct,
  pages         = {A33},
  primaryclass  = {astro-ph.IM},
  title         = {{Astropy: A community Python package for astronomy}},
  volume        = 558,
  year          = 2013
}

@article{astropy:2018,
  author        = {{Astropy Collaboration} and {Price-Whelan}, A.~M. and {Sip{\H{o}}cz}, B.~M. and {G{\"u}nther}, H.~M. and {Lim}, P.~L. and {Crawford}, S.~M. and {Conseil}, S. and {Shupe}, D.~L. and {Craig}, M.~W. and {Dencheva}, N. and {Ginsburg}, A. and {Vand erPlas}, J.~T. and {Bradley}, L.~D. and {P{\'e}rez-Su{\'a}rez}, D. and {de Val-Borro}, M. and {Aldcroft}, T.~L. and {Cruz}, K.~L. and {Robitaille}, T.~P. and {Tollerud}, E.~J. and {Ardelean}, C. and {Babej}, T. and {Bach}, Y.~P. and {Bachetti}, M. and {Bakanov}, A.~V. and {Bamford}, S.~P. and {Barentsen}, G. and {Barmby}, P. and {Baumbach}, A. and {Berry}, K.~L. and {Biscani}, F. and {Boquien}, M. and {Bostroem}, K.~A. and {Bouma}, L.~G. and {Brammer}, G.~B. and {Bray}, E.~M. and {Breytenbach}, H. and {Buddelmeijer}, H. and {Burke}, D.~J. and {Calderone}, G. and {Cano Rodr{\'\i}guez}, J.~L. and {Cara}, M. and {Cardoso}, J.~V.~M. and {Cheedella}, S. and {Copin}, Y. and {Corrales}, L. and {Crichton}, D. and {D'Avella}, D. and {Deil}, C. and {Depagne}, {\'E}. and {Dietrich}, J.~P. and {Donath}, A. and {Droettboom}, M. and {Earl}, N. and {Erben}, T. and {Fabbro}, S. and {Ferreira}, L.~A. and {Finethy}, T. and {Fox}, R.~T. and {Garrison}, L.~H. and {Gibbons}, S.~L.~J. and {Goldstein}, D.~A. and {Gommers}, R. and {Greco}, J.~P. and {Greenfield}, P. and {Groener}, A.~M. and {Grollier}, F. and {Hagen}, A. and {Hirst}, P. and {Homeier}, D. and {Horton}, A.~J. and {Hosseinzadeh}, G. and {Hu}, L. and {Hunkeler}, J.~S. and {Ivezi{\'c}}, {\v{Z}}. and {Jain}, A. and {Jenness}, T. and {Kanarek}, G. and {Kendrew}, S. and {Kern}, N.~S. and {Kerzendorf}, W.~E. and {Khvalko}, A. and {King}, J. and {Kirkby}, D. and {Kulkarni}, A.~M. and {Kumar}, A. and {Lee}, A. and {Lenz}, D. and {Littlefair}, S.~P. and {Ma}, Z. and {Macleod}, D.~M. and {Mastropietro}, M. and {McCully}, C. and {Montagnac}, S. and {Morris}, B.~M. and {Mueller}, M. and {Mumford}, S.~J. and {Muna}, D. and {Murphy}, N.~A. and {Nelson}, S. and {Nguyen}, G.~H. and {Ninan}, J.~P. and {N{\"o}the}, M. and {Ogaz}, S. and {Oh}, S. and {Parejko}, J.~K. and {Parley}, N. and {Pascual}, S. and {Patil}, R. and {Patil}, A.~A. and {Plunkett}, A.~L. and {Prochaska}, J.~X. and {Rastogi}, T. and {Reddy Janga}, V. and {Sabater}, J. and {Sakurikar}, P. and {Seifert}, M. and {Sherbert}, L.~E. and {Sherwood-Taylor}, H. and {Shih}, A.~Y. and {Sick}, J. and {Silbiger}, M.~T. and {Singanamalla}, S. and {Singer}, L.~P. and {Sladen}, P.~H. and {Sooley}, K.~A. and {Sornarajah}, S. and {Streicher}, O. and {Teuben}, P. and {Thomas}, S.~W. and {Tremblay}, G.~R. and {Turner}, J.~E.~H. and {Terr{\'o}n}, V. and {van Kerkwijk}, M.~H. and {de la Vega}, A. and {Watkins}, L.~L. and {Weaver}, B.~A. and {Whitmore}, J.~B. and {Woillez}, J. and {Zabalza}, V. and {Astropy Contributors}},
  title         = "{The Astropy Project: Building an Open-science Project and Status of the v2.0 Core Package}",
  journal       = {\aj},
  year          = 2018,
  month         = sep,
  volume        = {156},
  number        = {3},
  eid           = {123},
  pages         = {123},
  doi           = {10.3847/1538-3881/aabc4f},
  archiveprefix = {arXiv},
  eprint        = {1801.02634},
  primaryclass  = {astro-ph.IM},
  adsurl        = {https://ui.adsabs.harvard.edu/abs/2018AJ....156..123A}
}

@article{astropy:2022,
  author        = {{Astropy Collaboration} and {Price-Whelan}, Adrian M. and {Lim}, Pey Lian and {Earl}, Nicholas and {Starkman}, Nathaniel and {Bradley}, Larry and {Shupe}, David L. and {Patil}, Aarya A. and {Corrales}, Lia and {Brasseur}, C.~E. and {N{"o}the}, Maximilian and {Donath}, Axel and {Tollerud}, Erik and {Morris}, Brett M. and {Ginsburg}, Adam and {Vaher}, Eero and {Weaver}, Benjamin A. and {Tocknell}, James and {Jamieson}, William and {van Kerkwijk}, Marten H. and {Robitaille}, Thomas P. and {Merry}, Bruce and {Bachetti}, Matteo and {G{"u}nther}, H. Moritz and {Aldcroft}, Thomas L. and {Alvarado-Montes}, Jaime A. and {Archibald}, Anne M. and {B{'o}di}, Attila and {Bapat}, Shreyas and {Barentsen}, Geert and {Baz{'a}n}, Juanjo and {Biswas}, Manish and {Boquien}, M{'e}d{'e}ric and {Burke}, D.~J. and {Cara}, Daria and {Cara}, Mihai and {Conroy}, Kyle E. and {Conseil}, Simon and {Craig}, Matthew W. and {Cross}, Robert M. and {Cruz}, Kelle L. and {D'Eugenio}, Francesco and {Dencheva}, Nadia and {Devillepoix}, Hadrien A.~R. and {Dietrich}, J{"o}rg P. and {Eigenbrot}, Arthur Davis and {Erben}, Thomas and {Ferreira}, Leonardo and {Foreman-Mackey}, Daniel and {Fox}, Ryan and {Freij}, Nabil and {Garg}, Suyog and {Geda}, Robel and {Glattly}, Lauren and {Gondhalekar}, Yash and {Gordon}, Karl D. and {Grant}, David and {Greenfield}, Perry and {Groener}, Austen M. and {Guest}, Steve and {Gurovich}, Sebastian and {Handberg}, Rasmus and {Hart}, Akeem and {Hatfield-Dodds}, Zac and {Homeier}, Derek and {Hosseinzadeh}, Griffin and {Jenness}, Tim and {Jones}, Craig K. and {Joseph}, Prajwel and {Kalmbach}, J. Bryce and {Karamehmetoglu}, Emir and {Ka{l}uszy{'n}ski}, Miko{l}aj and {Kelley}, Michael S.~P. and {Kern}, Nicholas and {Kerzendorf}, Wolfgang E. and {Koch}, Eric W. and {Kulumani}, Shankar and {Lee}, Antony and {Ly}, Chun and {Ma}, Zhiyuan and {MacBride}, Conor and {Maljaars}, Jakob M. and {Muna}, Demitri and {Murphy}, N.~A. and {Norman}, Henrik and {O'Steen}, Richard and {Oman}, Kyle A. and {Pacifici}, Camilla and {Pascual}, Sergio and {Pascual-Granado}, J. and {Patil}, Rohit R. and {Perren}, Gabriel I. and {Pickering}, Timothy E. and {Rastogi}, Tanuj and {Roulston}, Benjamin R. and {Ryan}, Daniel F. and {Rykoff}, Eli S. and {Sabater}, Jose and {Sakurikar}, Parikshit and {Salgado}, Jes{'u}s and {Sanghi}, Aniket and {Saunders}, Nicholas and {Savchenko}, Volodymyr and {Schwardt}, Ludwig and {Seifert-Eckert}, Michael and {Shih}, Albert Y. and {Jain}, Anany Shrey and {Shukla}, Gyanendra and {Sick}, Jonathan and {Simpson}, Chris and {Singanamalla}, Sudheesh and {Singer}, Leo P. and {Singhal}, Jaladh and {Sinha}, Manodeep and {Sip{H{o}}cz}, Brigitta M. and {Spitler}, Lee R. and {Stansby}, David and {Streicher}, Ole and {{{S}}umak}, Jani and {Swinbank}, John D. and {Taranu}, Dan S. and {Tewary}, Nikita and {Tremblay}, Grant R. and {Val-Borro}, Miguel de and {Van Kooten}, Samuel J. and {Vasovi{'c}}, Zlatan and {Verma}, Shresth and {de Miranda Cardoso}, Jos{'e} Vin{'i}cius and {Williams}, Peter K.~G. and {Wilson}, Tom J. and {Winkel}, Benjamin and {Wood-Vasey}, W.~M. and {Xue}, Rui and {Yoachim}, Peter and {Zhang}, Chen and {Zonca}, Andrea and {Astropy Project Contributors}},
  title         = "{The Astropy Project: Sustaining and Growing a Community-oriented Open-source Project and the Latest Major Release (v5.0) of the Core Package}",
  journal       = {\apj},
  year          = 2022,
  month         = aug,
  volume        = {935},
  number        = {2},
  eid           = {167},
  pages         = {167},
  doi           = {10.3847/1538-4357/ac7c74},
  archiveprefix = {arXiv},
  eprint        = {2206.14220},
  primaryclass  = {astro-ph.IM},
  adsurl        = {https://ui.adsabs.harvard.edu/abs/2022ApJ...935..167A}
}

@article{2007CSE.....9c..21P,
  author        = {{Perez}, Fernando and {Granger}, Brian E.},
  title         = "{IPython: A System for Interactive Scientific Computing}",
  journal       = {Computing in Science and Engineering},
  year          = "2007",
  month         = "Jan",
  volume        = {9},
  number        = {3},
  pages         = {21--29},
  doi           = {10.1109/MCSE.2007.53},
  adsurl        = {https://ui.adsabs.harvard.edu/abs/2007CSE.....9c..21P}
}

@conference{kluyver2016jupyter,
  title         = {Jupyter Notebooks -- a publishing format for reproducible computational workflows},
  author        = {Thomas Kluyver and Benjamin Ragan-Kelley and Fernando P{\'e}rez and Brian Granger and Matthias Bussonnier and Jonathan Frederic and Kyle Kelley and Jessica Hamrick and Jason Grout and Sylvain Corlay and Paul Ivanov and Dami{\'a}n Avila and Safia Abdalla and Carol Willing},
  booktitle     = {Positioning and Power in Academic Publishing: Players, Agents and Agendas},
  editor        = {F. Loizides and B. Schmidt},
  organization  = {IOS Press},
  pages         = {87--90},
  year          = {2016}
}

@article{Hunter:2007,
  author        = {Hunter, J. D.},
  title         = {Matplotlib: A 2D graphics environment},
  journal       = {Computing in Science \& Engineering},
  volume        = {9},
  number        = {3},
  pages         = {90--95},
  publisher     = {IEEE COMPUTER SOC},
  doi           = {10.1109/MCSE.2007.55},
  year          = 2007
}

@article{numpy,
  title         = {Array programming with {NumPy}},
  author        = {Charles R. Harris and K. Jarrod Millman and St{\'{e}}fan J. van der Walt and Ralf Gommers and Pauli Virtanen and David Cournapeau and Eric Wieser and Julian Taylor and Sebastian Berg and Nathaniel J. Smith and Robert Kern and Matti Picus and Stephan Hoyer and Marten H. van Kerkwijk and Matthew Brett and Allan Haldane and Jaime Fern{\'{a}}ndez del R{\'{i}}o and Mark Wiebe and Pearu Peterson and Pierre G{\'{e}}rard-Marchant and Kevin Sheppard and Tyler Reddy and Warren Weckesser and Hameer Abbasi and Christoph Gohlke and Travis E. Oliphant},
  year          = {2020},
  month         = sep,
  journal       = {Nature},
  volume        = {585},
  number        = {7825},
  pages         = {357--362},
  doi           = {10.1038/s41586-020-2649-2},
  publisher     = {Springer Science and Business Media {LLC}},
  url           = {https://doi.org/10.1038/s41586-020-2649-2}
}

@book{python,
  author        = {Van Rossum, Guido and Drake, Fred L.},
  title         = {Python 3 Reference Manual},
  year          = {2009},
  isbn          = {1441412697},
  publisher     = {CreateSpace},
  address       = {Scotts Valley, CA}
}

@article{software-citation-station-paper,
  author        = {{Wagg}, Tom and {Broekgaarden}, Floor S.},
  title         = "{Streamlining and standardizing software citations with The Software Citation Station}",
  journal       = {arXiv e-prints},
  year          = 2024,
  month         = jun,
  eid           = {arXiv:2406.04405},
  pages         = {arXiv:2406.04405},
  archiveprefix = {arXiv},
  eprint        = {2406.04405},
  primaryclass  = {astro-ph.IM},
  adsurl        = {https://ui.adsabs.harvard.edu/abs/2024arXiv240604405W}
}

@misc{software-citation-station-zenodo,
  author       = {Tom Wagg and
                  Floor Broekgaarden and
                  Phil Van-Lane and
                  Kai Wu and
                  Kayhan Gültekin},
  title        = {TomWagg/software-citation-station: v1.4},
  month        = nov,
  year         = 2025,
  publisher    = {Zenodo},
  version      = {v1.4},
  doi          = {10.5281/zenodo.17654855},
  url          = {https://doi.org/10.5281/zenodo.17654855},
  swhid        = {swh:1:dir:9a009430037c791424a572f542e9a5d5c1fb44ff
                   ;origin=https://doi.org/10.5281/zenodo.13225526;vi
                   sit=swh:1:snp:ef11f058d718d691f0661c9445c2251328cb
                   ac95;anchor=swh:1:rel:84cde4e532032537b7f014c4467c
                   102430a53fa2;path=TomWagg-software-citation-
                   station-61a588a
                  },
}

@ARTICLE{Dokkum1996,
       author = {{van Dokkum}, P.~G. and {Franx}, M.},
        title = "{The Fundamental Plane in CL 0024 at z = 0.4: implications for the evolution of the mass-to-light ratio.}",
      journal = {\mnras},
         year = 1996,
        month = aug,
       volume = {281},
       number = {3},
        pages = {985-1000},
          doi = {10.1093/mnras/281.3.985},
archivePrefix = {arXiv},
       eprint = {astro-ph/9603063},
 primaryClass = {astro-ph},
       adsurl = {https://ui.adsabs.harvard.edu/abs/1996MNRAS.281..985V}
}

@ARTICLE{Dokkum2001,
       author = {{van Dokkum}, Pieter G. and {Franx}, Marijn},
        title = "{Morphological Evolution and the Ages of Early-Type Galaxies in Clusters}",
      journal = {\apj},
         year = 2001,
        month = may,
       volume = {553},
       number = {1},
        pages = {90-102},
          doi = {10.1086/320645},
archivePrefix = {arXiv},
       eprint = {astro-ph/0501236},
 primaryClass = {astro-ph},
       adsurl = {https://ui.adsabs.harvard.edu/abs/2001ApJ...553...90V}
}

\begin{appendix}

\section{Comparison between instantaneous and time-averaged circularity}
\label{instantaneous_time-averagedCircularity}

\begin{figure*}[t]
     \sidecaption
     \includegraphics[width=12cm]{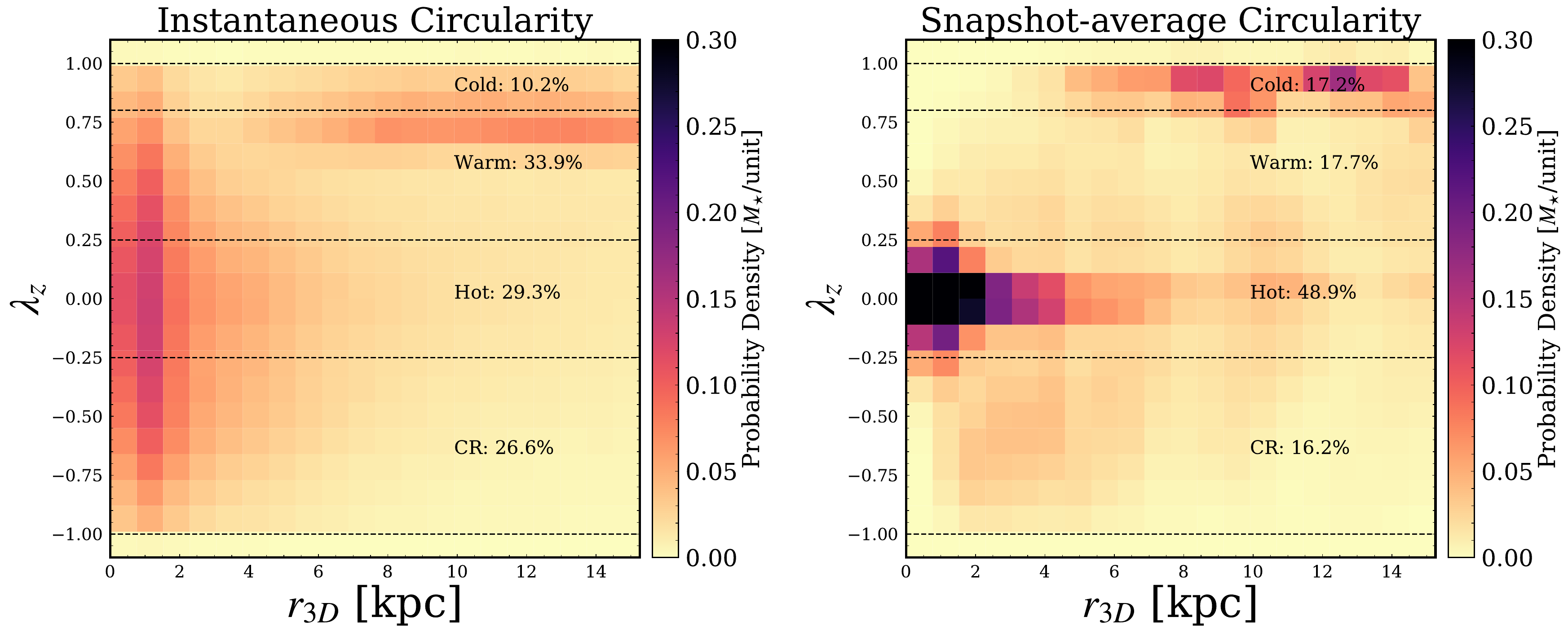}
     \caption{Comparison of stellar orbital distributions for \texttt{SubhaloID: 468590} from the \texttt{TNG50-1} simulation at $z = 0$: instantaneous circularity (left), computed from a single snapshot, and time-averaged circularity (right), computed as the average over 11 snapshots. Colours indicate the probability density of stellar mass in $(r_{3D}, \overline{\lambda_z})$ space. The fraction of each orbital component (cold, warm, hot, and CR) is shown in its respective region for both approaches.}
     \label{fig:lambda_z_instan_time_averaged}
\end{figure*}

Figure \ref{fig:lambda_z_instan_time_averaged} compares the stellar orbit distribution of a subhalo (\texttt{SubhaloID: 468590}) from \texttt{TNG50-1} at ( z = 0 ), obtained using instantaneous circularity (left) and time-averaged circularity (right). As explained in Sect. \ref{sec:orbitaldecomposition}, the time-averaged circularity is computed by averaging over 11 snapshots, from \texttt{SnapNum = 99} to \texttt{SnapNum = 89}. The differences between the two approaches are evident in the fractional contributions of the orbital components. In the instantaneous case, box orbits are spread across the $r_{3D} - \overline{\lambda_z}_{z}$ space, whereas in the time-averaged case, they are concentrated near $\overline{\lambda_z}_{z} \sim 0 $. As a result, the fraction of box orbits increases from 29\% in the instantaneous circularity to 48.9\% in the snapshot-averaged circularity, due to the reassignment of box orbits to their corrected orbital component.

\section{Comparison between snapshot and phase-space averaging}
\label{ComparisonPhase-Space}

In this work, the averaged orbital circularities are computed using snapshot-based averaging, as described in Section \ref{sec:orbitaldecomposition}. Some previous studies using the \texttt{IllustrisTNG} simulations have instead adopted a phase-space averaging approach to calculate circularities \citep{Xu2019MNRAS, Zhang2025}. To assess the sensitivity of our results to the adopted averaging scheme, we perform an additional check using phase-space averaging and compare the results to our fiducial snapshot-averaging.

For the phase-space averaging, the circularity of each stellar particle is defined as $\overline{\lambda_z} \equiv \frac{L_z}{J_c(E)},$ where $L_z$ is the specific angular momentum along the galaxy’s symmetry axis and $J_c(E)$ is the maximum specific angular momentum of a circular orbit at the same binding energy $E$. The binding energy of each particle is computed as the sum of its kinetic and gravitational potential energy. To estimate $J_c(E)$, all stellar particles are first ranked according to their binding energy. For a given particle, we then identified its 200 nearest neighbors in energy-momentum space and adopted the maximum absolute value of $L_z$ among these neighbors as $J_c(E)$. This procedure closely follows the methodology used in earlier studies and provides a robust estimate of the circular orbit angular momentum as a function of energy, we refer \cite{Xu2019MNRAS} for more details.

\begin{figure*}[t]
    \sidecaption
    \includegraphics[width=12cm]{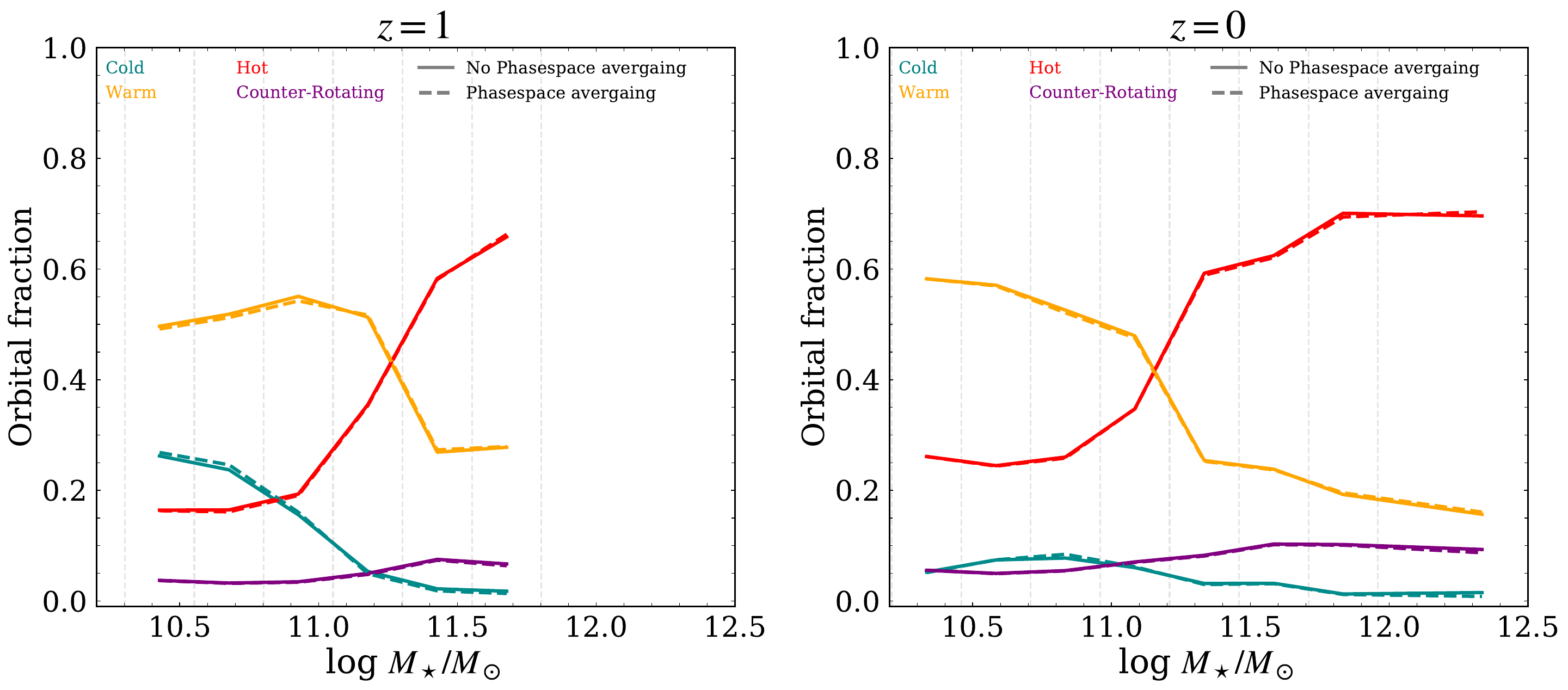}
    \caption{Comparison of orbital fractions derived with and without applying phase-space averaging prior to snapshot averaging. Shown are the median fractions of the cold, warm, hot, and CR components as a function of stellar mass at $z=1$ (top panels) and $z=0$ (bottom panels). Solid lines correspond to snapshot averaging without phase-space averaging, identical to Fig.~\ref{fig:fraction_plot_z1_z0}, while dashed lines show results obtained using snapshot averaging with phase-space-averaged circularities. The close agreement between the two sets of curves demonstrates that our results are insensitive to the adopted averaging scheme.}
    \label{fig:with_without_phasespace_averaging}
\end{figure*}

Using these phase-space-averaged circularities, we then applied the snapshot-averaging and recompute the orbital fractions corresponding to the cold, warm, hot, and CR components. Figure \ref{fig:with_without_phasespace_averaging} compares the resulting orbital fractions as a function of stellar mass at $z=1$ and $z=0$ for both snapshot averaging without  phase-space averaging (solid lines) and snapshot averaging with phase-space averaging (dashed lines).

Overall, we find excellent agreement between the two averaging schemes. The orbital fractions derived from phase-space averaging closely track those obtained from snapshot averaging across the full stellar mass range and at both redshifts considered. Minor differences are visible at the highest stellar masses, but these remain well within the intrinsic scatter and do not affect any of our qualitative or quantitative conclusions.

\section{Mergers: Major and minor}
\label{MajorMinorMerger}

\begin{figure*}[t]
\centering
    \includegraphics[width=1\textwidth]{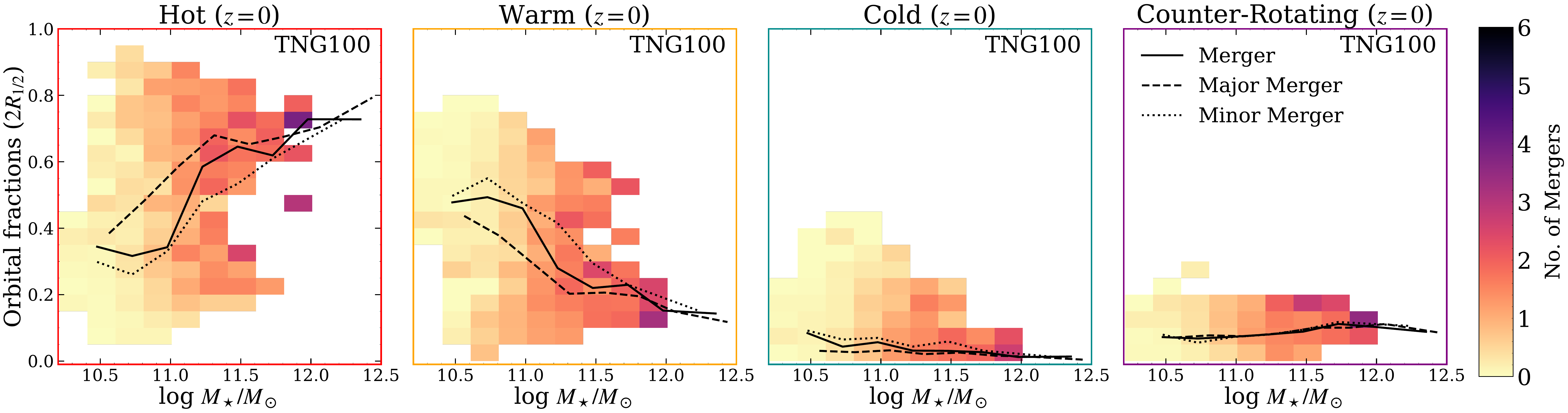}
    \caption{Orbital fractions measured within $2R_{1/2}$ as a function of stellar mass for galaxies at $z=0$. From left to right: Hot, warm, cold, and CR orbital components. The background 2D histograms are color-coded by the number of mergers in each stellar mass--orbital fraction bin, illustrating that more massive galaxies generally experience a larger number of mergers than lower-mass systems. Overplotted lines show the median trends for all mergers (solid), major mergers (dashed), and minor mergers (dotted). Major mergers have the strongest impact on the orbital structure, significantly increasing the hot orbital fraction while suppressing the warm and cold components. In contrast, minor mergers produce comparatively weaker changes. The CR component remains small across all stellar masses, with only a mild enhancement in galaxies that experience mergers.}

     \label{fig:MergersMajorMinor}
\end{figure*}

Figure~\ref{fig:MergersMajorMinor} shows the impact of major and minor mergers on the orbital components of the \texttt{merger-rich} galaxy sample at two epochs, $z=1$ (top panels) and $z=0$ (bottom panels). As in Figure~\ref{fig:Combined_mergers_comparison}, the orbital fractions are shown as a function of stellar mass, but the merger-rich population is now separated into major and minor merger contributions.

At both redshifts, major mergers have a stronger influence on the orbital structure than minor mergers across the full stellar mass range. At fixed stellar mass, major mergers are more efficient at increasing the hot and CR orbital fractions, while simultaneously reducing the cold and warm components. Minor mergers follow the same qualitative trends but with systematically weaker effects.

We note that a single galaxy can experience both major and minor mergers and therefore may contribute to both samples. We also tested exclusive merger samples containing only major or only minor mergers. While this enhances the contrast between the two populations, the overall trends with stellar mass and redshift remain unchanged.

\section{Orbital evolution: Stellar mass and scatter in orbital fraction}
\label{OrbitalEvolution_Stellar mass_Sigma}

\begin{figure*}[t]
    \centering
    \includegraphics[width=1\textwidth]{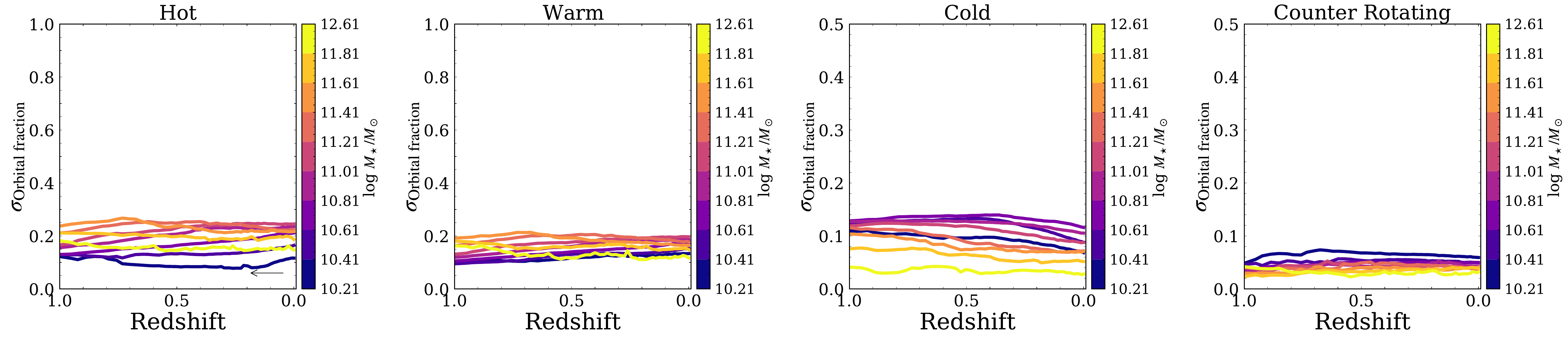}
    \caption{$1\sigma$ scatter in the fraction of the hot, warm, cold, and CR orbital components within each stellar-mass bin, corresponding to the orbital decomposition shown in Fig.~\ref{fig:orbital_evolution}.}
    \label{fig:orbital_evolution_std}
\end{figure*}

Figure~\ref{fig:orbital_evolution_std} shows the redshift evolution of the $1\sigma$ scatter in the fractions of hot, warm, cold, and CR orbits for galaxies binned by their descendant stellar mass at $z=0$. The corresponding median values are shown in Fig.~\ref{fig:orbital_evolution}. Across all redshifts, the scatter remains broadly constant, with low-mass (high-mass) bins exhibiting smaller (larger) scatter in the hot and warm (cold and CR) orbital components. Similarly, Fig.~\ref{fig:orbital_evolution} shows the evolution of the stellar mass of galaxies in each bin, where the solid lines represent the median values and the shaded regions indicate the $1\sigma$ scatter.

\begin{figure}[t]
    \centering
    \includegraphics[width=0.5\textwidth]{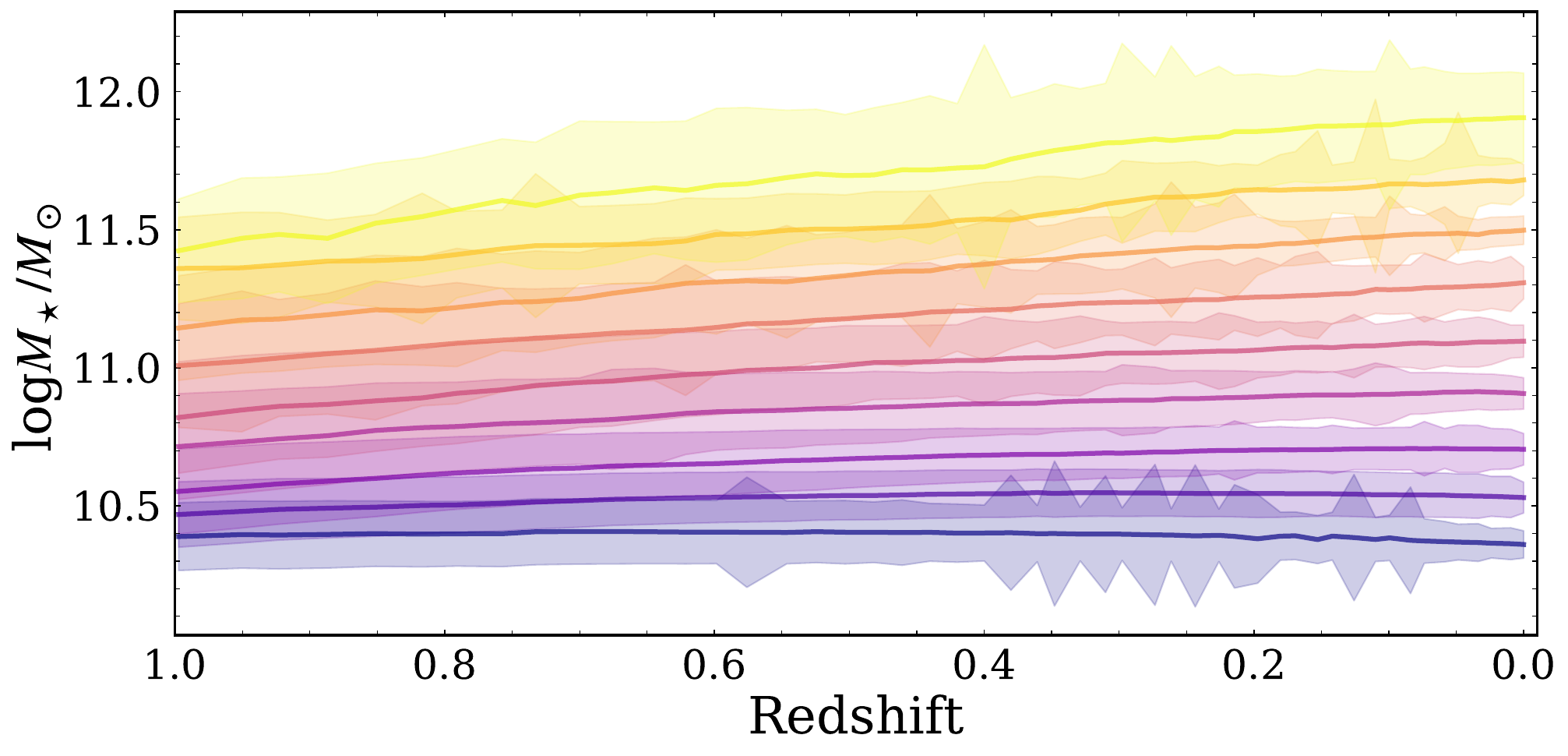}
    \caption{Evolution of the median stellar mass of galaxies as a function of redshift. Galaxies are binned by their stellar mass at $z=0$ and the shaded regions represent the $1\sigma$ scatter within each mass bin for the plot shown in Fig.~\ref{fig:orbital_evolution}.}
    \label{fig:Evolution_mass}
\end{figure}

\begin{figure}[t]
    \centering
    \includegraphics[width=0.5\textwidth]{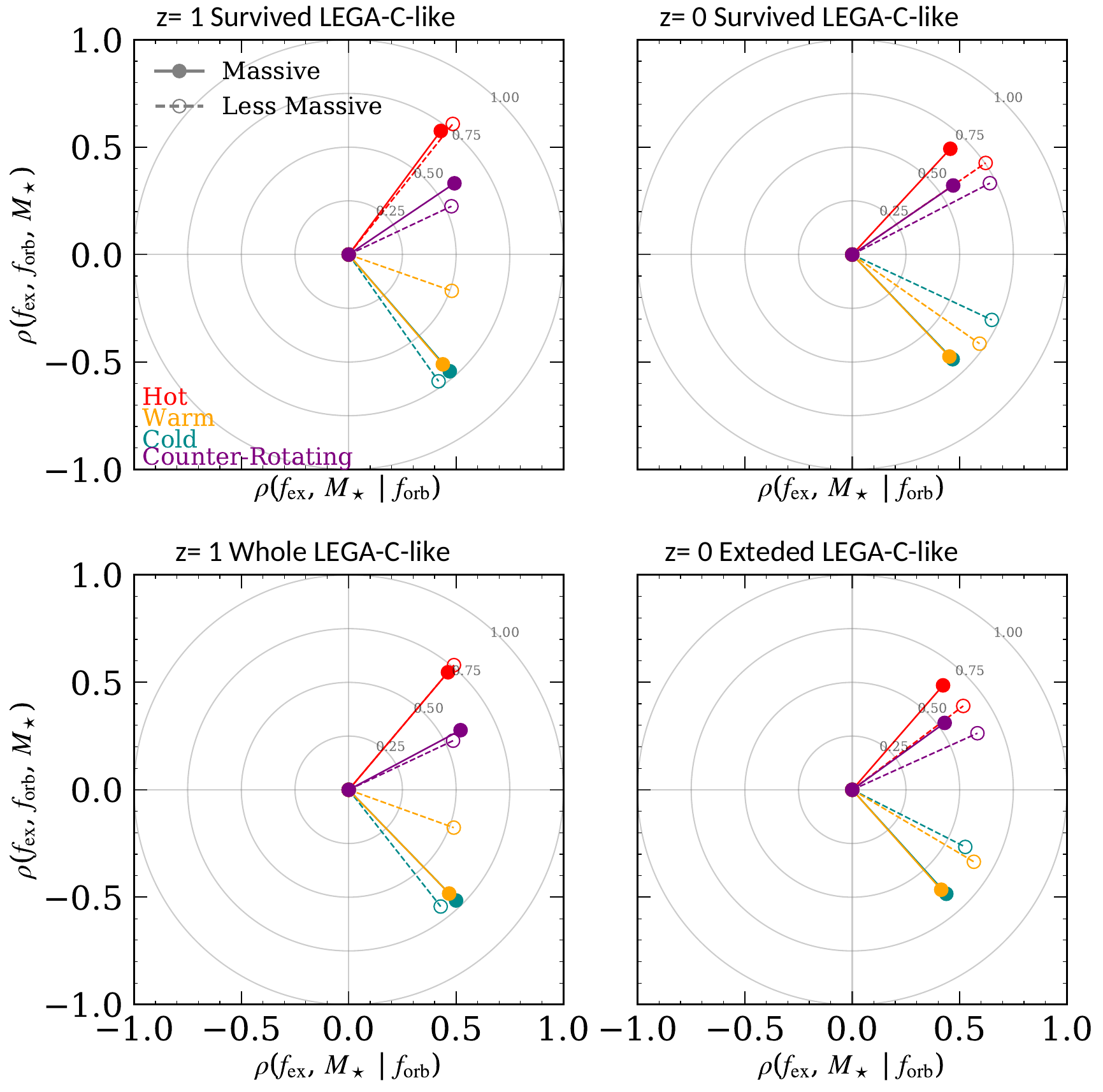}
    \caption{Polar diagrams of the partial correlation coefficients 
    (PCCs) between the ex situ stellar mass fraction, 
    $f_{\rm ex\text{-}situ}$, the orbital fractions, $f_{\rm orb}$, 
    and stellar mass, $M_\star$, shown separately for the 
    \texttt{survived LEGA-C-like} sample (top row) and 
    the \texttt{extended LEGA-C-like} sample bottom row), 
    at $z = 1$ (left) and $z = 0$ (right). In each 
    panel, filled circles connected by solid lines denote the massive 
    subsample ($\log(M_\star/M_\odot) \geq 11$), while open circles 
    connected by dashed lines denote the less massive subsample 
    ($\log(M_\star/M_\odot) < 11$). The horizontal axis shows 
    $\rho_{f_{\rm ex\text{-}situ},\,M_\star\,|\,f_{\rm orb}}$ and the 
    vertical axis shows 
    $\rho_{f_{\rm ex\text{-}situ},\,f_{\rm orb}\,|\,M_\star}$, for 
    the hot (red), warm (orange), cold (teal), and CR 
    (purple) orbital components. The radial distance from the origin 
    encodes the amplitude of the combined partial correlation. Vectors 
    pointing rightward reflect a strong residual dependence of 
    $f_{\rm ex\text{-}situ}$ on $M_\star$, while the vertical 
    component captures the intrinsic correlation between 
    $f_{\rm ex\text{-}situ}$ and the orbital fraction at fixed stellar 
    mass.}
    \label{fig:pcc_polar}
\end{figure}

\section{Mass-dependent partial correlations between ex situ fraction 
and orbital components}
\label{appendix:pcc_mass_split}

In Section~\ref{sec:ExsituMass}, we presented the partial correlation 
coefficients (PCCs) between the ex situ stellar mass fraction, 
$f_{\rm ex\text{-}situ}$, the orbital fractions, $f_{\rm orb}$, and 
stellar mass, $M_\star$, for the full \texttt{survived LEGA-C-like} and 
\texttt{extended LEGA-C-like} samples. However, galaxies above and below 
$\log(M_\star/M_\odot) = 11$ display qualitatively different behaviour, 
owing to the steep increase in $f_{\rm ex\text{-}situ}$ above this mass. 
Here, we present a more detailed breakdown of the PCC analysis by 
splitting each sample into a massive subsample 
($\log(M_\star/M_\odot) \geq 11$, filled circles) and a less massive 
subsample ($\log(M_\star/M_\odot) < 11$, open circles), shown as polar 
diagrams in Fig.~\ref{fig:pcc_polar}.

In each panel, the radial coordinate measures the amplitude of the 
partial correlation, while the angular position encodes its direction. 
The horizontal axis corresponds to 
$\rho_{f_{\rm ex\text{-}situ},\,M_\star\,|\,f_{\rm orb}}$, the partial 
correlation between $f_{\rm ex\text{-}situ}$ and $M_\star$ controlling 
for $f_{\rm orb}$, and the vertical axis corresponds to 
$\rho_{f_{\rm ex\text{-}situ},\,f_{\rm orb}\,|\,M_\star}$, the partial 
correlation between $f_{\rm ex\text{-}situ}$ and $f_{\rm orb}$ 
controlling for $M_\star$. Vectors pointing toward the upper right 
therefore indicate that $f_{\rm ex\text{-}situ}$ correlates positively 
with both $M_\star$ and the orbital fraction in question, whereas 
vectors pointing toward the lower right indicate that the ex situ 
fraction correlates positively with stellar mass but negatively with 
that orbital component at fixed mass.

At both $z=1$ and $z=0$, the hot (and, to a lesser extent, CR) orbital 
fraction shows the strongest positive correlation with 
$f_{\rm ex\text{-}situ}$ across all samples and mass ranges, whereas 
the cold and warm fractions are anti-correlated with 
$f_{\rm ex\text{-}situ}$ in every case. Galaxies with larger accreted 
stellar mass fractions therefore systematically contain larger 
fractions of dynamically hot orbits, even after the primary dependence 
on stellar mass is removed.

Comparing the two mass regimes, the massive subsample generally shows 
stronger positive correlations between the hot (and CR) fractions and 
$f_{\rm ex\text{-}situ}$ than the less massive subsample at both 
redshifts. Conversely, the anti-correlation of the cold fraction with 
$f_{\rm ex\text{-}situ}$ is typically stronger in the less massive 
subsample, particularly at $z=0$. At $z=1$, this difference is 
marginal in both the \texttt{survived} and \texttt{whole} samples, 
with the massive subsample showing a slightly stronger 
anti-correlation. The warm fraction, in contrast, is more strongly 
anti-correlated with $f_{\rm ex\text{-}situ}$ in the massive subsample.
\end{appendix}
\end{document}